\documentclass{aa}  
\usepackage[flushleft]{threeparttable}
\usepackage{float}
\usepackage{graphics}
\usepackage{graphicx}
\usepackage{amssymb, amsmath}
\usepackage{tikz}
\usetikzlibrary{graphs, positioning, quotes, shapes.geometric}
\usepackage{amstext}
\usepackage[T1]{fontenc}
\usepackage{mathrsfs}
\usepackage[normalem]{ulem}
\usepackage{natbib,twoopt}
\usepackage{hyperref} 
\usepackage{soul,xcolor}
\usepackage{booktabs}
\usepackage{rotating}
\usepackage{txfonts}
\makeatletter
\def\@to{to}
\def\as     {\ifmmode {\rlap.}$\,$''$\,$\! \else ${\rlap.}$\,$''$\,$\!$\fi}
     \def\decsec  {\ifmmode {\rlap.}$\,$^{\rm s}$\,$\! \else ${\rlap.}$\,$^{\rm s}$\,$\!$\fi}\def\decss  {\ifmmode {\rlap.}$\,$^{\rm s}$\,$\! \else ${\rlap.}$\,$^{\rm s}$\,$\!$\fi}
\makeatother

\begin{document}

   \title{Initial conditions of star formation at $\lesssim$2000\,au: physical structure and NH$_{3}$ depletion of three early-stage cores}


   \author{Y. Lin         
          \inst{1}
                          \and S. Spezzano
   \inst{1}
\and J. E. Pineda
          \inst{1}
          \and J. Harju
          \inst{2}
          \and A. Schmiedeke
          \inst{3}
          \and S. Jiao
          \inst{4}
          \and H. B. Liu
          \inst{5}
          \and P. Caselli\inst{1}}

   \institute{Max-Planck-Institut f{\"u}r Extraterrestrische Physik, Giessenbachstr. 1, D-85748 Garching bei M{\"u}nchen, Germany\\
              \email{ylin@mpe.mpg.de}
         \and
          Department of Physics, P.O. Box 64, FI-00014, University of Helsinki, Finland
          \and 
          Green Bank Observatory, PO Box 2, Green Bank, WV 24944, USA          
          \and
          National Astronomical Observatories, Chinese Academy of Sciences, Beijing 100101, China
          \and
          Department of Physics, National Sun Yat-Sen University, No. 70, Lien-Hai Road, Kaohsiung City 80424, Taiwan, R.O.C.
             }

   \date{Received ; accepted }

 
  \abstract
{Pre-stellar cores represent a critical evolutionary phase in low-mass star formation. Characterisations of the physical conditions of pre-stellar cores provide important constraints to star and planet formation theory, and are pre-requisite for establishing the dynamical evolution and the related chemical processes.}
{We aim to unveil the detailed thermal structure and density distribution of three early-stage cores, starless core L1517B, and prestellar core L694-2 and L429, with the high angular resolution observations of the NH$_{3}$ (1,1) and (2,2) inversion transitions obtained with VLA and GBT. In addition, we explore where/if NH$_{3}$ depletes in the central regions of the cores.}
{We calculate the physical parameter maps of gas kinetic temperature, NH$_{3}$ column density, line-width and centroid velocity of the three cores, utilising the NH$_{3}$(1,1) and (2,2) lines.
We apply the mid-infrared extinction method to the {\it{Spitzer}} 8$\,\mu$m map to obtain a high angular resolution hydrogen column density map. We examine the correlation between the derived parameters, and the properties of individual cores. We derive the gas density profile from the column density maps and assess the variation of NH$_{3}$ abundance as a function of gas volume density.} 
{The measured temperature profiles of the cores L429 and L1517B show a minor decrease towards the core center, dropping from $\sim$9\,K to below 8\,K, and $\sim$11 K to 10 K, while L694-2 has a rather uniform temperature distribution around $\sim$9 K. Among the three cores, L429 has the highest central gas density, close to sonic velocity line-width, and largest localised velocity gradient, all indicative of an advanced evolutionary stage. We resolve that the abundance of NH$_{3}$ becomes 2 times lower in the central region of L429, occurring around a gas density of 4.4$\times$10$^{4}$\,cm$^{-3}$. Compared to Ophiuchus/H-MM1 (\citealt{Pineda22a}) which shows an even stronger drop of the NH$_{3}$ abundance at 2$\times$10$^{5}$\,cm$^{-3}$, the abundance variations of the three cores plus Ophiuchus/H-MM1 suggest a progressive NH$_{3}$ depletion with increasing central density in pre-stellar cores.} 
{}
\keywords{ISM: pre-stellar core -- ISM: L429, L694-2, L1517B -- ISM: structure -- stars: formation}
\maketitle
%

\section{Introduction}


Pre-stellar cores represent a critical stage in the process of low-mass star formation: the molecular gas has reached adequate density for self-gravity to balance or even surpass the outward forces (thermal 
and turbulent pressure, rotation and magnetic field, 
 see e.g. \citealt{MB83}, \citealt{BT07}, \citealt{Pineda22b}). Compared to its prior stage of starless cores, these cores are denser and at the verge of forming protostars. The significance of studying the physical conditions of pre-stellar cores is twofold: (1) it sheds light on the physical mechanisms at play in the imminence of star formation (\citealt{KC10}, \citealt{Keto15}), (2) it provides essential constraints on understanding the chemical processes that influence the properties of gas and dust at a critical phase of interstellar medium evolution (\citealt{CC12}).

Along with the special physical status, pre-stellar cores are characterised by strong molecular depletion including accretion/freezing onto dust grains, and appear chemically distinctive to the preceding and more evolved core phases. The strong molecular freeze-out enhances the deuterium fractionation, e.g., deuterated isotopologues of NH$_{3}$ can form through reactions of NH$_{3}$ with deuterated ions in the gas phase (\citealt{RC01}, \citealt{Roueff05}, \citealt{Sipila15}), the process of which also affects the abundance of NH$_{3}$, composing another source of NH$_{3}$ depletion. However, among various molecules, nitrogen-bearing species (e.g., N$_{2}$H$^{+}$, NH$_{3}$) show relative longevity in the gas (\citealt{Caselli99}, \citealt{Caselli02b}, \citealt{Aikawa05}, \citealt{Flower06}, \citealt{BT07}, \citealt{Sipila15}) and have been important molecular tracers of the inner core region 
\citep[e.g.,][]{Bergin06,Friesen09,Pineda10, Pineda11,Chit14,Pineda15}.
%
While previous observations show marginal evidence that ammonia depletion happens at very high gas densities (e.g., \citealt{Tafalla04}, \citealt{Crapsi05}), recent work based on the deep interferometric observations towards pre-stellar core Ophiuchus/H-MM1 (hereafter H-MM1) uncover direct evidence that NH$_{3}$ depletes already at a few times 10$^{5}\,$cm$^{-3}$ (\citealt{Pineda22a}); this showcases the necessity of high angular observations for revealing the depletion of NH$_{3}$ and further elucidating the chemical properties related to NH$_{3}$ formation, e.g., the volatility of atomic and molecular nitrogen. 

The inversion transitions of NH$_{3}$ in metastable rotational levels are an important thermometer for dense gas (\citealt{HT83}, \citealt{WU83}), making ammonia a valuable tracer for molecular clouds. Towards low-mass cores in particular,  NH$_{3}$ is a reliable tracer of line-of-sight mass averaged temperature (\citealt{Juvela12}), when the cores can be approximated by a $\sim$1\,$M_{\odot}$ Bonnor-Ebert sphere without being too opaque. Due to photo-electric heating, high-density ($\sim$10$^{5}\,$cm$^{-3}$) starless cores immersed in interstellar radiation field are predicted to exhibit a gas temperature increase toward the core edge (\citealt{Galli02}).
Based on measurement of NH$_{3}$ lines, while some starless cores have a rather constant gas temperature profile (variations of $\lesssim$1 K, \citealt{Tafalla04}, \citealt{R11}, \citealt{Chit14}, \citealt{Spear21}), centrally decreasing temperature profile has been revealed towards a handful of starless and pre-stellar cores (\citealt{Hotzel01}, \citealt{Crapsi07}, \citealt{PB07}, \citealt{Harju17}, \citealt{Pineda22a}). In particular, \citealt{Crapsi07} unveil a remarkable temperature drop down to $\sim$6 K inside a late-stage pre-stellar core L1544. This deviation from isothermality and the corresponding varying core density conform with an evolutionary view for starless cores immediately prior to protostar formation (\citealt{EW01}, \citealt{Zucconi01}, \citealt{KetoField05}, \citealt{KetoCaselli08}).

In this work, we utilise high angular resolution NH$_{3}$ (1,1) and (2,2) observations of one starless core L1517B and two pre-stellar cores L694-2 and L429, to investigate the physical conditions and the variations of NH$_{3}$ abundance in pre-stellar cores. 
We adopt mid-infrared extinction methods to characterise the density structure of the three cores, achieving a similar angular resolution of 5$''$ with the NH$_{3}$ observations to allow a direct comparison. This enables us to provide a detailed picture of temperature and density distribution, and the NH$_{3}$ abundance mapping of the three cores. 

In Sect. \ref{sec:obs} we describe the observations, data reduction and combination procedure. Sect. \ref{sec:extinction}
-\ref{sec:nprofiles} detail the calculation of hydrogen column density maps with extinction methods, and the derived radial density profiles. In Sect. \ref{sec:nh3_model}-\ref{sec:paramaps} we elaborate on the fitting procedure of the NH$_{3}$ lines and present the obtained parameter maps. In Sect. \ref{sec:dis} we compare the derived physical parameters of the three cores, focusing primarily on the thermal structure and the NH$_{3}$ abundance variations.  

\section{Observations and data reduction}\label{sec:obs}
The $K$-band, single-pointing NH$_{3}$ (1,1) and (2,2) image cubes of the three cores (Table \ref{tab:ss}, more in Appendix A) were taken with the Jansky Very Large Array (JVLA) in 2013 (Project ID: 13A-394, PI: S. Chitsazzadeh). The JVLA correlator was configured to
use the 8-bit sampler and two basebands, of 8 MHz bandwidth and 2048 channels in dual polarisation mode, achieving a velocity resolution of 0.05 km\,s$^{-1}$. The quasars 3C286 and 3C48 are used as the flux and passband calibrators, and J1743-0350, J1925+2106, J0414+3418 as phase calibrators for the target sources L429, L694-2 and L1517B, respectively.
The raw data was calibrated with the pipeline of Common Astronomy Software Applications (CASA) version 6.2.1.
Single-dish data was obtained additionally from the Green Bank Telescope (GBT) using the $K$-band Focal Plane Array (KFPA), achieving an angular resolution of $\sim$33$''$ (Project ID: GBT10B-020, GBT10C-055, GBT11A-052, PI: S. Chitsazzadeh). The raw data was reduced with the GBT pipeline. 

We utilise a hybrid method for combining the VLA and GBT data, following the procedure elaborated in \citet{Liu15} {\footnote{\url{https://github.com/baobabyoo/almica}}} and used by e.g., \citet{Monsch18}, \citet{Lin22}, which is based on imaging tasks provided in the {\tt{miriad}} software (\citealt{Sault95}). The procedure consists of two essential parts: generating pseudo-visibility from the single-dish data with {\tt{uvrandom}} and {\tt{uvmodel}} tasks, and joint deconvolution of the pseudo visibitility with the interferometric visibilities to obtain a clean datacube; the datacube is further merged with the single-dish data with {\tt{immerge}} task, to preserve the total flux. The following analysis and results are based on the final datacubes produced by this combination method. With the {\tt{invert}} task in joint deconvolution step, we apply a taper function to obtain a gaussian beam of $\sim$5$''$ for the data cubes, which is preserved after {\tt{immerge}}.       

We additionally adopt the combination method elaborated in \citet{Pineda22a} which is essentially a model-assisted deconvolution procedure of the interferometric data. 
The Common Astronomy Software Applications (CASA) software is used for this approach. Specifically, the single-dish data is used as a {\tt{startmodel}} in {\tt{tclean}} to facilitate image reconstruction of the VLA data. This combination method essentially adopts the single-dish image as prior information to model the missing extended emission solely from the interferometry-only constraints. We use multi-scale deconvolution algorithm with natural weighting and a common restoring beam of 5$''$ after applying {\tt{uvtaper}}. The {\tt{scales}} parameter is set to be 0$''$, 5$''$, 15$''$ and 45$''$, reflecting the typical size scales of dominant image features. As a final step, we use {\tt{feather}} to add this model-assisted deconvolved datacube with the single-dish data, obtaining the final data products.   

Both methods yield a final datacube regridded to a channel width of 0.1 km\,s$^{-1}$. The achieved noise level is $\sim$3 mJy beam$^{-1}$ per channel. We compare the obtained NH$_{3}$ column density of the two data products in Appendix \ref{app:comb_nh3_app}, which shows that the analysis is not biased by the different combination methods, for a robust comparison with the results in \citet{Pineda22a}.

\begin{table}[htb]
\centering
\begin{threeparttable}
\footnotesize
\caption{The observed cores.}
 \label{tab:ss}
 \begin{tabular}{llllll}
\toprule
Source  & R.A. & Dec. &$\rm{v}_{\mathrm{LSR}}$&Distance$^{a}$\\
   & (J2000)&(J2000)&(km\,s$^{-1}$)&(pc)\\
\midrule
L694-2&19:41:04.5&$+$10:57:02.0&9.5&203 \\
L429&18:17:05.8&$-$08:14:05.3&6.7&436\\
L1517B&04:55:18.3&$+$30:37:47.0&5.8&159\\

\bottomrule
\end{tabular}
 \begin{tablenotes}
      \small
      \item $^{a}$ The distance references are \citet{Kim22}, \citet{OL18}, and \citet{Galli19}. 
      \end{tablenotes}
  \end{threeparttable}
\end{table}

\section{Results}

\begin{figure*}[htb]
\centering
\hspace{-.5cm}
\includegraphics[height=5cm]{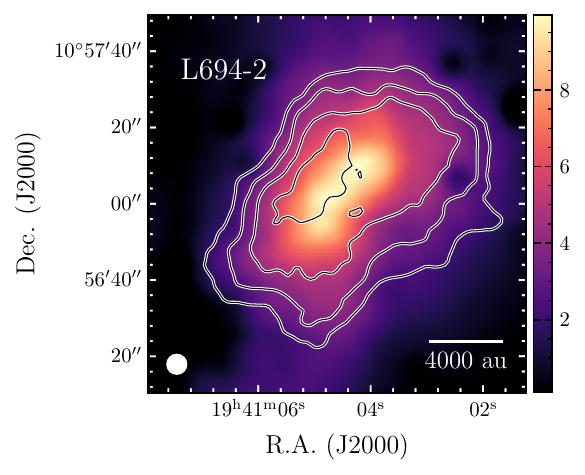}
\includegraphics[height=5cm]{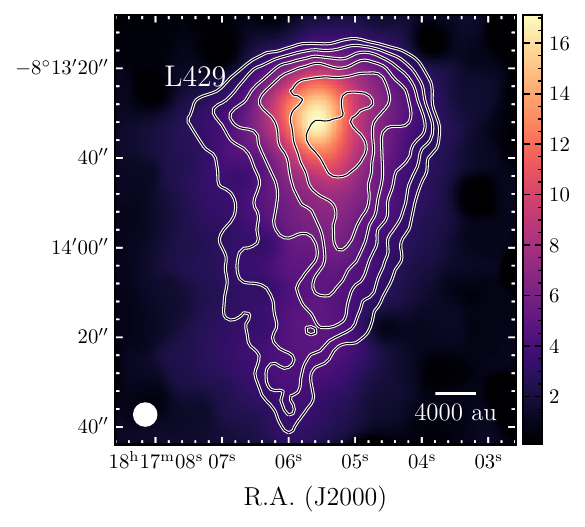}
\includegraphics[height=5cm]{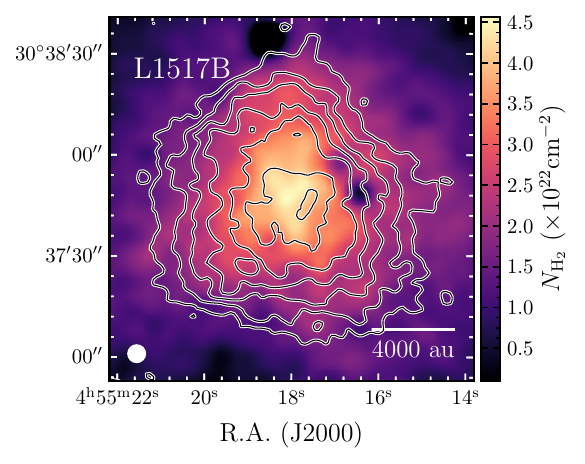}
\caption{Hydrogen column density maps ($N(\mathrm{H_{2}})$) derived from {\it{Spitzer}} 8\,$\mu m$ extinction. The contours show the integrated intensity map of the main component of NH$_{\mathrm{3}}$ (1,1) line. The void regions in L694-2 and L1517B of very low $N(\mathrm{H_{2}})$ correspond to nearby bright point 8\,$\mu m$ sources. The beam size is indicated as a white filled circle in each figure. For L694-2 and L429 the contours start from 1\,K\,km\,s$^{-1}$ and 1.5 \,K\,km\,s$^{-1}$, respectively, with a spacing of 0.5 \,K\,km\,s$^{-1}$. For L1517B the contours start from 0.5 \,K\,km\,s$^{-1}$ with a spacing of 0.3 \,K\,km\,s$^{-1}$.}
\label{fig:NH2}
\end{figure*}

\subsection{H$_{2}$ column density maps from {\it{Spitzer}} 8\,$\mu m$}\label{sec:extinction}
Owing to high column densities of cold dust, pre-stelelar cores often appear as dark structures on mid-infrared surface brightness maps.
Following the mid-infrared extinction calculations elaborated in \citet{Harju20}, we adopted the 850\,$\mu$m emission 
map observed by SCUBA-2 {\footnote{\url{https://www.cadc-ccda.hia-iha.nrc-cnrc.gc.ca/en/jcmt/}}}{\footnote{The James Clerk Maxwell Telescope is operated by the East Asian Observatory on behalf of The National Astronomical Observatory of Japan; Academia Sinica Institute of Astronomy and Astrophysics; the Korea Astronomy and Space Science Institute; the National Astronomical Research Institute of Thailand; Center for Astronomical Mega-Science (as well as the National Key R\&D Program of China with No. 2017YFA0402700). Additional funding support is provided by the Science and Technology Facilities Council of the United Kingdom and participating universities and organizations in the United Kingdom and Canada.}} and the 250\,$\mu$m, 350\,$\mu$m and 500\,$\mu$m emission maps{\footnote{We adopt the level 2.0 processed data products; the observational ID (OBSID) for L694-2 is 1342230846, for L429 is 1342239787, and for L1517B is 1342204843/1342204844.}} from the {\it{Herschel}}/SPIRE telescope\thanks{{\it Herschel} is an ESA space observatory with science instruments provided by European-led Principal Investigator consortia and with important participation from NASA.}(\citealt{Griffin10}) to facilitate the derivation of the high angular resolution, hydrogen column density maps ($N(\mathrm{H_{2}})$) from the {\it{Spitzer}}/IRAC 8\,$\mu$m data. 

With a modified black-body emission model, we first employed {\it{Herschel}}/SPIRE maps to derive a dust temperature map ($T_{\mathrm{dust}}$) at $\sim$21$''$ resolution (with the high angular resolution SPIRE maps in the archive) from Spectral Energy Distribution (SED) fitting. The temperature map is later used to derive the dust opacity map at 850\,$\mu$m ($\tau_{\mathrm\scriptsize{850\,\mu m}}$). We adopted dust emissivities for unprocessed (not coagulated) dust grains with thin ice mantles from \citet{OH94} (OH4 model); in this model, the dust emissivities in the submm regime can be approximated by a power-law form with index of $\beta$ = 2.0 with a reference value for the cross section per unit mass of gas at 850$\,\mu$m, $\kappa_{850\,\mu m}$, of 0.011\,cm$^{2}$ g$^{-1}$, adopting a gas-to-dust mass ratio of 100. The corresponding $\kappa_{8\,\mu m}$ is 8.85\,cm$^{2}$ g$^{-1}$. 
With this assumption, we neglect the change in dust emissivities in the cores due to the possibly varying dust grain properties. For example, the $\kappa_{8\mu m}$/$\kappa_{850\,\mu m}$ decrease with coagulation with increasing gas densities and the formation of ice mantles, which is likely happening in the innermost, denser region of these pre-stellar cores (e.g., \citealt{Bergin06}, \citealt{CT17, CT19a}). We discuss possible bias and hints for dust opacity variations below and in Appendix \ref{app:unc_NH2}. 

The mid-infrared extinction calculation in \citet{Harju20} adopts the $\tau_{\mathrm\scriptsize{850\,\mu m}}$ map to estimate the foreground (and zero point correction) and background emission level for the 8 $\mu$m map. Specifically, the 8$\mu m$ map is first masked out of the regions of high $\tau_{\mathrm\scriptsize{850\,\mu m}}$ (area of dense core) and of high 8$\mu m$ emission level (compact bright sources). The background image is then constructed by linearly interpolating over the masked regions using triangulation. For a relatively uniform 8 $\mu$m emission field this method produces a rather smooth background image. In case of highly locally varying 8$\mu m$ field, e.g., with bright compact sources immersed in relatively strong large-scale emission showing low contrast between spatial scales, the masking of bright sources based on constant threshold can cause visible defect to the interpolated background image. We therefore also adopt the small-scale median filter method (\citealt{ButlerTan09}) to estimate the emission background, which first effectively smooths out the local inhomogeneity and then conducts interpolation for the masked region. The detailed procedure is elaborated in Appendix \ref{app:unc_NH2}. At any rate, the process of background estimation is subject to the difficulty of distinguishing between small-scale background emission variations and the genuine absorbing components (\citealt{ButlerTan09}). 

Estimation of the foreground emission level $I_{\mathrm{fg}}$ is guided by the dust opacity map obtained from SED fitting. The foreground emission is assumed to be spatially constant and is estimated in an iterative way such that the resultant peak $\tau_{\mathrm\scriptsize{8\,\mu m}}$ after smoothing achieves better consistency with that predicted by SED, adopting the dust opacity relations mentioned before. Naturally, the upper limit of $I_{\mathrm{fg}}$ of the extinction method is the minimum flux level toward the core center, i.e. when the core absorbs all the background emission and that the observed flux solely comes from the foreground emission. 

Although the $\tau_{\mathrm\scriptsize{850\,\mu m}}$ is of higher angular resolution compared to 
opacity maps from {\it{Herschel}} data, lack of extended emission inherent to ground-based bolometric observations of SCUBA2 can easily result in underestimates of the true column densities, which is not necessarily only affecting the extended region of the core. In our specific case, the 850\,$\mu$m emission 
maps available for the two cores L429 and L694-2 are also rather shallow, achieving an rms level of $\sigma$$\sim$0.15 mJy/arcsec$^{-2}$ (mass sensitivity per beam corresponds to 0.05 $M_{\odot}$ for a source at 200 pc of temperature 10 K). 
To improve the quality of the SCUBA2 850\,$\mu$m emission map and preserve the extended emission, we resort to a continuum combination method applicable to emission maps obtained from ground-based bolometers and space telescopes, which is first proposed by \citealt{Liu15}, and further developed in \citet{Jiao22}. We utilised the PLANCK 353 GHz ($\lambda$\,=850\,$\mu$m) continuum data for compensating the extended structures; the PLANCK image was first deconvolved with a model image of extrapolated 850 $\mu$m emission from the SED of the {\it{Herschel}} maps. 
We give a brief summary of the method in Appendix \ref{app:850cb}. 

After obtaining a combined 850\,$\mu$m image (Fig. \ref{fig:scuba2_comb}), we derived a 14$''$ $N(\mathrm{H_{2}})$ map by applying the $T_{\mathrm{d}}$ map to the 850\,$\mu$m flux. This $N(\mathrm{H_{2}})$ is relatively free of unexpected uncertainties compared to the high resolution $N(\mathrm{H_{2}})$ derived from extinction method. 
With the assumed dust opacity values, the peak from the derived $\tau_{\mathrm\scriptsize{8\,\mu m}}$ should match that predicted by $\tau_{\mathrm\scriptsize{850\,\mu m}}$ map ($\tau_{\mathrm\scriptsize{8\,\mu m}, \mathrm{850\,\mu m\ pred}}$ = $\tau_{\mathrm\scriptsize{850\,\mu m}}\frac{\kappa_{\mathrm{\scriptsize{8\,\mu m}}}}{\kappa_{\mathrm{\scriptsize{850\,\mu m}}}}$). For L1517B which lacks SCUBA2 measurements, we instead use the $\tau_{\mathrm\scriptsize{250\,\mu m}}$ to gauge the $\tau_{\mathrm\scriptsize{8\,\mu m}}$. Practically, we find that even if we set the foreground emission level to be the lowest flux level of the 8$\mu$m, the derived peak $\tau_{\mathrm\scriptsize{8\,\mu m}}$ is lower than that predicted  based on $\tau_{\mathrm\scriptsize{850\,\mu m}}$ (or $\tau_{\mathrm\scriptsize{250\,\mu m}}$ in the case of L1517B). This is likely due to (1) underestimate of background level based on the interpolation method (2) possible variations of dust opacity values in the center of the core. Comparing the extended regions of the core, we find a similar mismatch, i.e. a systematic underestimate of the derived $\tau_{\mathrm\scriptsize{8\,\mu m}}$. Empirically, we find by applying a constant scaling factor to the derived $\tau_{\mathrm\scriptsize{8\,\mu m}}$, the consistency with predicted $\tau_{\mathrm\scriptsize{8\,\mu m}}$ can be achieved at the peak region and most of the extended region we are interested in (see more in Appendix \ref{app:unc_NH2}).  The scaling factor for L429, L694-2 and L1517B is determined to be 2.5, 2 and 1.5, respectively.  

In the final step, using these factors, we manually scaled up the $\tau_{\mathrm\scriptsize{8\,\mu m}}$ map to match that of $\tau_{\mathrm\scriptsize{8\,\mu m}, \mathrm{850\,\mu m\ pred}}$ (or $\tau_{\mathrm\scriptsize{8\,\mu m}, \mathrm{250\,\mu m\ pred}}$). 
The scaling essentially adds back the contribution of absorption from gas structures in front of the core, related to the parental cloud where the foreground emission level is most likely to be underestimated. The obtained $N(\mathrm{H_{2}})$ maps were smoothed to 5$''$ to match the NH$_{3}$ observations. The smoothed maps are shown in Fig. \ref{fig:NH2}.

\begin{table}
\centering
\begin{threeparttable}
\footnotesize
\caption{Density profile parameters from fitting the $N(\mathrm{H_{2}})$ map derived from 8$\,\mu m$ extinction.}
 \label{tab:dprofile}
 \begin{tabular}{llllllll}
\toprule
Source  & $n_{\mathrm{c}}$& $r_{\mathrm{flat}}$ &R$^{a}$&Mass$^{b}$\\
   & (10$^{5}$\,cm$^{-3}$)&(au)&(pc)& ($M_{\odot}$)\\
\midrule
L694-2&8.3(0.2)&2700(100)&0.1&8.3\\

L429&9.5(0.5)&4000(400)&0.1&4.0\\
L1517B&3.2(0.1)&3200(100)&0.1&2.0\\

\bottomrule
\end{tabular}
\begin{tablenotes}
      \small
      \item $^{a}$ The maximum outer radius $R$ in the fits is set to 0.10 pc for all three cores.
      \item $^{b}$ The total mass is calculated within a common radius of 0.05 pc.
      \item Uncertainties of parameter $n_{\mathrm{c}}$ and $r_{\mathrm{flat}}$ are listed in parentheses as absolute values.
      \end{tablenotes}
  \end{threeparttable}
\end{table}

\subsection{Core density profiles constrained from $N(\mathrm{H_{2}})$ maps}\label{sec:nprofiles}

\begin{figure}[htb]
\begin{tabular}{p{0.85\linewidth}}

\hspace{0.1cm}\includegraphics[scale=0.52]{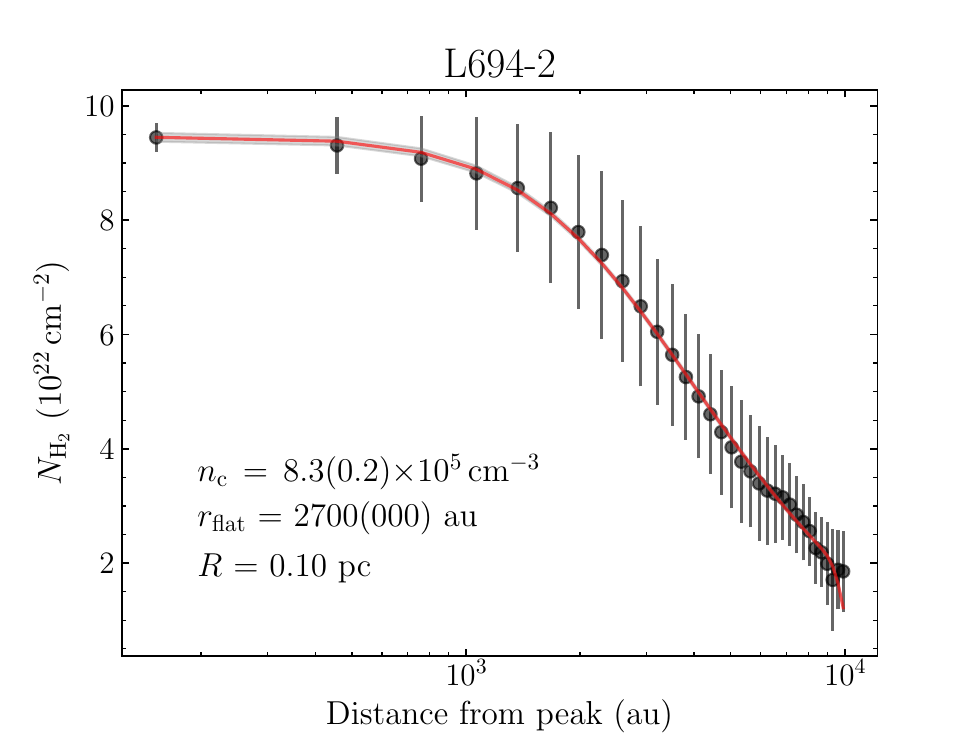}\\
\hspace{0.1cm}\includegraphics[scale=0.52]{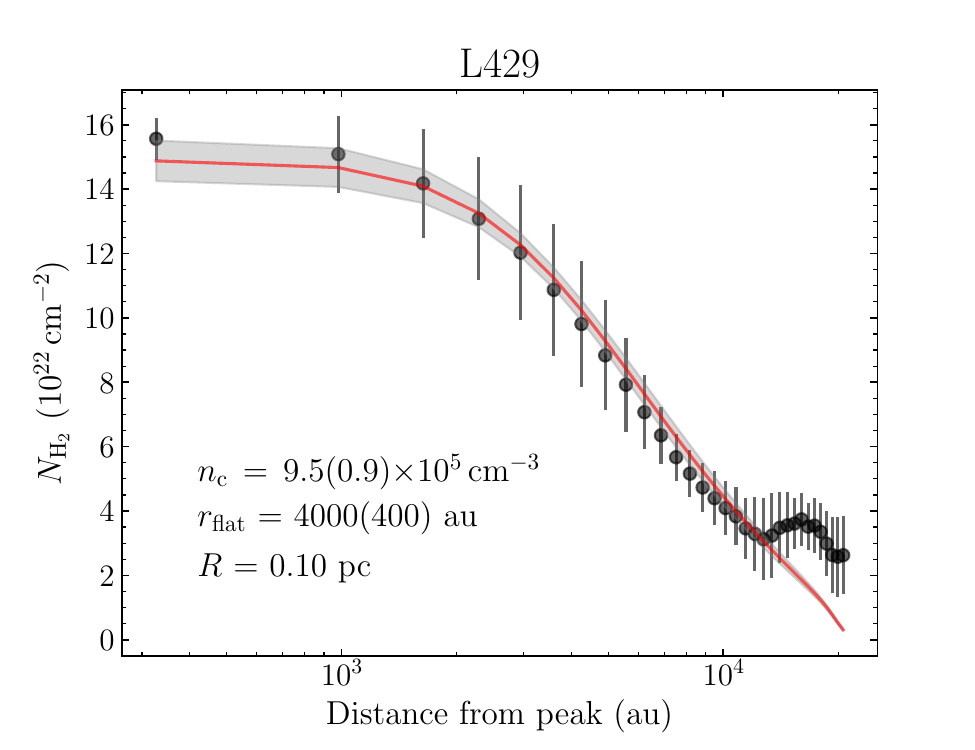}\\
\hspace{0.1cm}\includegraphics[scale=0.52]{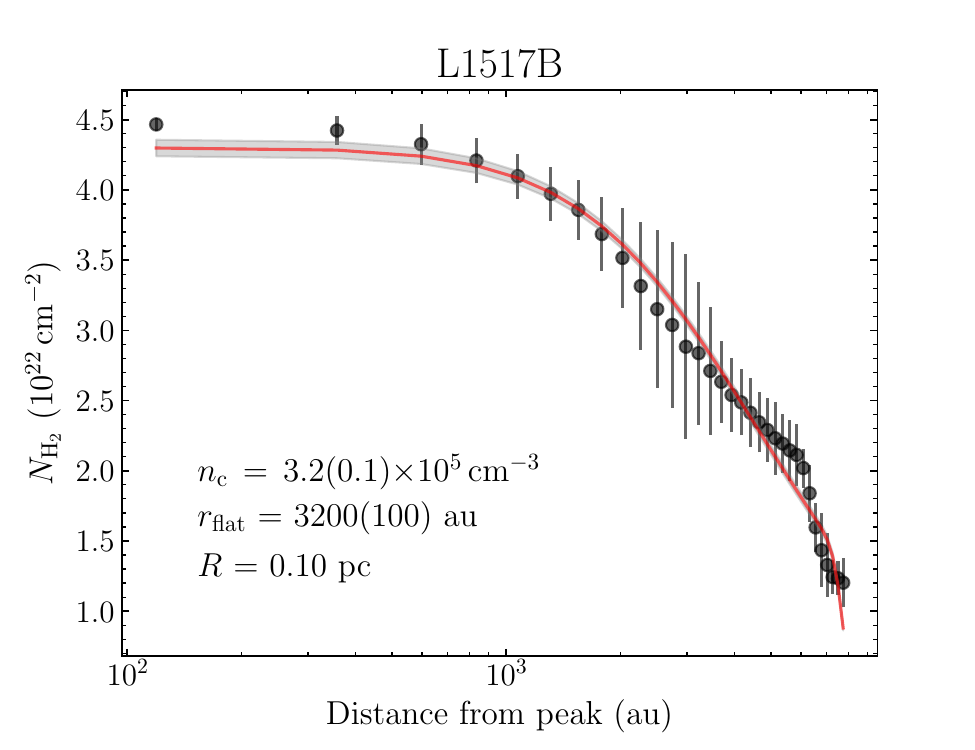}\\
\end{tabular}
\caption{The azimuthally averaged column density radial profiles, derived from the $N(\mathrm{H_{2}})$ map (Fig. \ref{fig:NH2}). The solid red line
shows the curve of the best-fit model following Eq. \ref{eq:rhor}. The best-fit parameters of central density, $n_{c}$, and radius of inner flat region, $r_{\mathrm{flat}}$, are indicated in each subplot.}

\label{fig:profile_DB}
\end{figure}

\begin{figure}[htb]
    \centering
    \includegraphics[height=7cm]{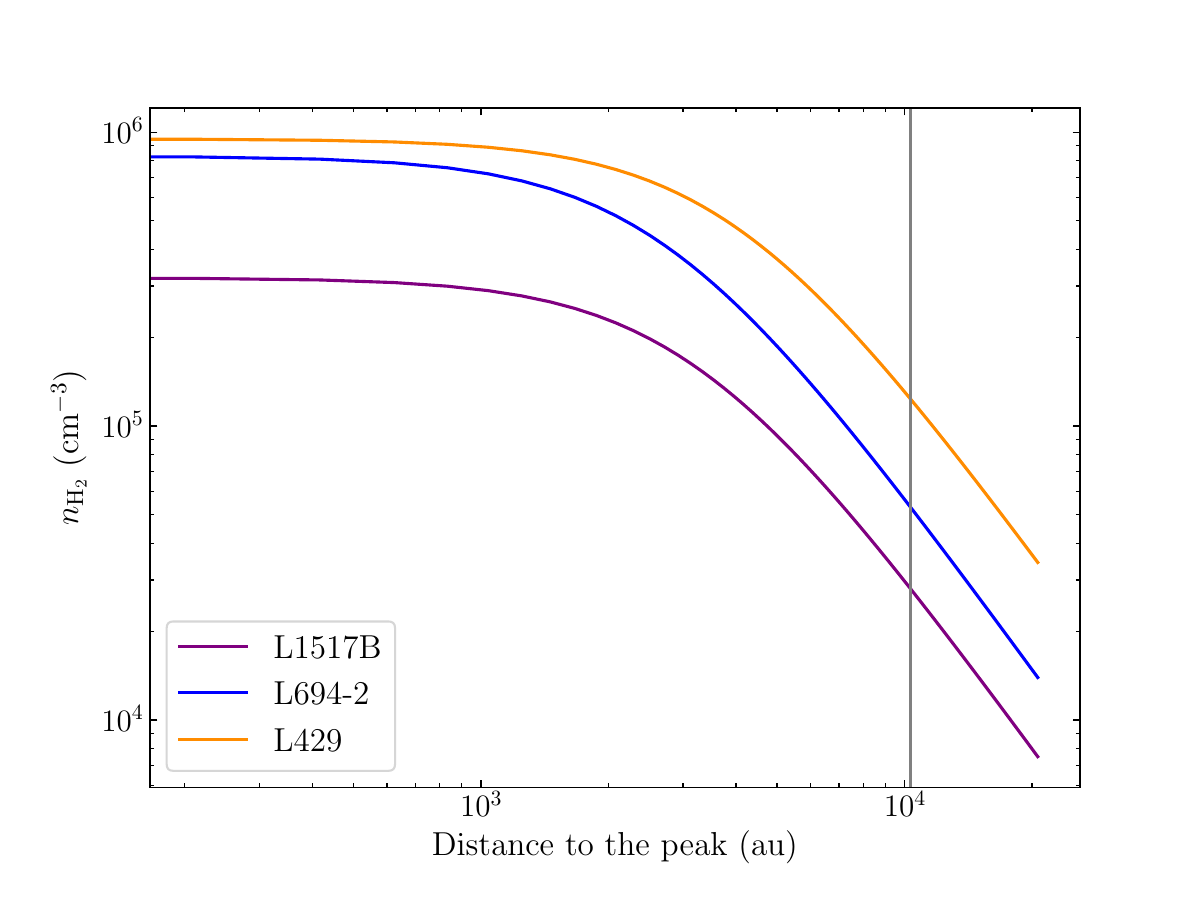}
    \caption{The fitted density profiles of the three cores following the form of \citet{DB09}. The vertical line marks a radius of 0.05 pc.}
    \label{fig:nprofile_threefunc}
\end{figure}

We assume that the volume density profile, $n(r)$, of the cores follow a Bonnor-Ebert like model (\citealt{DB09}), 
\begin{equation}
 n(r) = \frac{n_{\mathrm{c}}}{(r/r_{\mathrm{flat}})^2+1}\label{eq:rhor}~.
 \end{equation}
Specifically, the radial profile has a central flat region of radius $r_{\mathrm{flat}}$, i.e. a ``plateau", plus a power-law decline $\propto\,r^{-2}$ at the outer radii up to the radius of the core, $R$. Since $R$ is mostly degenerate with $n_{\mathrm{c}}$, we fix $R$ to 0.1 pc for the three cores. The corresponding functional form of column density profile is derived by integrating from the center of the core to a certain projected radius, $x$. 
We obtained the best-fit $n_{\mathrm{c}}$ and $r_{\mathrm{flat}}$ for the three cores, listed in Table \ref{tab:dprofile}. The fitted profile is shown in Fig. \ref{fig:profile_DB} comparing the fitted form with the measurements for the three cores separately, and in Fig. \ref{fig:nprofile_threefunc} for the three fitted profiles in conjunction. L429 shows the highest central density of $\sim$10$^{6}$\,cm$^{-3}$. L1517B and L694-2 have similar central plateau size of $\sim$3000 au, with L694-2 showing two times higher central density, of 8$\times$10$^{5}$\,cm$^{-3}$. The enclosed core mass within 0.05 pc of L429 and L694-2 are $\sim$8$M_{\odot}$ and $\sim$4$M_{\odot}$, while that of L1517B is $\sim$2$M_{\odot}$. In Fig. \ref{fig:nprofile_threefunc}, it can be seen that the density profiles rise from L1517B to L429, so that the density in L429 is at all radii higher than that in the other two cores.

\subsection{Fits of the NH$_{3}$ lines}\label{sec:nh3_model}
To derive the physical parameters, we fit the NH$_{3}$ (1,1) and (2,2) lines simultaneously with functions in {\tt{pyspecnest}}{\footnote{\url{https://github.com/vlas-sokolov/pyspecnest}}} package (\citealt{Sokolov20}). The method adopts Bayesian model selection to determine the spectral multiplicity, i.e. number of velocity components, for the observed spectrum. This has been done in the past to better fit multiple components (\citealt{Chen22}, Choudhury in prep.). A nested sampling algorithm is used to explore the parameter space ({\tt{pymultinest}}\footnote{\url{https://johannesbuchner.github.io/PyMultiNest/}}, \citealt{Buchner14}), ideal for parameter estimation in the case of multi-modal
distribution. We use the {\tt{cold\_ammonia}} model in {\tt{pyspeckit}} package {\footnote{\url{https://github.com/pyspeckit/pyspeckit}}}(\citealt{Ginsburg21}) which assumes that only the (1,1) and (2,2) states of para-NH$_{3}$ are populated rotational levels, which is valid for cold cores (see \citealt{Friesen17}). The fitting outputs best-fit parameters including system velocity, $\rm{v}_{\mathrm{LSR}}$, velocity dispersion, $\sigma_{\rm{v}}$, kinetic temperature, $T_{\mathrm{kin}}$, excitation temperature, $T_{\mathrm{ex}}$, and NH$_{3}$ column density, $N(\mathrm{NH_{3}})$. The ortho-to-para NH$_{3}$ ratio is set to 1.
We assume uniform priors for all the parameters with boundaries set to reflect reasonable ranges for respective source. 

We note that although the total optical depth of the NH$_{3}$(1,1) can be high in the central core area ($\tau_{\mathrm{1,1}}\,$$\gtrsim\,$30), the fact that satellite lines of (1,1) are only marginally optically thick, e.g., the two hyperfine components of F\,=\,0-1 at 23.6929 GHz have opacities of 0.074 and 0.15 times of $\tau_{\mathrm{1,1}}$ and appear to be isolated components given the typical velocity dispersions associated with these cores, and that the total optical depth of the NH$_{3}$(2,2) is $\lesssim$0.3 ensure that the derivation of $T_{\mathrm{kin}}$ and $N(\mathrm{NH_{3}})$ are not biased in the physical regimes in consideration. The uncertainties and degeneracies of the parameters that come with the assumptions of the model itself, i.e., a homogeneous gas layer of constant temperature along line-of-sight, are properly estimated by the posterior probability distributions from the nested-sampling method (\citealt{Sokolov20}).

\begin{figure*}
\begin{tabular}{p{0.485\linewidth}p{0.485\linewidth}}
\hspace{0.2cm}\includegraphics[scale=0.75]{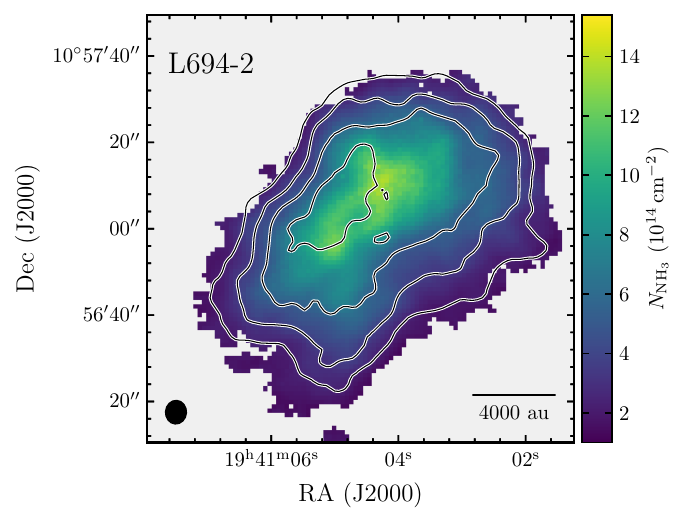}&\includegraphics[scale=0.75]{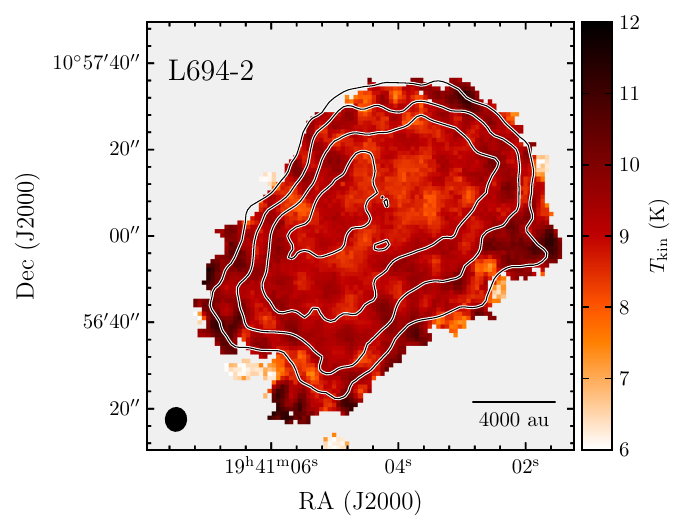}\\
\hspace{0.3cm}\includegraphics[scale=0.75]{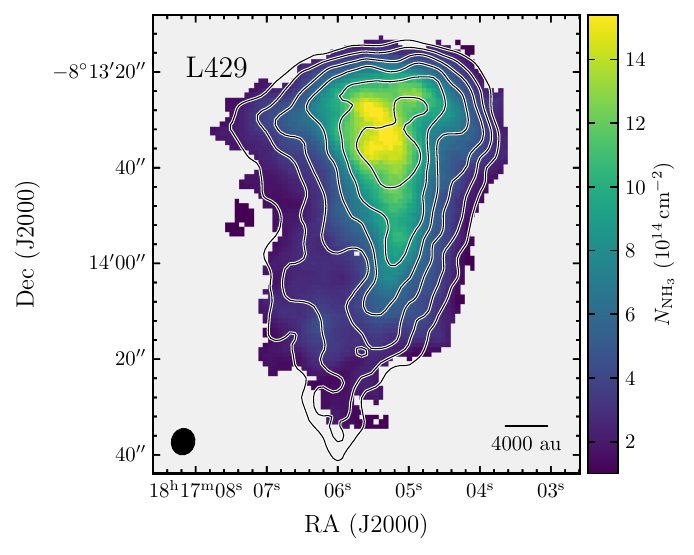}&\includegraphics[scale=0.75]{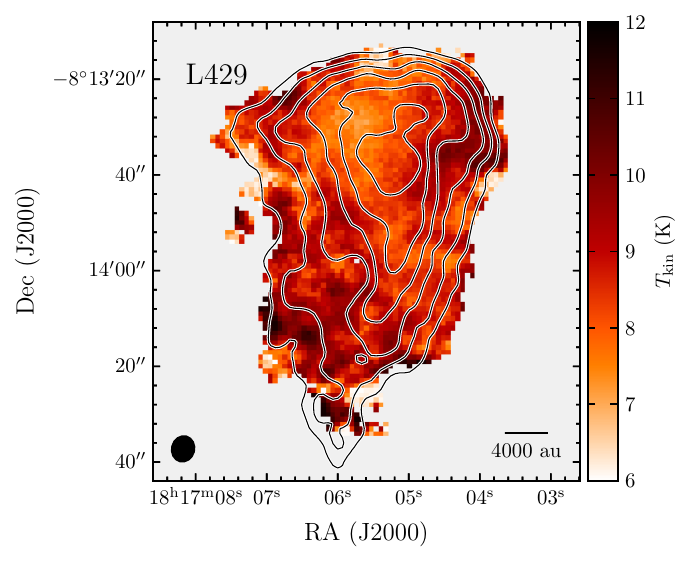}\\
\hspace{0.3cm}\includegraphics[scale=0.75]{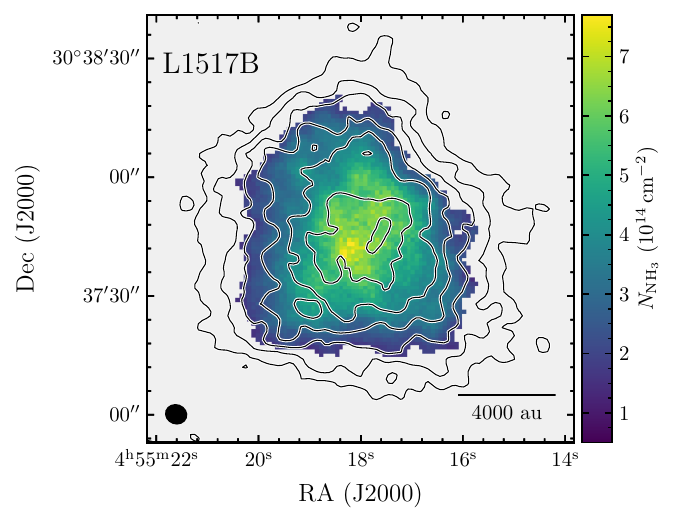}&\includegraphics[scale=0.75]{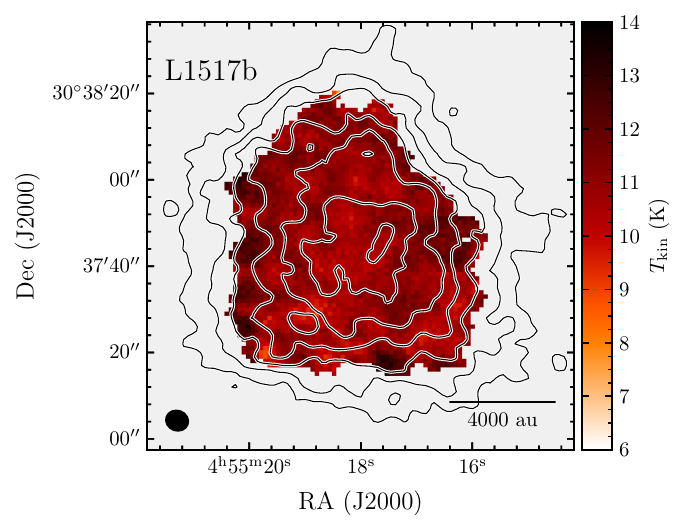}\\
\end{tabular}
\caption{The ammonia column density, $N(\mathrm{NH_{3}})$ and kinetic temperature, $T_{\mathrm{kin}}$ maps for the three cores, constrained from one-component fit. The beam size is indicated as a black filled circle in each image. The contours are the same as that defined in Fig. \ref{fig:NH2}, showing the $N(\mathrm{H_{2}})$ levels.}

\label{fig:Tkin_NNH3}
\end{figure*}

\begin{figure*}
\begin{tabular}{p{0.485\linewidth}p{0.485\linewidth}}
\hspace{0.2cm}\includegraphics[scale=0.75]{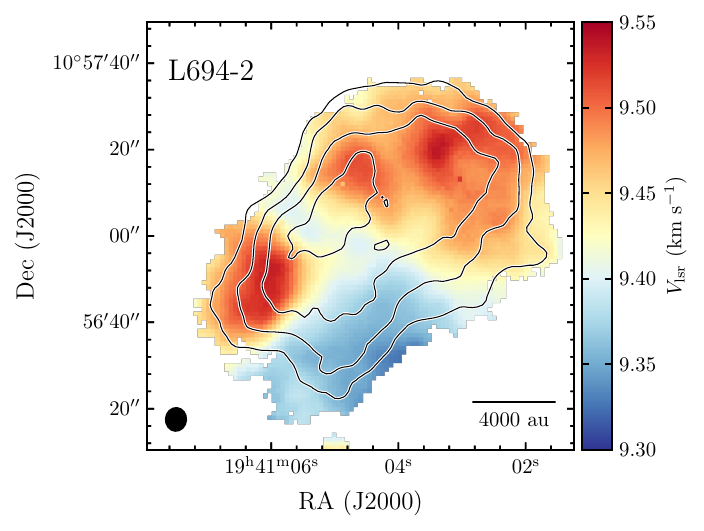}&\includegraphics[scale=0.75]{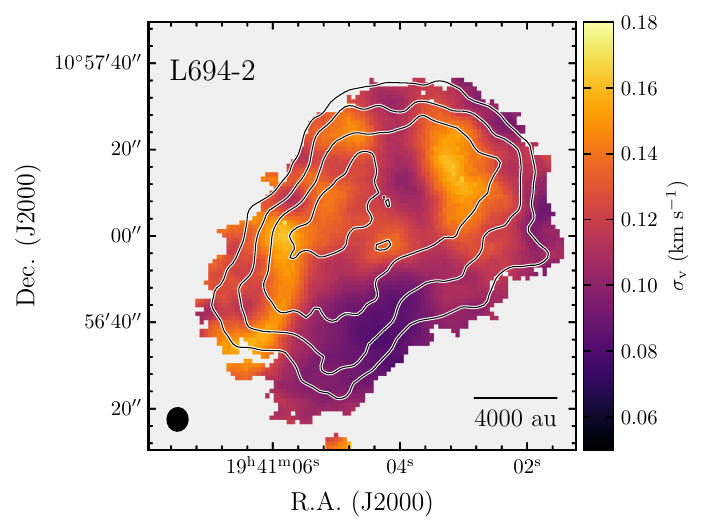}\\
\hspace{0.3cm}\includegraphics[scale=0.75]{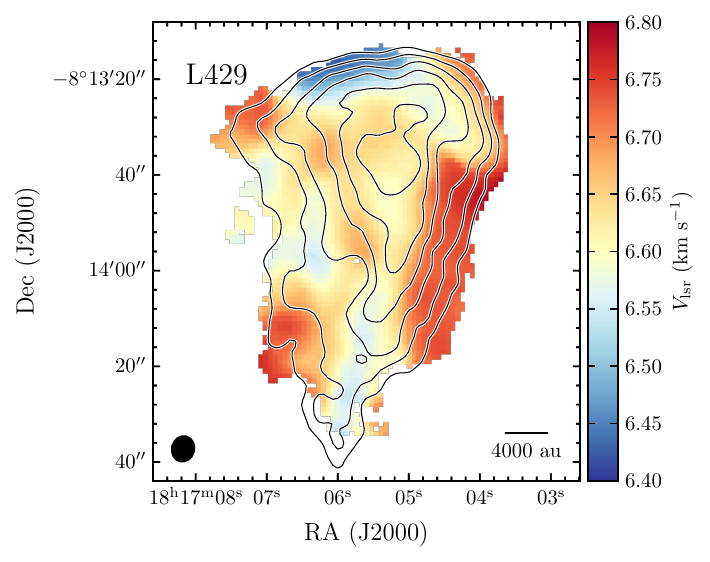}&\includegraphics[scale=0.75]{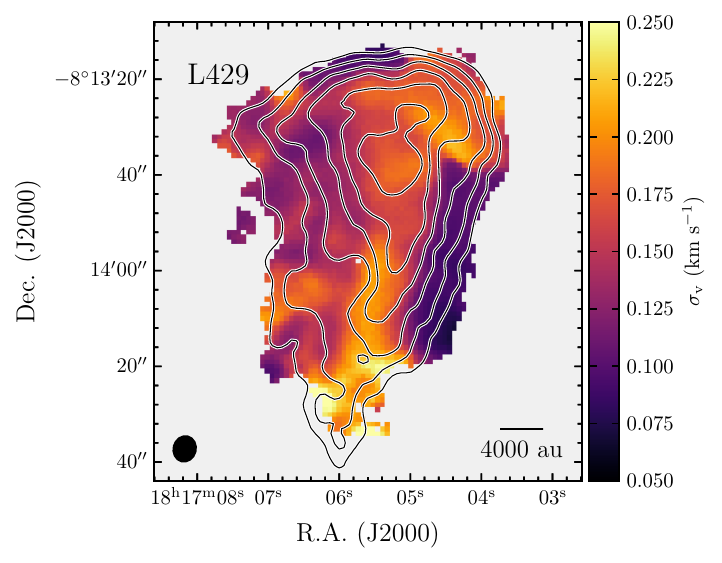}\\
\hspace{0.3cm}\includegraphics[scale=0.75]{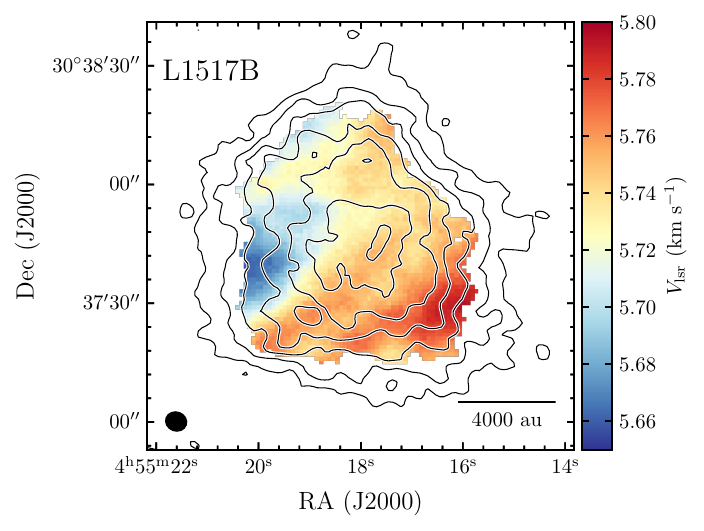}&\includegraphics[scale=0.75]{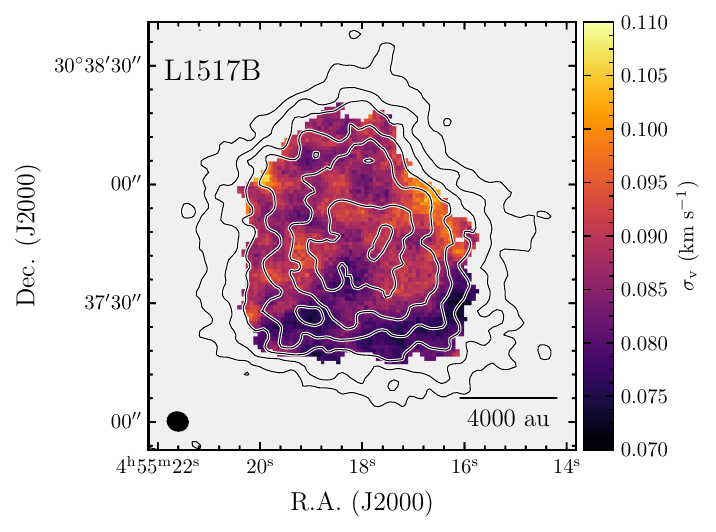}\\
\end{tabular}
\caption{The centroid velocity, $\rm{v}_{\mathrm{LSR}}$ and velocity dispersion, $\sigma_{\rm{v}}$ maps for the three cores, constrained from one-component fit. The beam size is indicated as a black filled circle in each image.}

\label{fig:vlsr_sigmav}
\end{figure*}

Following the logic elaborated in \citealt{Sokolov20}, we set the threshold of Bayes factor, ln\,$K^{2}_{1}$ $\geq$ 5, for identifying spectra that show an apparent secondary velocity component. Within our achieved rms, it turns out that for L1517B a one-component model can describe all spectra across the core, while for L694-2 and L429, there are localised sub-regions that show evidence of two velocity components. We show the ln\,$K^{2}_{1}$ maps in Appendix A. The nature of these sub-regions are discussed in Sect. \ref{sec:kinematics}. In finalising the fitted parameter maps, following \citet{Pineda15, Pineda22a}, we trimmed the pixels with uncertainties of $T_{\mathrm{kin}}$ or $T_{\mathrm{ex}}$ larger than 1 K, or uncertainties of $\rm{v}_{\mathrm{LSR}}$ or $\sigma_{\rm{v}}$ larger than two channel widths (0.2 km\,s$^{-1}$). These criteria also ensure that $N(\mathrm{NH_{3}})$ is well constrained, with uncertainties $\lesssim$20$\%$, as reflected from the posterior probability distributions. Small, isolated regions of less than a beam size are also removed from the final maps.
Since even for L694-2 and L429, the majority of the core region is characterised by one velocity component, we present the parameter maps from the one-component model for the three cores, $T_{\mathrm{kin}}$ and $N(\mathrm{NH_{3}})$ in Fig. \ref{fig:Tkin_NNH3}, $\rm{v}_{\mathrm{LSR}}$ and $\sigma_{\rm{v}}$ in Fig. \ref{fig:vlsr_sigmav}.

\subsection{NH$_{3}$ column density, kinetic and excitation temperature}\label{sec:paramaps}

Comparing the $N(\mathrm{NH_{3}})$ of the three cores in Fig. \ref{fig:Tkin_NNH3}, the range of $N(\mathrm{NH_{3}})$ of core L429 and L694-2 is similar, reaching up to $\sim$2$\times$10$^{15}$\,cm$^{-2}$ (higher than that of H-MM1, \citealt{Pineda22a}), while L1517B has a lower peak $N(\mathrm{NH_{3}})$, $\sim$7$\times$10$^{14}$\,cm$^{-2}$. The morphology of the higher $N(\mathrm{NH_{3}})$ structures of L694-2 and L429 appear elongated, along the north-west to south-east, and the North to the South direction, respectively, while that of L1517B appears more roundish. 

The $T_{\mathrm{kin}}$ spatial distribution of all three cores are mostly uniform (Fig. \ref{fig:Tkin_NNH3}), with a slight increase towards the outer region; this is similar to $T_{\mathrm{kin}}$ map of H-MM1 (\citealt{Pineda22a}). Compared to H-MM1 that has an average central $T_{\mathrm{kin}}$ of 11 K, L429 and L694-2 show an overall lower $T_{\mathrm{kin}}$, varying between 8-10 K. The temperature drop in the central core region is more obvious in the left plot of Fig. \ref{fig:kde11}: the average temperatures of L429 drops from $\sim$9 K to below 8 K, of L1517B from $\sim$11 K to 10 K, while L694-2 shows a more smooth temperature structure around $\sim$9 K. 

\begin{figure}[htb]
\begin{tabular}{p{0.85\linewidth}}
\hspace{-0.25cm}\includegraphics[scale=0.48]{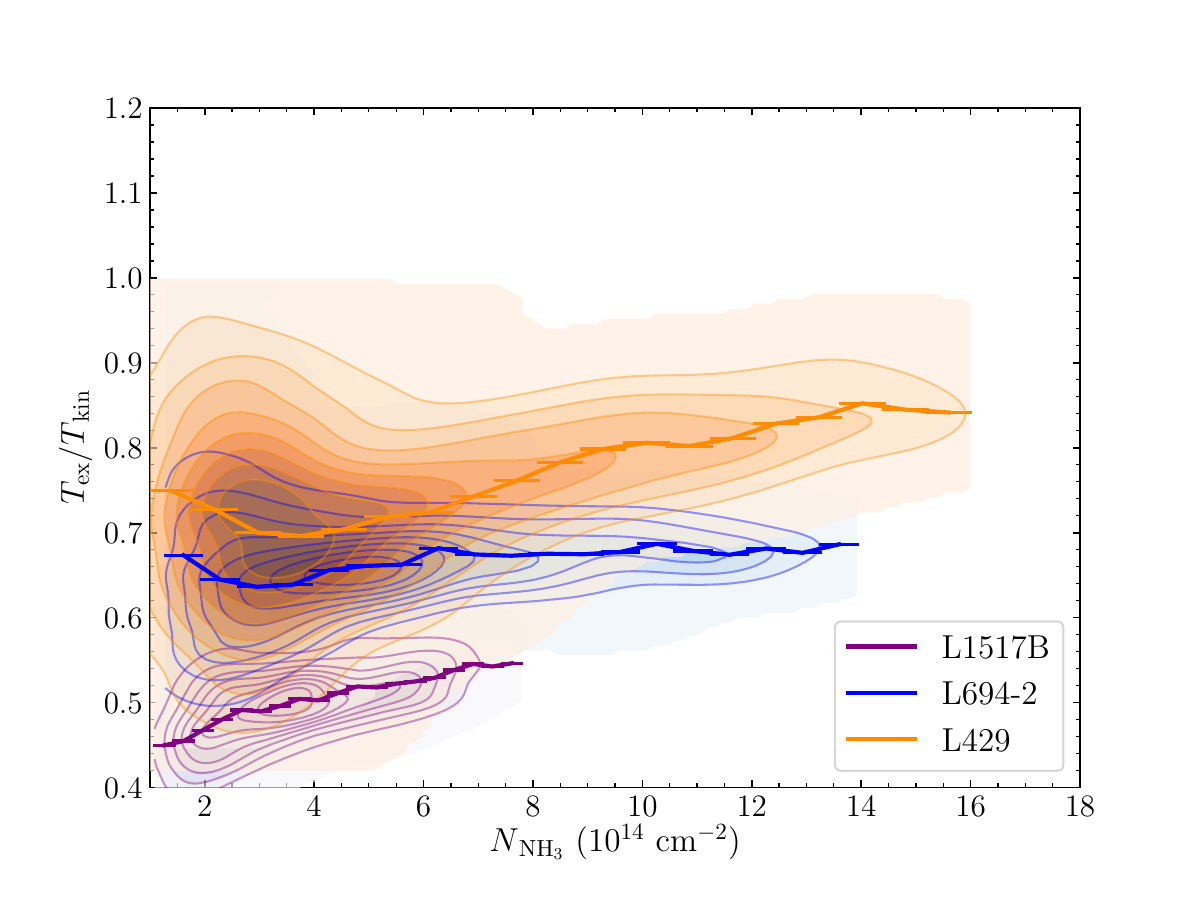}\\
\end{tabular}
\caption{The Gaussian KDE of the ratio of $T_{\mathrm{ex}}$/$T_{\mathrm{kin}}$ as a function of NH$_{\mathrm{3}}$ column density. The line segments correspond to mean values in  NH$_{\mathrm{3}}$ column density bins, with solid lines linked together.}

\label{fig:kde2}
\end{figure}

Comparing $T_{\mathrm{kin}}$ with $T_{\mathrm{ex}}$ (Fig. \ref{fig:kde2}), for L1517B, $T_{\mathrm{ex}}$ is systematically lower than $T_{\mathrm{kin}}$. This indicates that the gas densities are relatively low and NH$_{3}$ lines are mainly sub-thermally excited (the thermalisation density of NH$_{3}$(1,1) is $\sim$10$^{5.5}\,$cm$^{-3}$ for a gas kinetic temperature of $\sim$10 K, \citealt{Shirley15}). For L694-2, the $T_{\mathrm{ex}}$ is closer to $T_{\mathrm{kin}}$; L429 shows the least overall difference between $T_{\mathrm{ex}}$ and $T_{\mathrm{kin}}$. The comparison between $T_{\mathrm{kin}}$ and $T_{\mathrm{ex}}$ implies a progressively denser gas environment of core L1517B, L694-2 and L429, which is compatible with the density profiles determined in Sect. \ref{sec:nprofiles} (Fig. \ref{fig:nprofile_threefunc}).

\subsection{Velocity field, velocity dispersion and velocity gradient distribution}\label{sec:vfield}
The centroid velocities of the three cores show small variations (Fig. \ref{fig:vlsr_sigmav}), within $\sim$0.4 km\,s$^{-1}$. For L694-2 and L429, the variations are within $\sim$0.3\,km\,s$^{-1}$ and for L1517B within $\sim$0.15\,km\,s$^{-1}$. The velocities of L694-2 and L1517B are more continuous across the core dominated by mostly large-scale variations, while L429 is characterised by a more complex velocity field with localised variations, showing alternating blue- and red-shifted velocity sub-regions. For all three sources, the most red-shifted and blue-shifted velocities located at the outer regions of the core. 

The velocity dispersions, $\sigma_{\rm{v}}$, of the three cores are all below 0.2\,km\,s$^{-1}$, with L1517B showing even smaller values, of $\lesssim$0.1\~km\,s$^{-1}$. The smallest $\sigma_{\rm{v}}$ are located at the outer core regions (for L694-2 and L429 in the South-West, for L1517B in the South), coincident with continuous, either red-shifted or blue-shifted velocity field.
There are some prominent, arc-like features of high $\sigma_{\rm{v}}$ seen towards L429 and L694-2, which are also correlated with abrupt change of $\rm{v}_{\mathrm{LSR}}$ or/and appearance of secondary velocity component (Appendix \ref{app:2comp}). 
We discuss the specific gas kinematics for each core in Sect. \ref{sec:kinematics}.

Using the centroid velocity maps, we calculated the local (line-of-sight) velocity gradient (magnitude, $\mathscr{G}$, and orientation, $\Theta$) following the method elaborated in \citet{Sokolov19}, which was first presented by \citet{Goodman93} and \citealt{Caselli02a}. The chosen area for fitting the local velocity gradient and the pruning of the pixel-wise fits (based on significance of the fitted $\mathscr{G}$ and $\Theta$) follow that described in \citet{Sokolov19}. The sub-regions where a secondary velocity component is detected are trimmed for the local velocity gradient calculation. We first calculate an average global velocity gradient across the core, which takes into account all the valid pixels in the centroid velocity maps. The average global velocity gradient for L694-2 is 1.65$\pm$0.03\,km\,s$^{-1}$\,pc$^{-1}$, for L429 is 0.52$\pm$0.02\,km\,s$^{-1}$\,pc$^{-1}$, and for L1517B is 1.77$\pm$0.03\,km\,s$^{-1}$\,pc$^{-1}$. 
These small-scale velocity gradients are in general larger than those measured from the single-dish observations towards the parental filamentary structure of the cores (see for L1517B in \citet{Hacar11} and L694-2 in \citet{Kim22}). This reflects that the small-scale structure of the core is showing more variations of velocity field. 

\begin{figure}[htb]
\begin{tabular}{p{0.85\linewidth}}
\hspace{0.25cm}\includegraphics[scale=0.65]{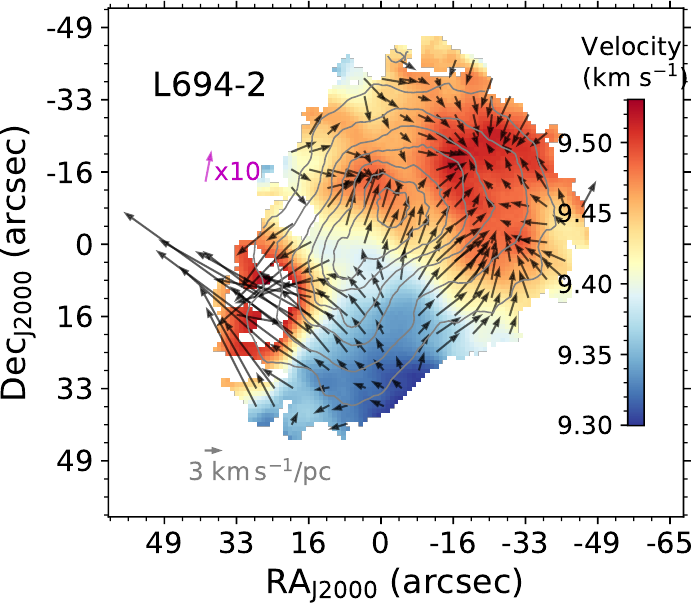}\\
\hspace{0.25cm}\includegraphics[scale=0.65]{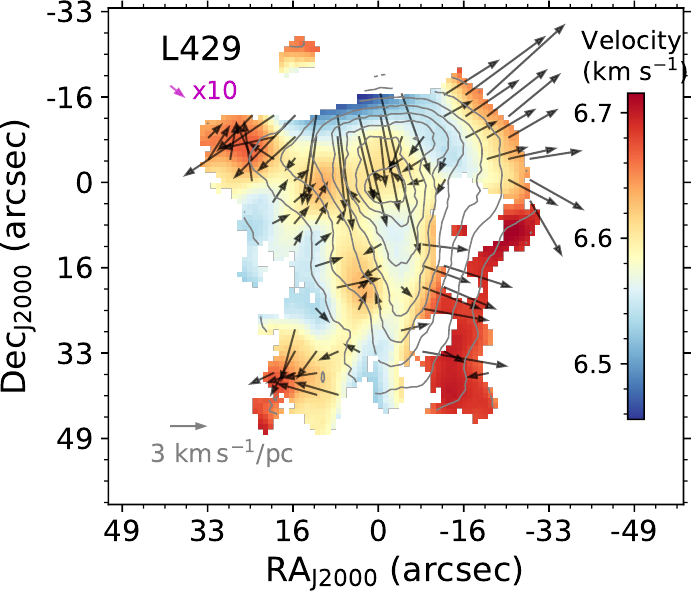}\\
\hspace{0.25cm}\includegraphics[scale=0.65]{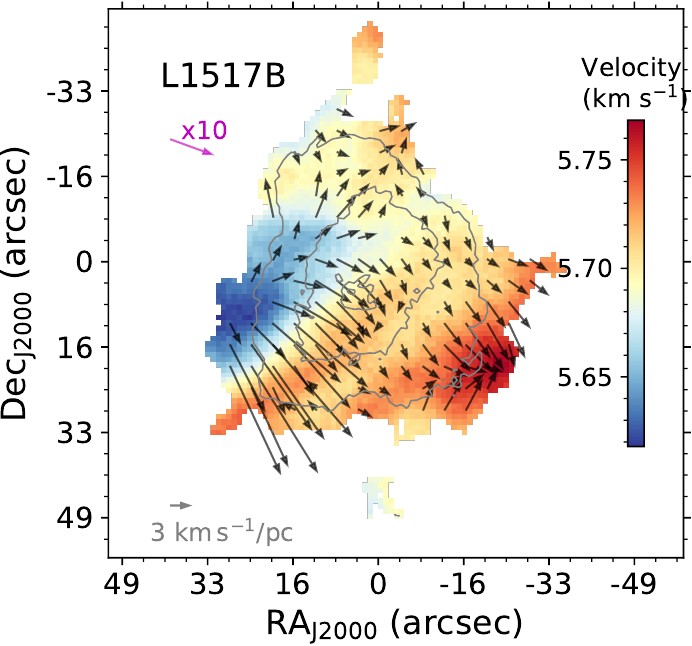}
\end{tabular}
\caption{Centroid velocity map overlaid with the velocity gradient (after subtracting the average overall velocity gradient) arrows. The sub-regions where a second velocity component is seen are excluded from the maps. The directions of the arrows points to the steepest velocity field change in the red-shifted direction, while the arrow lengths represent the relative vector magnitudes of the gradient, with the length of 3 km$\,$s$^{-1}$$\,$pc$^{-1}$ indicated in each subplot. The magenta arrow shows the average overall velocity gradient across the core; the length is artificially scaled up by a factor of ten for better representation. The contour levels indicate $N(\mathrm{NH_{3}})$ starting from 1.0$\times$10$^{14}\,$cm$^{-2}$ with an interval of 2.1$\times$10$^{14}\,$cm$^{-2}$.}

\label{fig:vel_gra}
\end{figure}

In Fig. \ref{fig:vel_gra} the obtained local velocity gradient maps of the three cores are shown. The velocity gradients have a wide range of directions, which can vary significantly over small scales, especially towards L694-2 and L429. The larger magnitudes of velocity gradient are not necessarily associated with higher $N(\mathrm{NH_{3}})$ regions. In fact, towards L429 in particular, the largest $\mathscr{G}$ mostly lie in the outer region of the core (in the North and West). Towards L694-2, despite the unpatterned overall variations of $\Theta$, some localised regions show converging velocity gradient, e.g. in the North-west direction, which may be driven by clumpy substructures that remain unresolved. 

\begin{figure}[htb]
\begin{tabular}{p{0.85\linewidth}}
\hspace{0.25cm}\includegraphics[scale=0.375]{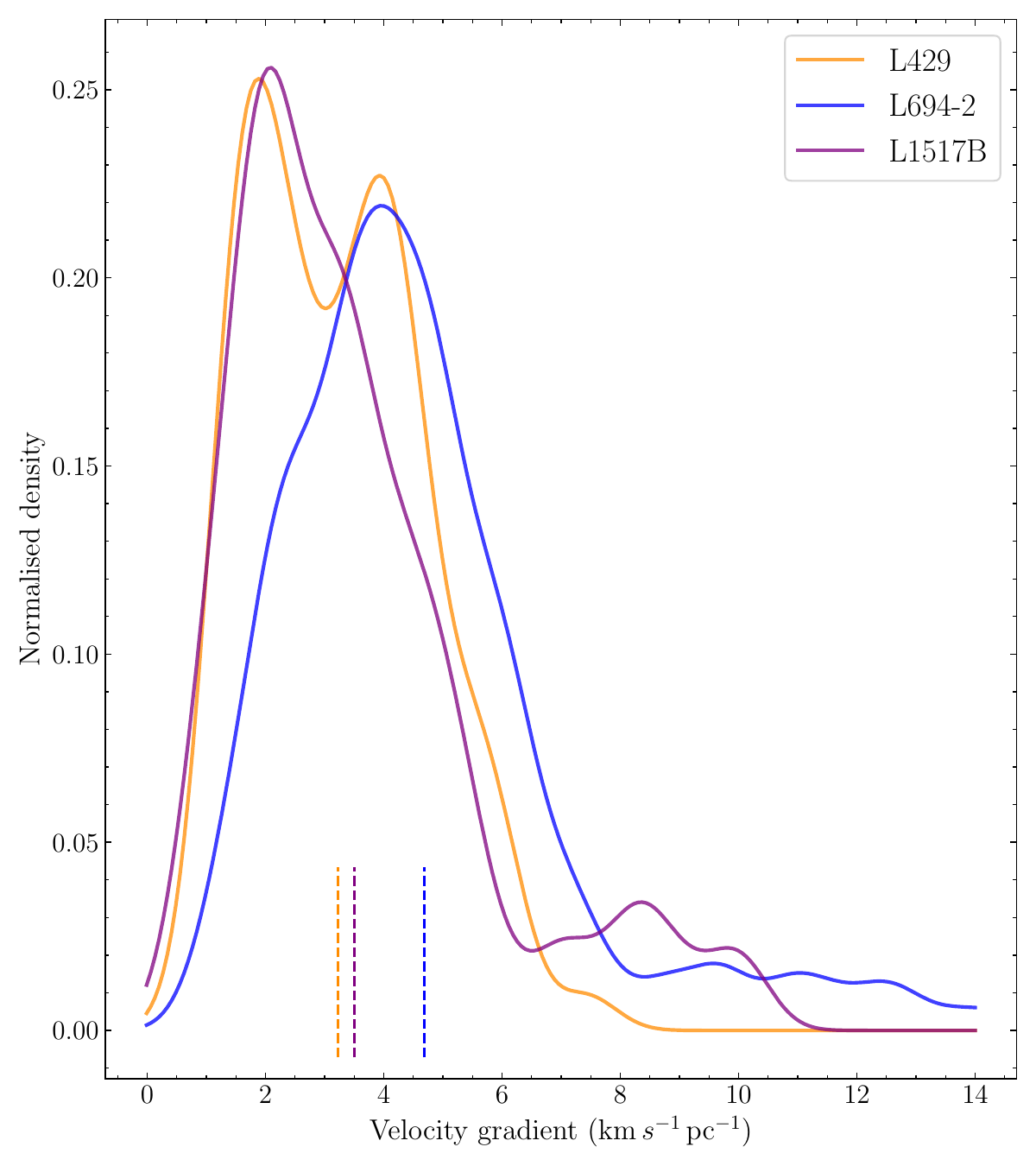}\\
\end{tabular}
\caption{Distribution of velocity gradient of the three cores.}

\label{fig:vel_gra_hist}
\end{figure}

Fig. \ref{fig:vel_gra_hist} compares the distribution of $\mathscr{G}$: among the three cores, L429 exhibits a bi-modal distribution of $\mathscr{G}$, and has an average $\mathscr{G}$ of 3.2$\pm$1.4 km$\,$s$^{-1}$$\,$pc$^{-1}$, similar to that measured toward L694-2 and L1517B, of 4.7$\pm$2.6 km$\,$s$^{-1}$$\,$pc$^{-1}$ and 3.5$\pm$2.1 km$\,$s$^{-1}$$\,$pc$^{-1}$, respectively. If we assume that the measured line-of-sight velocity gradients partially reflect the inward motions of the dense gas, then the higher average local velocity gradients of L694-2 and the higher velocity gradients in second peak of L429 imply independently that they are denser than L1517B, having a higher enclosed mass at similar scales. The more complex velocity gradient pattern of L429 also indicates it has more chaotic gas flows, than the other two cores.

\subsection{Variations of NH$_{\mathrm{3}}$ abundance}\label{sec:x_nh3}

In the left plot of Fig. \ref{fig:kde1_sctau}, the comparisons between $N(\mathrm{NH_{3}})$ and $N(\mathrm{H_{2}})$ of the three cores are shown. L694-2 and L1517B do not show any strong evidence of abundance drop of NH$_{3}$; L429, on the other hand, show a continuous flattening of $N(\mathrm{NH_{3}})$ as $N(\mathrm{H_{2}})$ increases. In order to account for different background levels of $H\mathrm{_{2}}$ column density, we estimate the average $N(\mathrm{H_{2}})$ offset, $N_{\mathrm{H_{2}\,off}}$, from the outermost region ($\sim$3 beam sizes in width) of the $N(\mathrm{NH_{3}})$ map (``zero-level'' of $N(\mathrm{NH_{3}})$, see also below), and plot the $N(\mathrm{NH_{3}})$ as a function of $N(\mathrm{H_{2}})$-$N_{\mathrm{H_{2}\,off}}$ in Fig. \ref{fig:kde1_sctau}, right panel. It can be seen that the slopes of L429 and L694-2 are similar in the lower $N(\mathrm{H_{2}})$-$N_{\mathrm{H_{2}\,off}}$ regime of up to $\sim$5$\times$10$^{22}$\,cm$^{-3}$, with L1517B showing a higher abundance of NH$_{3}$; the positive offsets of NH$_{3}$ at $N(\mathrm{H_{2}})$ = $N_{\mathrm{H_{2}\,off}}$ are similar among the three cores of $\sim$2$\times$10$^{14}$\,cm$^{-2}$, indicating similar NH$_{3}$ column density in their embedding molecular clouds. 

To describe the variation of $N(\mathrm{NH_{3}})$ as a function of $N(\mathrm{H_{2}})$, we fit a broken line model following the form detailed in \citet{Pineda22a}. The functional form includes two linear slopes $a$ and $c$, which apply at below and above the cutoff $N(\mathrm{H_{2}})$, $x_{0}$, where the variation happens, in addition to an intercept $b$ representing the $N(\mathrm{NH_{3}})$ at $N(\mathrm{H_{2}})$ = 0. In the case of L1517B and L694-2, the best-fit cutoff $N(\mathrm{H_{2}})$ are above the maximum $N(\mathrm{H_{2}})$ measured, indicating they are better described by a single linear relation. The fitted parameters are listed in Table \ref{tab:bl_sctau} for the three cores: in the case of L1517B and L694-2, the best-fit cutoff $N(\mathrm{H_{2}})$ are close to the maximum $N(\mathrm{H_{2}})$ measured, indicating they are better described by a single linear relation. Compared to the case of H-MM1, values of $a$ are similar. The cutoff $N(\mathrm{H_{2}})$ value in L429 is $\sim$8$\times$10$^{22}$cm$^{-2}$, which is higher than that in H-MM1.  
Concerning the variations of the abundance, H-MM1 shows a negative slope $c$ at higher $N(\mathrm{H_{2}})$ which strongly indicates ammonia depletion, towards L429 we have a more flattened but still positive slope $c$, indicating that there is only moderate ammonia depletion happening in the center.

\begin{table*}[htb]
\centering
\begin{threeparttable}
\footnotesize
\caption{Best-fit parameters of the broken line model of the three cores and H-MM1 (\citealt{Pineda22a})}
 \label{tab:bl_sctau}
 \begin{tabular}{llllll}
\toprule
          &L694-2$^{\star}$&L429& L1517B$^{\star}$&  H-MM1&\\
Parameter & && & &Unit\\
\midrule
$a$&1.3(0.4)&1.3(1.1)&1.6(0.6)&2.0& $\times$10$^{-8}$\\
$b$&-0.5(6.0)&-1.6(4.6)&-1.0(3.3)&1.1& $\times$10$^{14}$\,cm$^{-2}$\\
$c$&-&0.73(6.6)&-&$-$1.1&$\times$10$^{-8}$\\
$x_{0}$&  - &7.5(2.0)&-&2.6&$\times$10$^{22}$\,cm$^{-2}$\\
\bottomrule
\end{tabular}
 \begin{tablenotes}
      \small
      \item $^{\star}$ A linear relation is fitted to L1517B and L694-2 instead of the broken line model. \\
      \item Uncertainty of the parameter is listed in brackets, in percentage unit with respect to the best-fit value.
      \end{tablenotes}
 \end{threeparttable}
\end{table*}

\section{Discussion}\label{sec:dis}
\subsection{Temperature structure}\label{sec:temp}
We resolve a slight decrease of gas temperature towards L429 and L1517B, varying from $\sim$9 K to below 8 K for L429, and $\sim$11 K to 10 K for L1517B. The situation resembles the case of H-MM1 (\citealt{Pineda22a}), while H-MM1 shows an overall higher temperature, ranging from $\sim$12 K to 11 K. 
L694-2 shows a rather uniform temperature around 9 K, which is consistent with previous findings that suggest it is a prolate core slightly inclined to the line-of-sight (\citealt{Harvey03}). Compared to L1544 that displays a strong temperature drop down to $\sim$6 K (\citealt{Crapsi07}), these temperature variations are minor. Previous single-dish observations of NH$_{3}$ towards starless core L1517B report a constant temperature of 9.5 K (\citealt{Tafalla02, Tafalla06}), while with our improved angular resolution, we still only resolve a minor decrease of temperature, which is similar to that resolved in starless core CB17 (\citealt{Spear21}) and the northern Ophiuchus D core (\citealt{R11}). The nearly isothermal nature of these cores suggests that they are at an earlier stage of evolution (\citealt{Keto08}), characterised by a low gas density $\lesssim$10$^{5}\,$cm$^{-3}$ (line-of-sight averaged density seen by NH$_{3}$) and cooling for the bulk gas is governed by molecular line emission.   

The temperature drop in our sampled three cores is less significant than that in the prototypical pre-stellar core L1544, which is likely at a more evolved stage of evolution. In L1544, there is a drastic temperature drop seen in the inner $\sim$2500\,au down to 6\,$K$ (\citealt{Crapsi07}), while L429 has a similar central density and evolutionary stage as L1544, the fact that it locates further away with an achieved angular resolution similar to 2500\,au may also have hindered the significant temperature drop to be probed. One note is that \citet{Crapsi07} used VLA-only data to estimate $T_{\mathrm{kin}}$; in Appendix \ref{app:tkin_interf} we present the variations of $T_{\mathrm{kin}}$ derived from VLA-only data cubes,  \citet{Crapsi07}, which also only show minor drop of temperature.   

\subsection{Charactersing the temperature structure of the cores: radiative transfer modeling with {\tt{RADMC-3D}}}

The central drop of gas and dust temperature in pre-stellar cores is due to shielding and irradiation of their embedded radiation field. We note that the derivation of a temperature map assuming homogeneous layer(s) is subject to line-of-sight averaging effect and cannot fully recover the possible underlying temperature gradient. A full radiative transfer (RT) modeling incorporating the density, temperature and NH$_{3}$ abundance profiles tailored for the sampled cores may be more sensitive to the subtle change of the temperature structure, which is beyond the scope of this work. To understand the environment interstellar radiation field of the three cores and to gauge the gas temperature variations from NH$_{\mathrm{3}}$ lines, we conducted radiative transfer modelling with {\tt{RADMC-3D}} (\citealt{Dullemond}), assuming that the density structure of the cores follows the 1-dimensional form derived in Sect. \ref{sec:nprofiles}. We start the modeling by assuming a standard interstellar radiation field (ISRF) following the form in \citet{Hocuk17}, which consists of six modified black body components (\citealt{Zucconi01}) and a UV contribution (\citealt{Draine78}). We adjust the scaling of the standard ISRF and arrive at different dust temperature profiles. The mass averaged temperature profile is then calculated, assuming the threshold above which the gas structures are seen by NH$_{3}$ emission are 1$\times$10$^{4}\,$cm$^{-3}$ for L1517B, and 2$\times$10$^{4}\,$cm$^{-3}$ for L429 and L694-2. The thresholds are determined by comparing the observed $T_{\mathrm{ex}}$ and $T_{\mathrm{kin}}$ relation for each source to the modelled 
$T_{\mathrm{ex}}$ vs. $T_{\mathrm{kin}}$ relation based on non-LTE calculations for NH$_{3}$ column density of 6$\times$10$^{14}$\,cm$^{-2}$ (\citealt{Shirley15}). 

\begin{figure}[htb]

\includegraphics[scale=0.4]{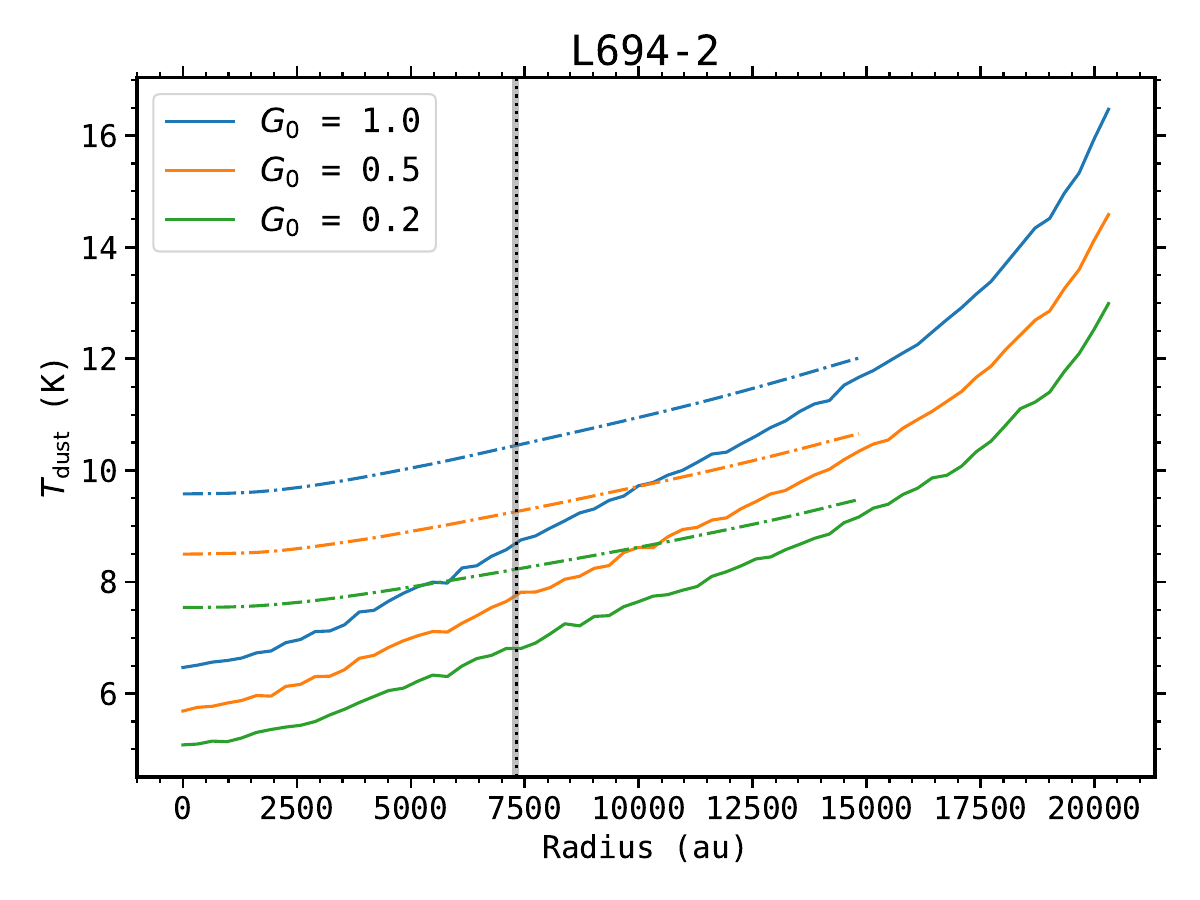}\\
\includegraphics[scale=0.4]{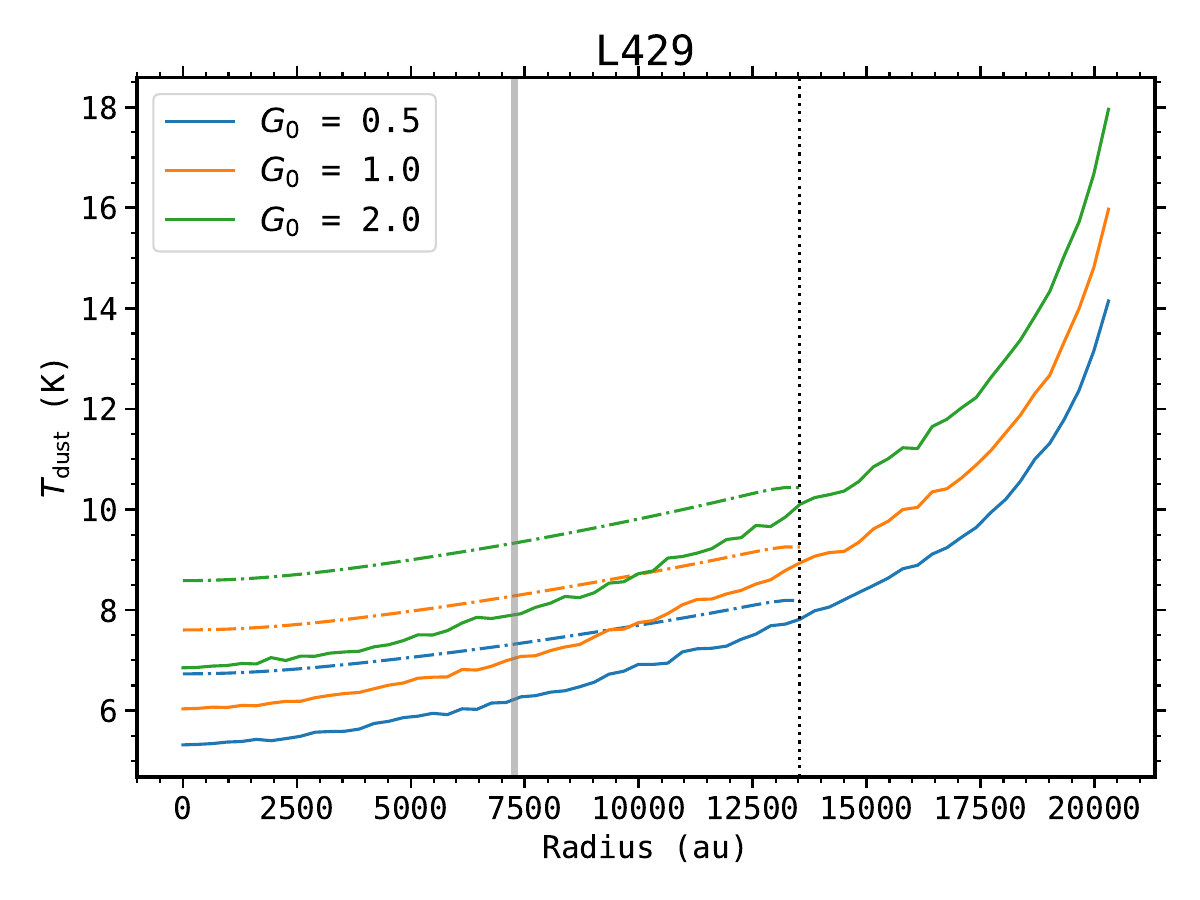}\\
\includegraphics[scale=0.4]{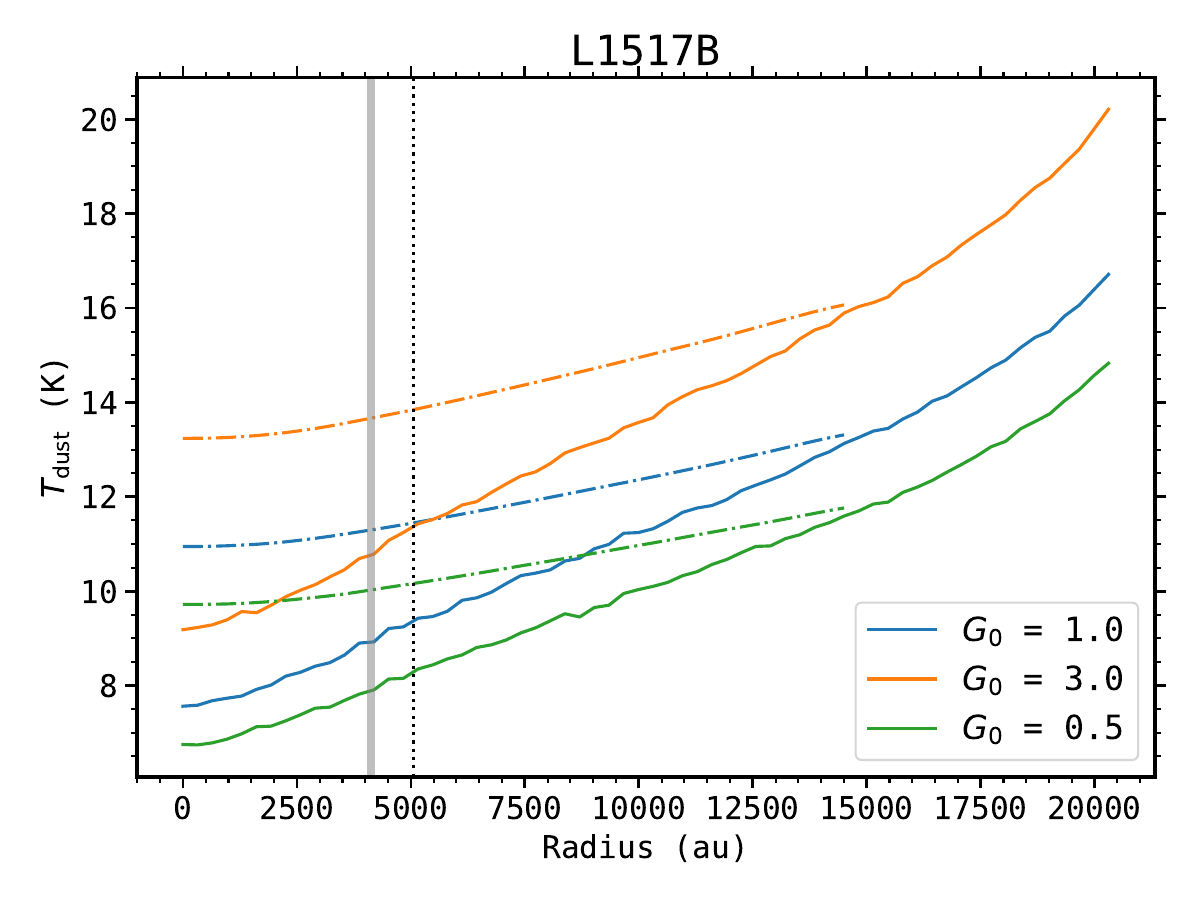}\\
\caption{Radial temperature profiles (solid lines) calculated by {\tt{RADMC-3D}} assuming different ISRF intensity, and the correspondant LOS mass averaged temperature profiles (dash-dotted lines).The vertical solid line in gray indicates the position within which gas density is above 10$^{5}$\,cm$^{-3}$ for each core. The vertical dotted line mark the effective radius of the obtained $T_{\mathrm{kin}}$ map from fitting the NH$_{3}$ (1,1) and (2,2) lines.}
\label{fig:radmc_temp}
\end{figure}

In Fig. \ref{fig:radmc_temp} we plot the output radial temperature profiles from {\tt{RADMC-3D}}, together with the (projected) radial profile after mass averaging along LOS considering the density profile constrained in Sect. \ref{sec:nprofiles}. The observed physical resolution is also taken into account in the mass averaging. If we assume that dust and gas are fully coupled above a gas density of 10$^{5}$\,cm$^{-3}$ (\citealt{Goldsmith01}, indicated by the vertical solid line in Fig. \ref{fig:radmc_temp}), then the resultant temperature profile of $G_{\mathrm{0}}$ = 1-2 for L429 matches well to the observed gas kinetic temperature derived from NH$_{3}$, showing a slight decrease from above 9 K to 8 K within the effective radius of the $T_{\mathrm{kin}}$ map derived from NH$_{3}$, while for L694-2, the $G_{\mathrm{0}}$ = 0.5 ISRF produces a rather invariant $\sim$9 K profile within the effective radius. Meanwhile, the NH$_{3}$ abundance drop in the central region of L429 can bring down further the mass averaged temperature to below 8 K, since there is non-uniform sampling of gas mass along LOS. For L1517B, the $G_{\mathrm{0}}$ = 1 ISRF produces a temperature around 11 K, although the drop is even more slight than what is observed. We therefore conclude that considering the high central density of the three cores, the gas kinetic temperatures seen by NH$_{3}$ are consistent with the dust temperatures assuming the cores are bathed in standard ISRF ($G_{\mathrm{0}}$$\sim$1-2). A more significant radial temperature variation towards the three cores may be resolved with a higher sensitivity NH$_{3}$ mapping, i.e. to increase the effective radius of the $T_{\mathrm{kin}}$ map.




\begin{figure*}[htb]
\begin{tabular}{p{0.485\linewidth}p{0.485\linewidth}}
\hspace{-0.15cm}\includegraphics[scale=0.49]{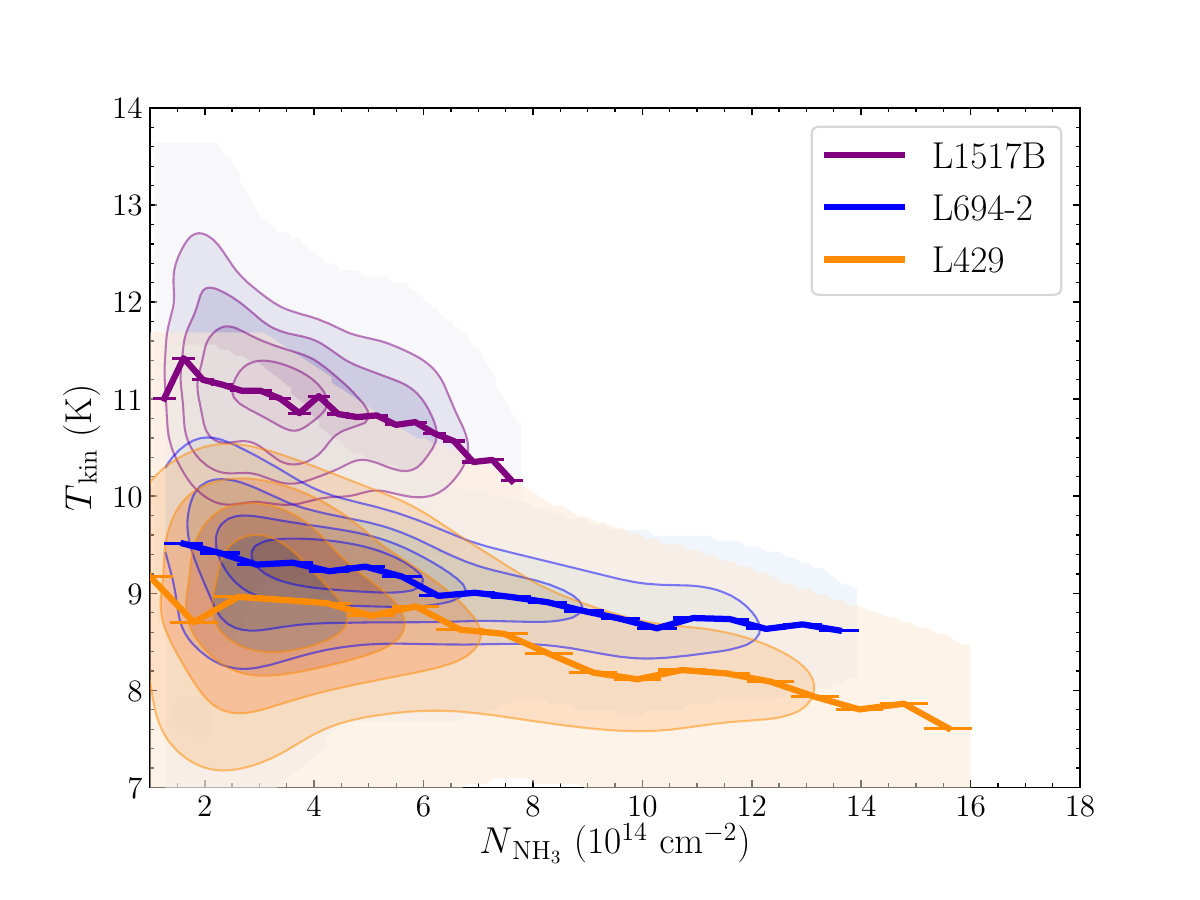}&\hspace{-0.25cm}\includegraphics[scale=0.49]{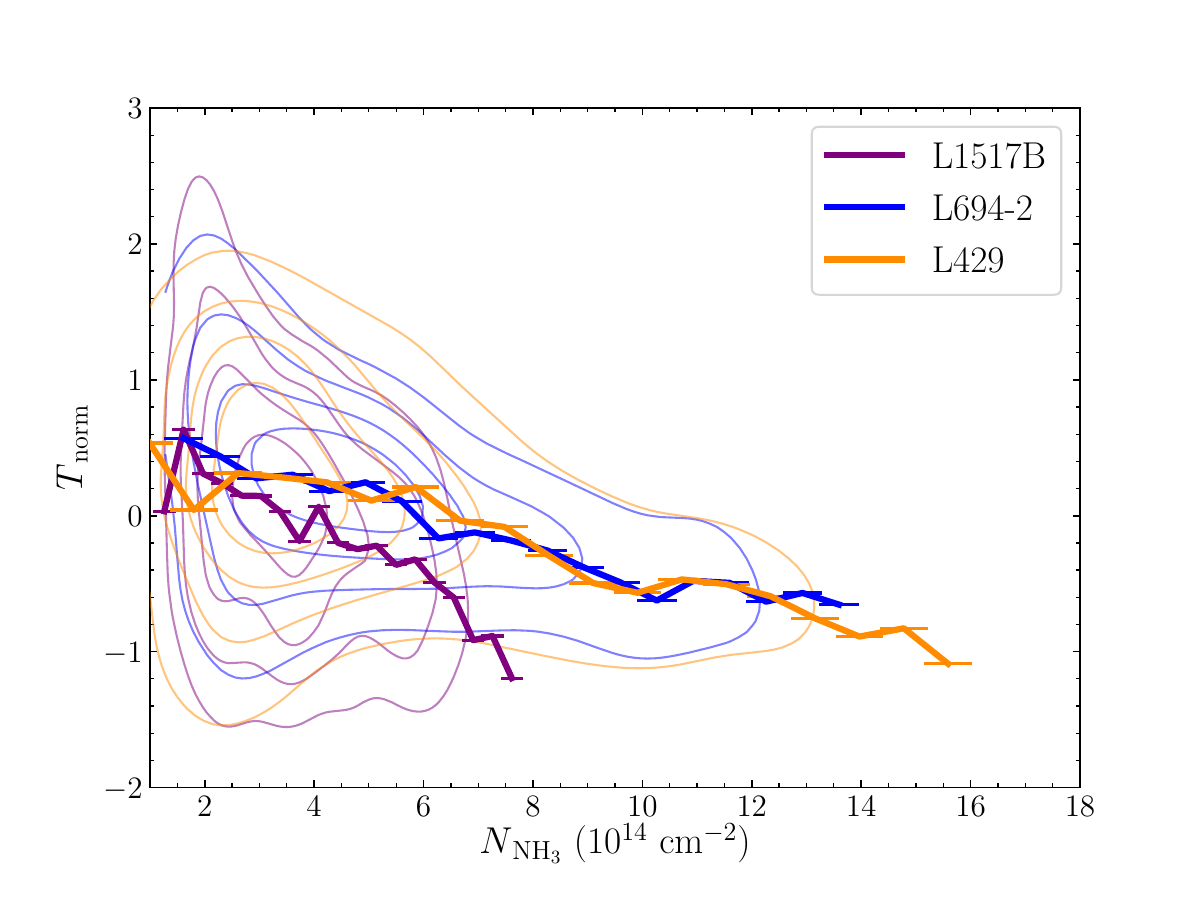}
\end{tabular}
\caption{KDE of kinetic temperature ({\it{left}}) and normalised kinetic temperature ({\it{right}}) as a function of NH$_{\mathrm{3}}$ column density. In the {\it{right}} figure, the kinetic temperatures of each source are transformed with $T_{\mathrm{norm}}$ = $T_{\mathrm{kin}}$-$\bar{T_{\mathrm{kin}}}$/$\sigma_{T_{\mathrm{kin}}}$.}
\label{fig:kde11}
\end{figure*}

\begin{figure*}[htb]
\begin{tabular}{p{0.485\linewidth}p{0.485\linewidth}}
\hspace{-0.25cm}\includegraphics[scale=0.49]{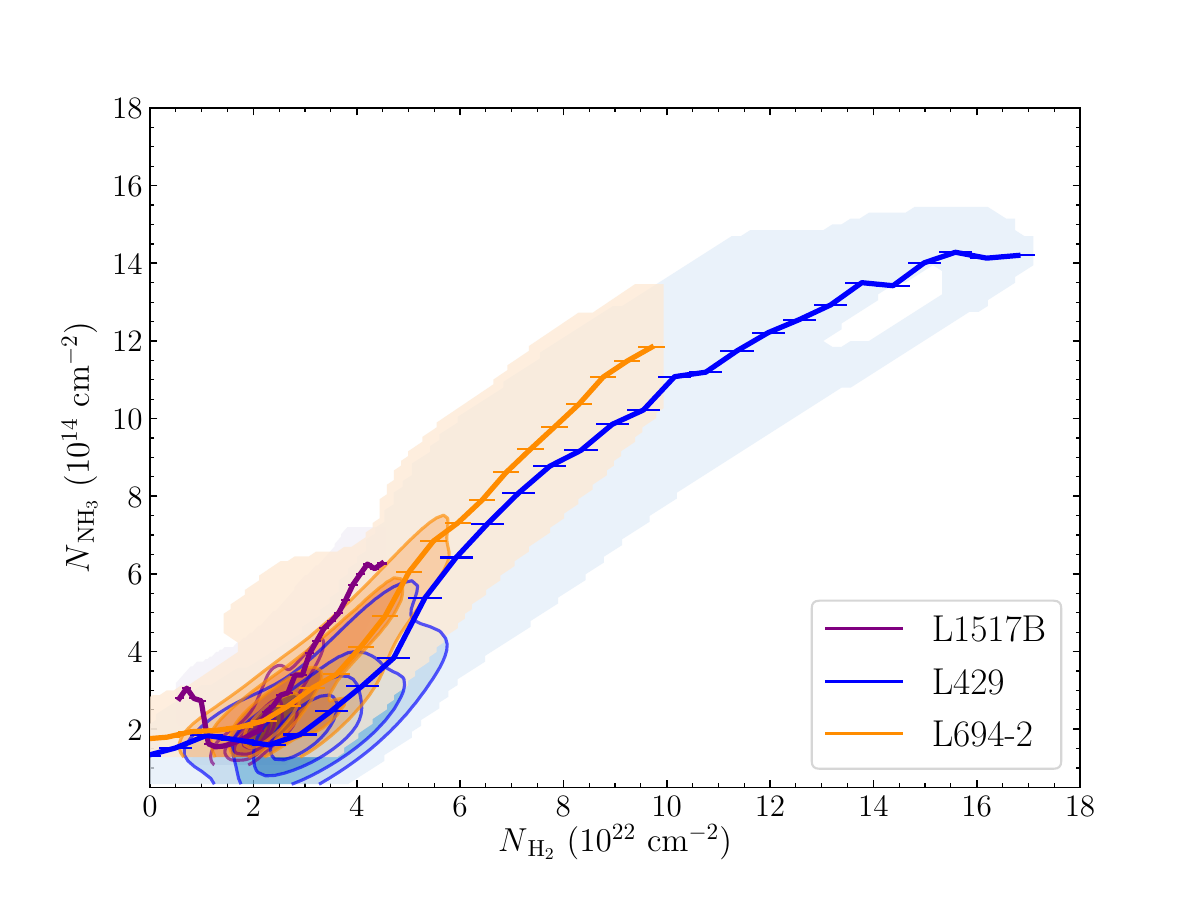}&\hspace{-0.15cm}
\includegraphics[scale=0.49]{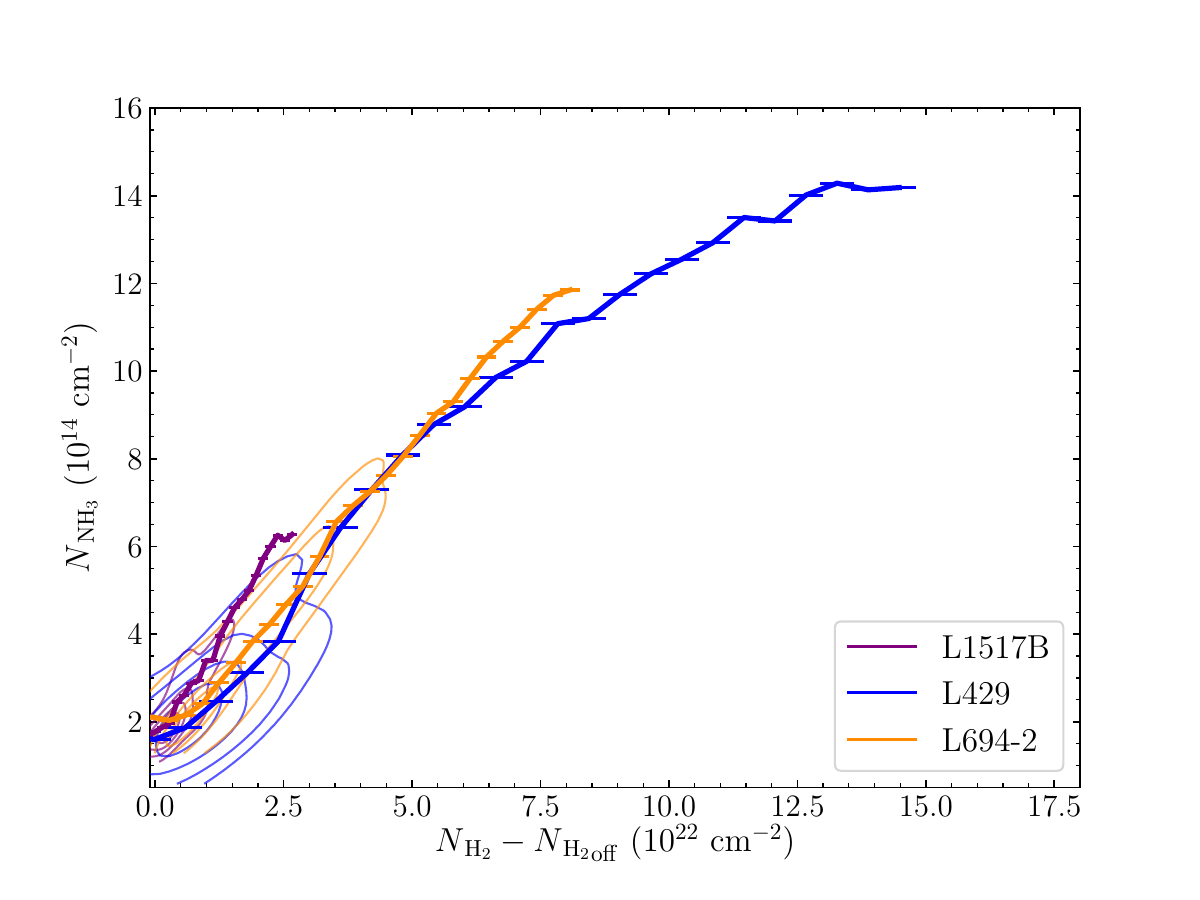}\\
\end{tabular}
\caption{The KDE of NH$_{\mathrm{3}}$ column density as a function of H$_{\mathrm{2}}$ column density. The line segments correspond to mean values in  H$_{\mathrm{2}}$ or NH$_{\mathrm{3}}$ column density bins, which are linked together. In the {\emph{right}} panel, H$_{\mathrm{2}}$ column density is corrected by H$_{\mathrm{2}}$ offset estimated based on ``zero-level" of NH$\mathrm{_{3}}$ column density. }
\label{fig:kde1_sctau}
\end{figure*}

\subsection{Abundance variations of NH$_{3}$ in pre-stellar cores}\label{sec:ab}

At the lower $N(\mathrm{H_{2}})$ end ($N(\mathrm{H_{2}})$<8$\times$10$^{22}$\,cm$^{-2}$), the abundance of NH$_{3}$ of the three cores are close, as indicated by their similar slopes between $N(\mathrm{NH_{3}})$ and $N(\mathrm{H_{2}})$ (Table \ref{tab:bl_sctau}), $\sim$1.5$\times$10$^{-8}$. This abundance value is close to previous interferometric measurements towards other pre-stellar cores (\citealt{Chit14}, \citealt{Spear21}, \citealt{Friesen17}, \citealt{Pineda22a}) within a factor of 2. 

Adopting the new distance measurements of the three cores (\citealt{Galli19}, \citealt{Kim22}, \citealt{OL18}), with the high-resolution extinction map, we derive the central densities of the three cores increase from L1517B to L429; L1517B has a central density of $\sim$3$\times$10$^{5}\,$cm$^{-3}$, $\sim$3 times lower than that towards L694-2 and L429. 
The fact that NH$_{3}$ abundances of L694-2 and L1517B do not vary as significantly as L429 in the core center is likely due to their overall lower density. From the relation of $N(\mathrm{NH_{3}})$ and $N(\mathrm{H_{2}})$ (Fig. \ref{fig:kde1_sctau}) of L429, the cutoff $N(\mathrm{H_{2}})$ where the abundance drops is fitted to be $\sim$7.5$\times$10$^{22}$\,cm$^{-3}$, and the abundance in the core center is 2 times lower than at core edge (comparing $a$ and $c$ in Table \ref{tab:bl_sctau}). This abundance variations are similar to L1689-SMM16 (\citealt{Spear21}). As we use a 1-D model to describe the core density profile, we can convert the cutoff $N(\mathrm{H_{2}})$ to LOS mass averaged gas volume density $\left\langle N(\mathrm{H_{2}})\right\rangle$, yielding 4.4$\times$10$^{4}$\,cm$^{-3}$ for L429.   
\citealt{Pineda22a} suggests a turnover gas density of 2.1$\times$10$^{5}\,$cm$^{-3}$, beyond which the NH$_{3}$ depletion is significant. With the conversion of $N(\mathrm{H_{2}})$ to $\left\langle N(\mathrm{H_{2}})\right\rangle$, none of the three cores has $\left\langle N(\mathrm{H_{2}})\right\rangle$ reaching 2.1$\times$10$^{5}\,$cm$^{-3}$ with the present angular resolution. This explains the moderate NH$_{3}$ depletion towards L429, compared to that resolved in H-MM1, as well as the rather invariant NH$_{3}$ abundance towards L694-2 and L1517B, which are due to their progressively lower gas densities compared to L429. This observational trend of NH$_{3}$ abundance variations is consistent with predictions of chemical models that show various degree of NH$_{3}$ depletion in the central region of the cores (\citealt{Sipila19}). 

Embedded in different parental molecular clouds (see Appendix \ref{app:source_info}), the three cores and H-MM1 have different environmental conditions, which are also reflective in their different average temperature (dominated by the outer parts of the core), and different levels of $N(\mathrm{H_{2}})$ offset (Sect. \ref{sec:x_nh3}). The fact that the NH$_{3}$ abundance variation of the four cores can be consistently explained by their different central densities seems to suggest that the only relevant physical property for the depletion of NH$_{3}$ is the evolutionary stage of the core.








\subsection{Gas kinematics}\label{sec:kinematics}

\begin{figure}[htb]
\begin{tabular}{p{0.85\linewidth}}

\hspace{0.1cm}\includegraphics[scale=0.52]{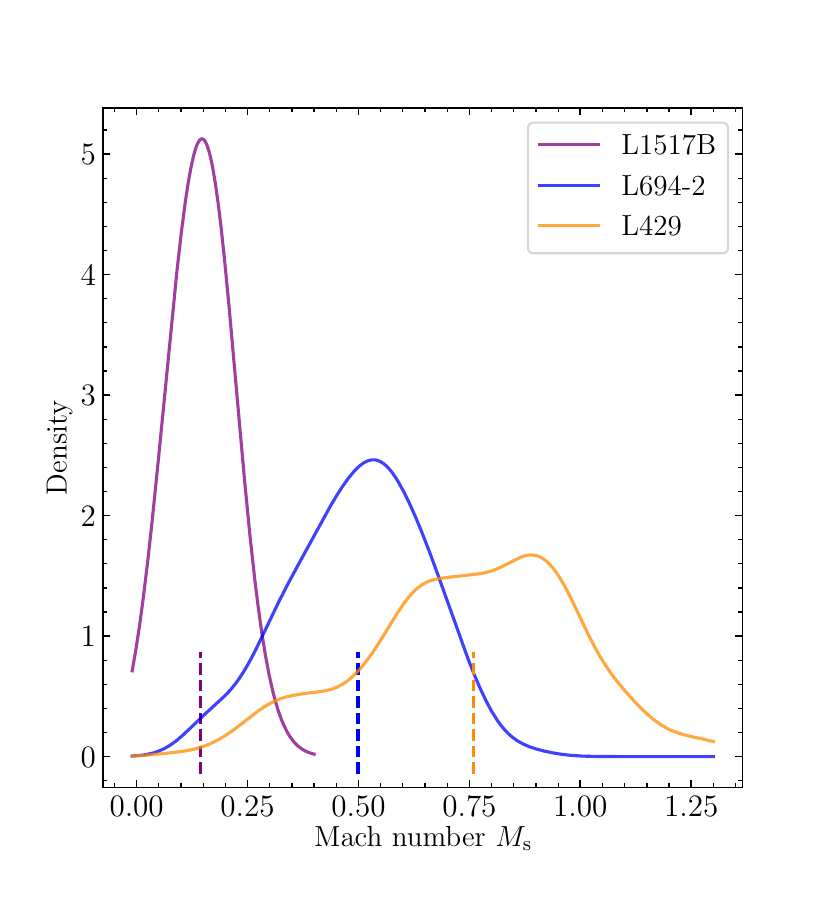}\\

\end{tabular}
\caption{The KDE of the Mach number of the three cores. The mean values are indicated with vertical line segments of respective colors. }

\label{fig:mach_hist}
\end{figure}

As described in Sect. \ref{sec:vfield}, the velocity dispersions, $\sigma_{\mathrm{v}}$ for the three cores show local variations, but are all $\lesssim$0.2$\,$km$\,$s$^{-1}$. We can estimate the thermal and non-thermal contributions using the derived $T_{\mathrm{kin}}$, while taking into account the instrumental broadening due to the channel-to-channel correlation (\citealt{Pineda10}, \citealt{Leroy16}, \citealt{Choudhury20}). The distribution of the sonic Mach number, $M_{\mathrm{s}}$, of the three cores is shown in Fig. \ref{fig:mach_hist}. The cores all appear predominantly subsonic ($M_{\mathrm{s}}$$\leqslant$1): L1517B shows the least non-thermal contribution with an average $M_{\mathrm{s}}$ of 0.2; L694-2 is intermediate with an average $M_{\mathrm{s}}$ of 0.5, and L429 shows the highest $M_{\mathrm{s}}$, on average of 0.8. The highest $M_{\mathrm{s}}$ towards L429 is consistent with its more complex velocity field and higher velocity gradient. 

Considering the stability of the cores, we can compare the scale of the inner plateau $a$ with the Jeans length estimated from central density $n_{c}$ and temperature; the scale ratio is denoted as $k$ in \citealt{DB09}. For similar gas density and temperature, the larger $a$ corresponds to a reduced pressure gradient, and the core is more likely collapsing with prevailing self-gravity. With central temperatures of 10 K of L1517B and 8 K for L694-2 and L429, we arrive at $k$ values of 0.8 of L1517B and 1.3 for both L694-2 and L429. Since the central temperature is estimated from the $T_{\mathrm{kin}}$ maps, which involve LOS averaging, which for a centrally dropping temperature profile, is higher than the real temperature, these $k$ values are likely upper limits. 
Nonetheless, this suggests that L429 and L694-2 are in a more evolved evolutionary phase than L1517B.

The embedding filament of L1517B shows signatures of velocity oscillation (\citealt{Hacar11}) and the core itself seems to reside at the position of converging velocities. The parental filamentary structure exhibits a similar velocity field among tracers sensitive to different density regimes \citep{Hacar11}, appearing to be a velocity coherent structure up to 0.5 pc with subsonic motions. Indeed, the resolved continuous variation of the velocity field inside L1517B, from north-east to south-west (Fig. \ref{fig:vlsr_sigmav}), roughly follows the large-scale filament structure where the velocity gradients may vary in direction due to the flow converging. The velocity turning from the large-scale filament may also originate partially from the filament rotation (e.g., \citealt{AG21}). This may reflect that the physical status of L1517B is close to quasi-static contraction, with the small-scale velocity gradient ($\sim$2 km\,s$^{-1}$\,pc$^{-1}$, Sec. \ref{sec:vfield}) slightly larger than that along the embedding filament ($\sim$1.4 km\,s$^{-1}$\,pc$^{-1}$).  

For L694-2 and L429, the sharp gradients and multiple velocity components (in sub-regions, Appendix \ref{app:2comp}) found at the edge of the cores might be the signature of the ongoing accretion of material from cloud to core (e.g., \citealt{Choudhury20, Choudhury21}, \citealt{Chen22}), which remains to be further studied with larger-scale maps and with better sensitivity. Cloud to core accretion has already been deduced with observations of HCO$^+$ (e.g. \citealt{Redaelli22}) and CS (e.g. \citealt{Mardones97}). How these cloud-core feeding flows link to the even smaller scale asymmetric accretion streamers surrounding the young stellar objects (core to disk flows, e.g., \citealt{Pineda20, Pineda22b}) carry essential clues of how protostars gain masses under the complex interplay of physical mechanisms over different spatial scales.

\section{Conclusions}
We present NH$_{3}$(1,1) and (2,2) data combined from VLA and GBT observations, for one starless core L1517B, and two pre-stellar cores L694-2 and L429. We derive the parameter maps from fitting the NH$_{3}$ lines and obtain the gas kinetic and excitation temperature, NH$_{3}$ column density, centroid velocity and velocity dispersion maps. We use {\it{Spitzer}} 8\,$\mu$m to derive high angular resolution hydrogen column density maps for the three cores. Our main findings are as follows.
\begin{itemize}
    \item A minor temperature drop is resolved towards L429 and L1517B, showing $\sim$9\,K at core edge to below 8 K in center, and $\sim$11 K to 10 K, respectively. L694-2 shows a uniform temperature structure, of $\sim$9\,K. These resolved gas radial temperature profiles are roughly consistent with each core's LOS mass averaged dust temperature profiles with a standard ISRF, considering their different gas density profiles and our effective mapping scales. 
    \item The Bonnor-Ebert like gas density profiles are constrained from the high-resolution extinction map (6$''$). The central density of L429 and L694-2 appears the largest, 0.8-1$\times$10$^{6}$\,cm$^{-3}$, while L1517B has a central density of 3.2$\times$10$^{5}$\,cm$^{-3}$. 

    \item The line-widths are predominantly subsonic with progressively overall increase from L1517B, L694-2 and L429, consistent with the increasing magnitude of the local velocity gradient. The velocity fields of L1517B and L694-2 show relatively large-scale variations, likely reflecting a mixture of contracting motions and acoustic oscillation. L429 exhibits more localised velocity changes, possibly reflecting an advanced collapse phase that involves more complex gas motions. 
    \item The NH$_{3}$ abundance becomes $\sim$2 times lower in the central region of L429, which happens around hydrogen column density of 7.5$\times$10$^{22}$\,cm$^{-2}$. The corresponding H$_{\mathrm{2}}$ volume density is 4.4$\times$10$^{4}$\,cm$^{-3}$. This is below the cutoff density of $\sim$2$\times$10$^{5}$\,cm$^{-3}$ where a stronger abundance drop is seen towards H-MM1 (\citealt{Pineda22a}).  
    L694-2 and L1517 show a rather invariant NH$_{3}$ abundance. 
    Overall, the three cores plus H-MM1 compose a consistent trend of 
    NH$_{3}$ depletion with increasing gas density in the central region of pre-stellar cores. 
\end{itemize}

\begin{acknowledgements}
      The authors acknowledge the financial support of the Max Planck Society. H.B.L. is supported by the National Science and Technology Council (NSTC) of Taiwan (Grant Nos. 111-2112-M-110-022-MY3).
\end{acknowledgements}

\bibliography{zref}

\begin{appendix}
\section{Target sources}\label{app:source_info}
\subsection*{L694-2}
L694-2 is a relatively isolated pre-stellar core (\citealt{Lee11}, \citealt{Spezzano16}). Based on {\it{Gaia}} Data Release 2 astrometric data, the new distance of L694-2 is updated to 203 pc (\citealt{Kim22}). Observations of N$_{2}$H$^+$ (1-0), DCO$^+$ (2-1) and HCO$^+$ (3-2), (4-3) lines all show blue-skewed profiles, which suggest infall motions from the extended region down to the inner core area (\citealt{Williams06}, \citealt{Keown16}, \citealt{Kim22}). Compared to the proto-typical late-stage pre-stellar core L1544 (\citealt{KC10}, \citealt{Caselli19, Caselli22}), its density gradient is smaller (\citealt{Williams06}), with larger infall velocities associated with outer layers (\citealt{Keown16}). These indicate L694-2 is a less evolved pre-stellar core than L1544. In addition, the core appears to be elongated and its density profile analysis based on near-infrared extinction suggests that it is likely a prolate structure with the major axis slightly inclined with respect to the line of sight (\citealt{Harvey03}). L694-2 has an embedding cloud that appears to be filamentary and evidence of gas flows originated from the parental cloud that feed the core has been found (\citealt{Kim22}). 

\subsection*{L429}
L429 is a late-stage pre-stellar core that holds high deuterium fraction among samples of starless cores (\citealt{Crapsi05}, \citealt{Bacmann03}, \citealt{Caselli08}). The core is likely associated with the Aquila Rift, located at a distance of 436\,pc (\citealt{OL18}). It appears to be a dark, absorption feature even at 70\,$\mu m$, reflecting that very dense gas is residing within the core, which may be already collapsing if there is not strong magnetic field support (\citealt{Stutz08}). This is compatible with the tentative infall signatures shown in the CS (3-2) line (as L429-1 in the survey of \citealt{Lee04}) and HCN (1-0) line (\citealt{Sohn07}). However, red-skewed line profile of CS lines are also detected towards the core (\citealt{Lee11}); according to the evolutionary sequence drawn by \citet{Lee11}, L429 is marked as an intermediate evolutionary stage, between L1517B and L694-2, showing expanding motions when a static core is being perturbed. 

\subsection*{L1517B}
L1517B is a starless core located in the molecular cloud Taurus-Auriga cloud (\citealt{Elias78}, \citealt{Galli19}), which is close to a state of thermal equilibrium (\citealt{Crapsi05}, \citealt{Tafalla04}, \citealt{Kirk06}). The chemical modeling of \citet{Maret13} suggests the core is at an evolved stage that may quickly collapse to form a protostar. Single-dish observations towards L1517B reveal systematically red-skewed line profiles (\citealt{Tafalla04}, \citealt{Sohn07}, \citealt{Fu11}). Analytical works of \citet{Fu11} suggest that L1517B is a typical core that shows coexistence of envelope expansion (or oscillating motions, e.g., \citealt{LouGao11}) and core collapse, which produce the observed spectra. The deuteration fraction derived from N$_{2}$D$^+$ and H$_{2}$D$^+$ of L1517B is much lower than that of L429 and L694-2 (\citealt{Crapsi05}, \citealt{Caselli08}, \citealt{Koumpia20}), indicating L1517B is at an earlier evolutionary stage than L429 and L694-2 (see also \citealt{Schnee13}).

\section{Derivation of the combined 850\,$\mu$m map}\label{app:850cb}
We retrieve the raw data of SCUBA2 850\,$\mu$m of L429 and L694-2 from the Canadian Astronomy Data Center (CADC{\footnote{https://www.cadc-ccda.hia-iha.nrc-cnrc.gc.ca/en/}}) archive. {\tt{SMURF}} software (\citealt{Chapin13}) implemented in {\tt{Starlink}} package {\footnote{The Starlink software (\citealt{Currie14}) is currently supported by the East Asian Observatory.}} is adopted for the data reduction. We use the {\tt{Makemap}} command with the configuration file{\footnote{The \texttt{dimmconfig\_bright\_extended$.$lis} file provided in {\tt{SMURF}} package}} suited for bright extended sources; compared to the parameters used for the generic pipeline products, a less aggressive spatial filtering of the raw map is conducted in the initial data cleaning to preserve the extended structures. 

\begin{figure*}[htb]
\centering
\hspace{-.5cm}
\includegraphics[height=8cm]{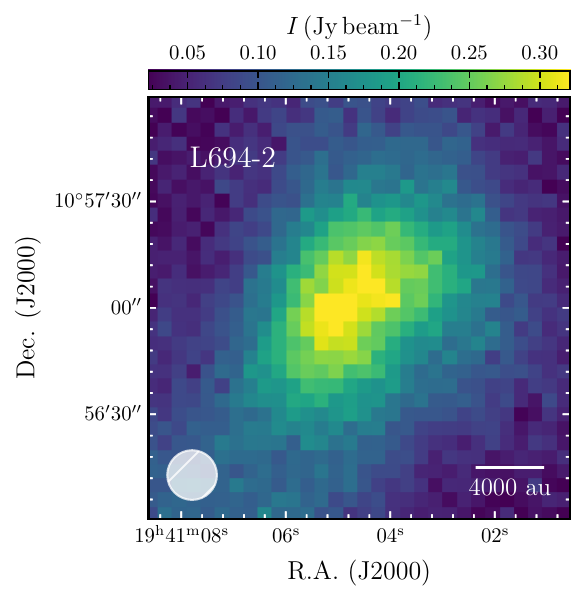}
\includegraphics[height=8cm]{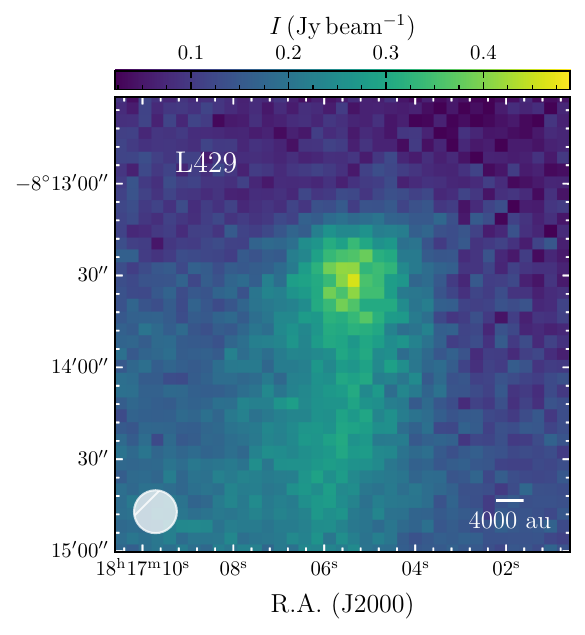}
\caption{The 850$\,\mu$m maps combining SCUBA2 850$\,\mu$m and deconvolved {\it{Planck}} images, of L694-2 and L429.}
\label{fig:scuba2_comb}
\end{figure*}

\begin{figure*}[htb]
\centering
\hspace{-.5cm}
\includegraphics[height=8cm]{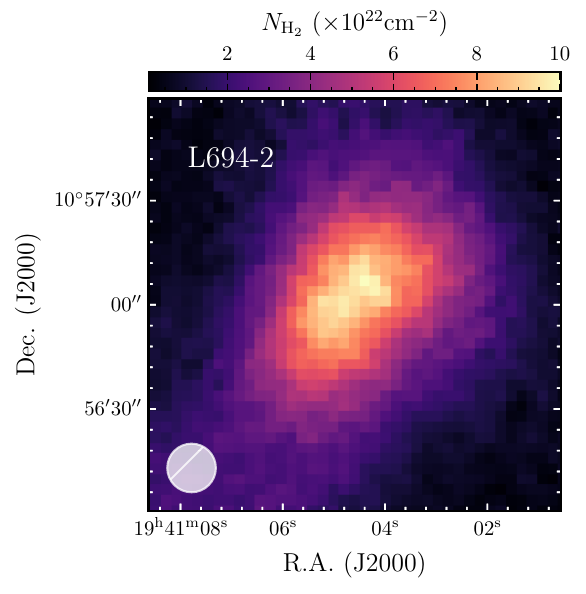}
\includegraphics[height=8cm]{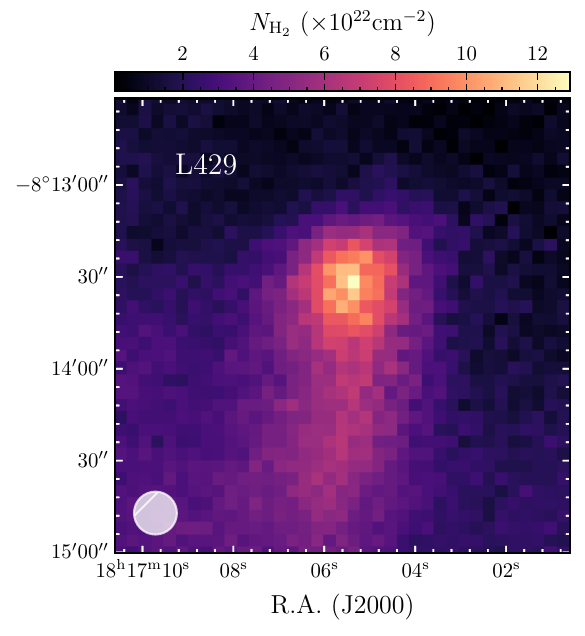}
\caption{The 14$''$ hydrogen column density map derived from combined 850$\,\mu$m map shown in Fig. \ref{fig:scuba2_comb} and $T_{\mathrm{d}}$ map from SED of {\it{Herschel}} data.}
\label{fig:scuba2_comb_NH2}
\end{figure*}

We extrapolate a 850$\,\mu$m flux map from SED of {\it{Herschel}} SPIRE data, and use this map as a model image to deconvolve the {\it{Planck}} image  
with Lucy-Richardson algorithm (\citealt{Lucy74}). The obtained deconvolved {\it{Planck}} image has an angular resolution close to the SPIRE 500$\,\mu$m image and preserves the flux level of the {\it{Planck}} image. The latter property is essential as the SCUBA2 maps suffer from different level of missing fluxes, which can be source sensitive, dependent on the observing weather condition and data reduction parameters. The deconvolved image which contains the extended emission structures is then combined with the SCUBA2 850$\,\mu$m image in the Fourier domain using {\tt{J-comb}} algorithm (\citealt{Jiao22}). The method has proved to be superior than conventional linear image combination methods implemented in e.g. {\tt{immerge}} in {\tt{miriad}} and {\tt{feather}} in {\tt{CASA}}, by largely suppressing imaging defects inherent to the ground-based bolometric observations as well as conserving the high-resolution Gaussian beam pattern (\citealt{Jiao22}). The obtained 14$''$ combined 850$\,\mu$m map of L694-2 and L429 is shown in Fig. \ref{fig:scuba2_comb}. Applying the $T_{\mathrm{d}}$ map from SED of {\it{Herschel}} SPIRE data, we calculate the $N(\mathrm{H_{2}})$ map shown in Fig. \ref{fig:scuba2_comb_NH2}.

\section{Comparison between adopted methods for the combination of VLA and GBT data}\label{app:comb_nh3_app}

We adopt both the joint deconvolution method using {\tt{Miriad}} and the model-assisted clean method using {\tt{CASA}}, for the combination of VLA and GBT data. The last step of both methods is linearly adding the single-dish data with the previous, intermediate combined product in the Fourier domain with {\tt{feather}} and {\tt{immerge}} task implemented in each software. In particular, the mathematical form used in the {\tt{immerge}} task ensures the total flux (within the primary beam) of the final combined image preserves that of the single-dish data (see e.g., \citet{Jiao22}). 

\begin{figure*}[htb]
\centering
\hspace{-.75cm}
\includegraphics[height=5.25cm]{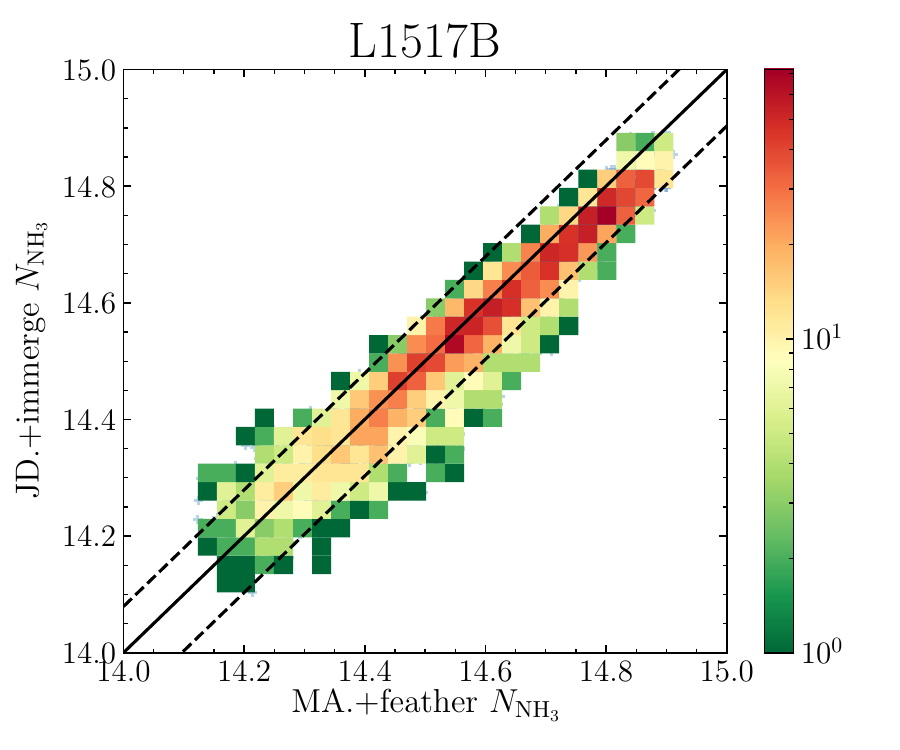}
\includegraphics[height=5.25cm]{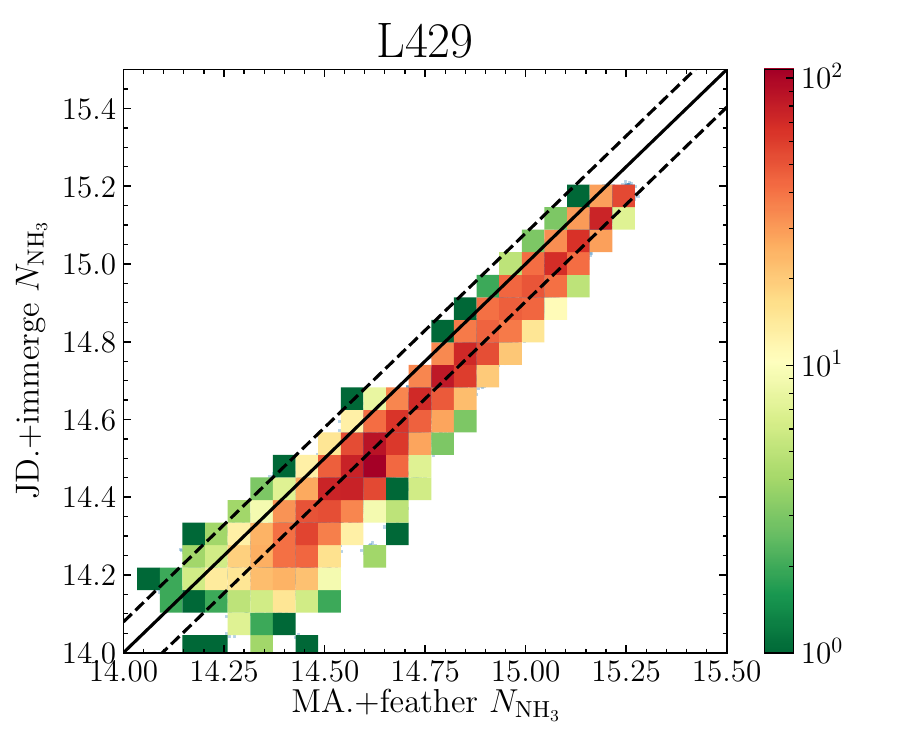}
\includegraphics[height=5.25cm]{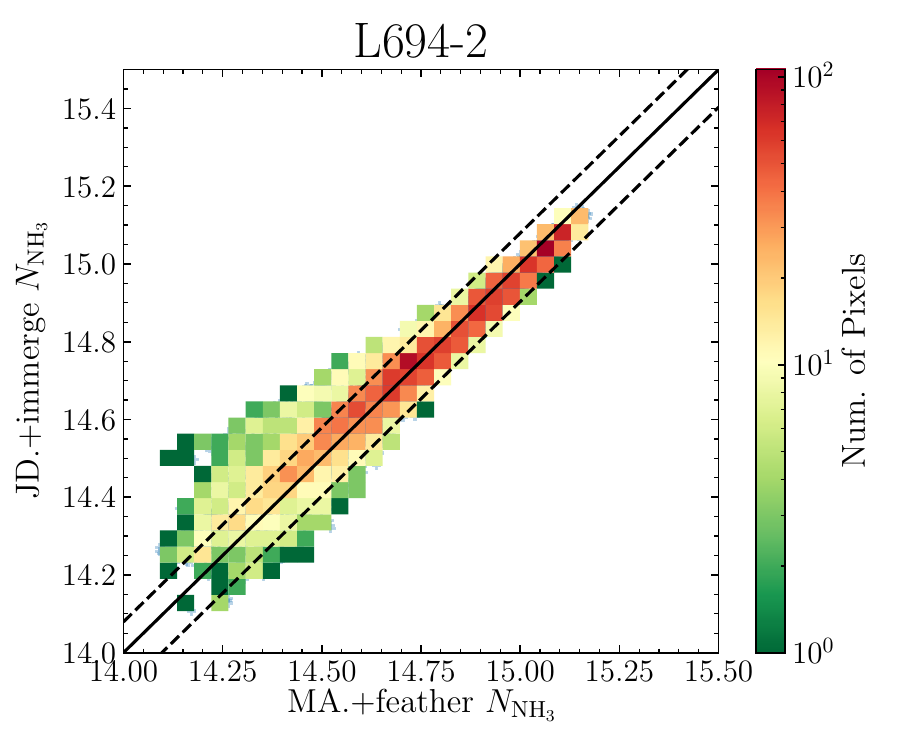}
\caption{2D histogram of comparison between fitted $N(\mathrm{NH_{3}})$ of different combined data products: one based on joint-deconvolution (JD) and {\tt{immerge}} using {\tt{Miriad}} and the other model-assisted (MA) multi-scale clean plus {\tt{feather}} using {\tt{CASA}}. The diagonal dashed and solid lines indicate a factor of 1.2, 1 and 0.8 difference.}
\label{fig:NNH3_comp}
\end{figure*}

A thorough benchmark between the two methods, dependent on the input parameters (e.g., the weighting of the single-dish data in the first method) and the characteristics of the observational data (e.g, sparse or decent {\it{uv}} sampling of the interferometric data, overlapping scales between the single-dish and interferometric data, etc.), is beyond the scope of this paper. In what follows, we simply compare the fitted NH$_{3}$ column density ($N(\mathrm{NH_{3}})$) using the combined image products from the two methods, shown in Fig. \ref{fig:NNH3_comp}. We found there is no significant discrepancy or systematic bias of the two set of $N(\mathrm{NH_{3}})$ values, which means that the different combination approaches should not affect the analysis and comparison of our results with respect to \citet{Pineda22a}.   

\section{Derivation and uncertainty of $N(\mathrm{H_{2}})$}\label{app:unc_NH2}

We use the SMF method in \citet{ButlerTan09} for background emission level estimates. After testing a set of different sized filters, we settled with a filter size of 90$''$, which minimizes the visual defect of interpolation while still preserving the median- to large-scale background emission pattern.  
The difference of the average background emission level for a filter size ranging between 60-270$''$ is within 10$\%$. 

\begin{figure*}[htb]
\centering
\hspace{-1cm}
\includegraphics[width=5.5cm]{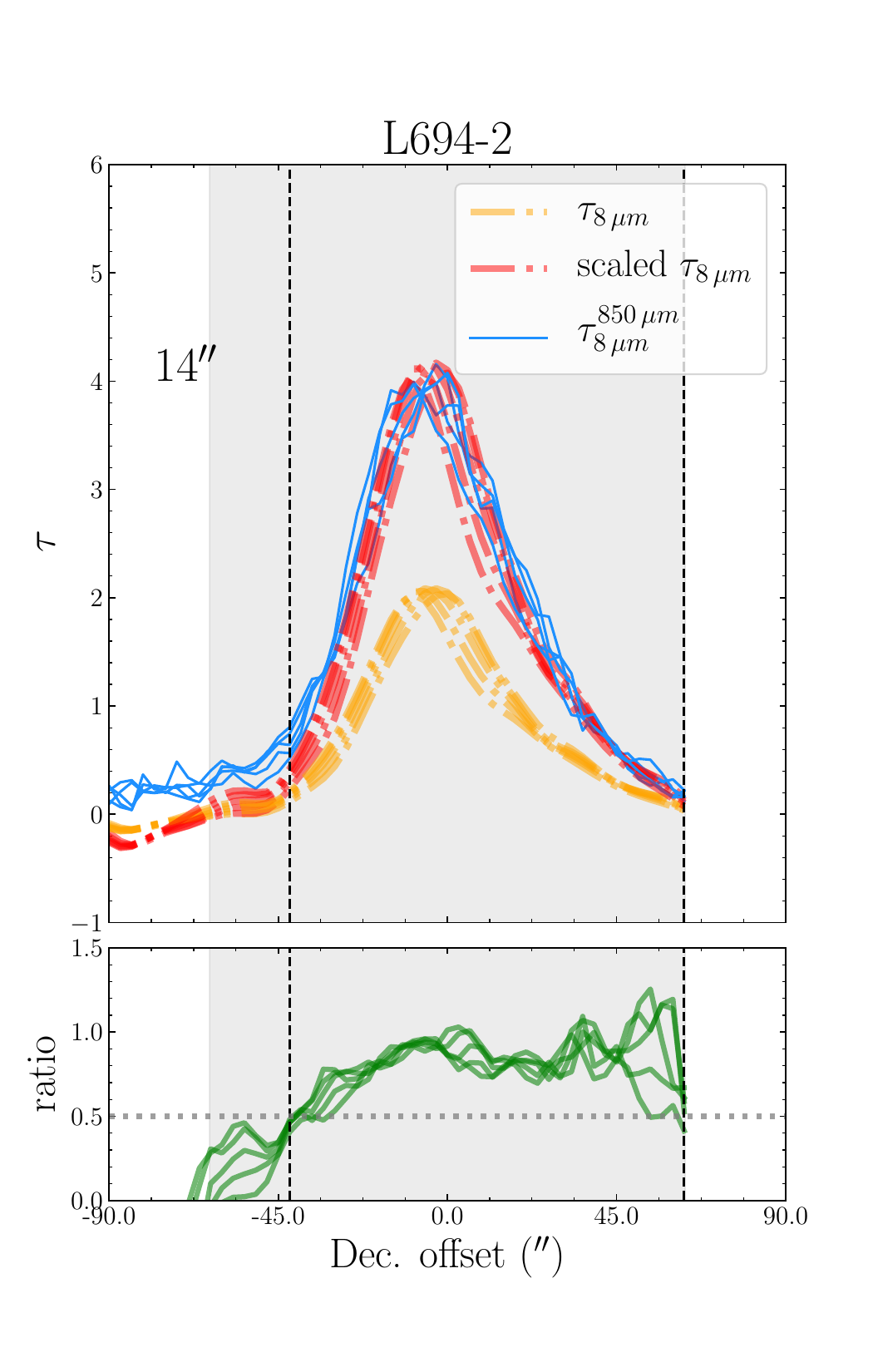}
\includegraphics[width=5.5cm]{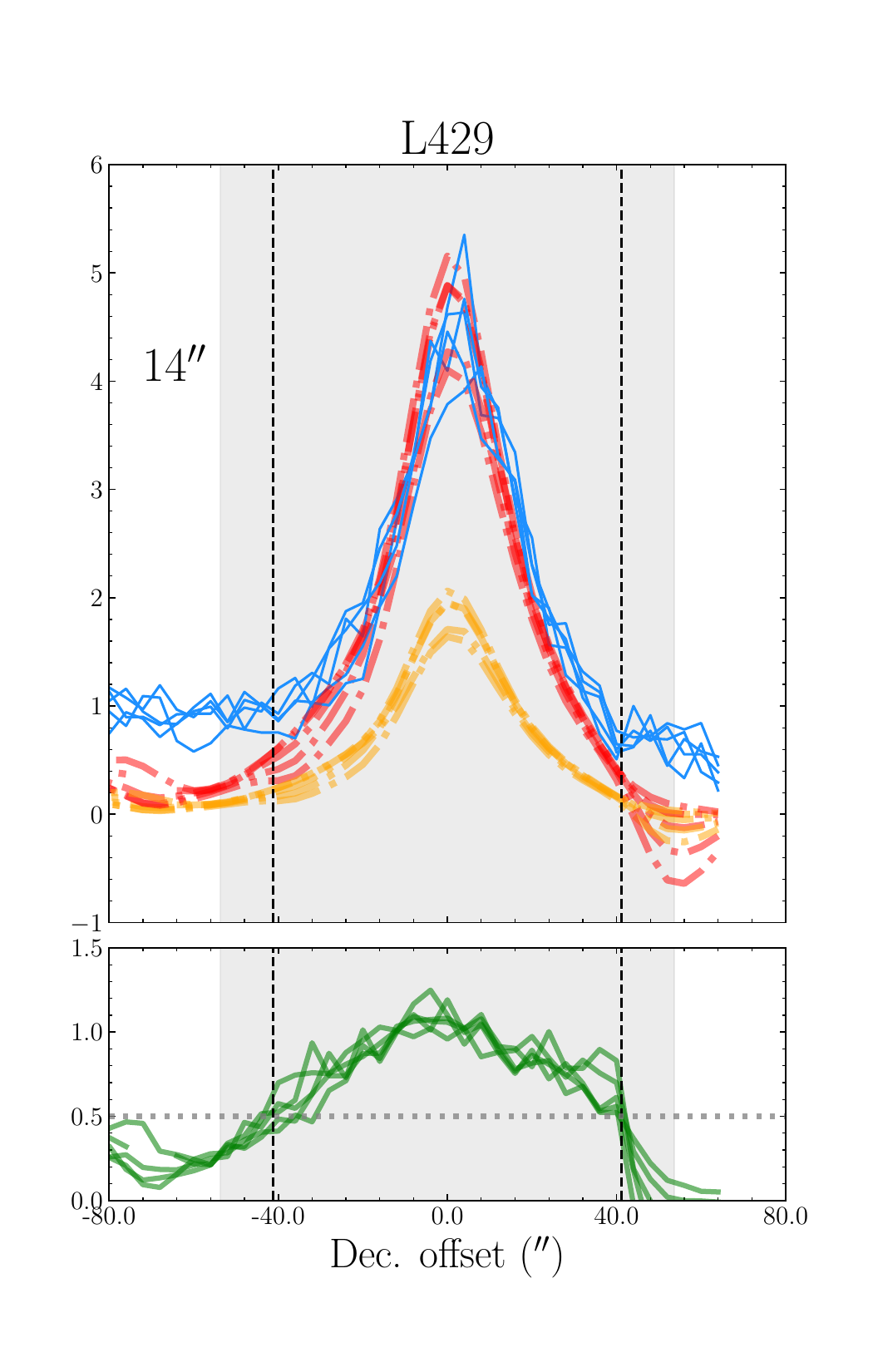}
\includegraphics[width=5.5cm]{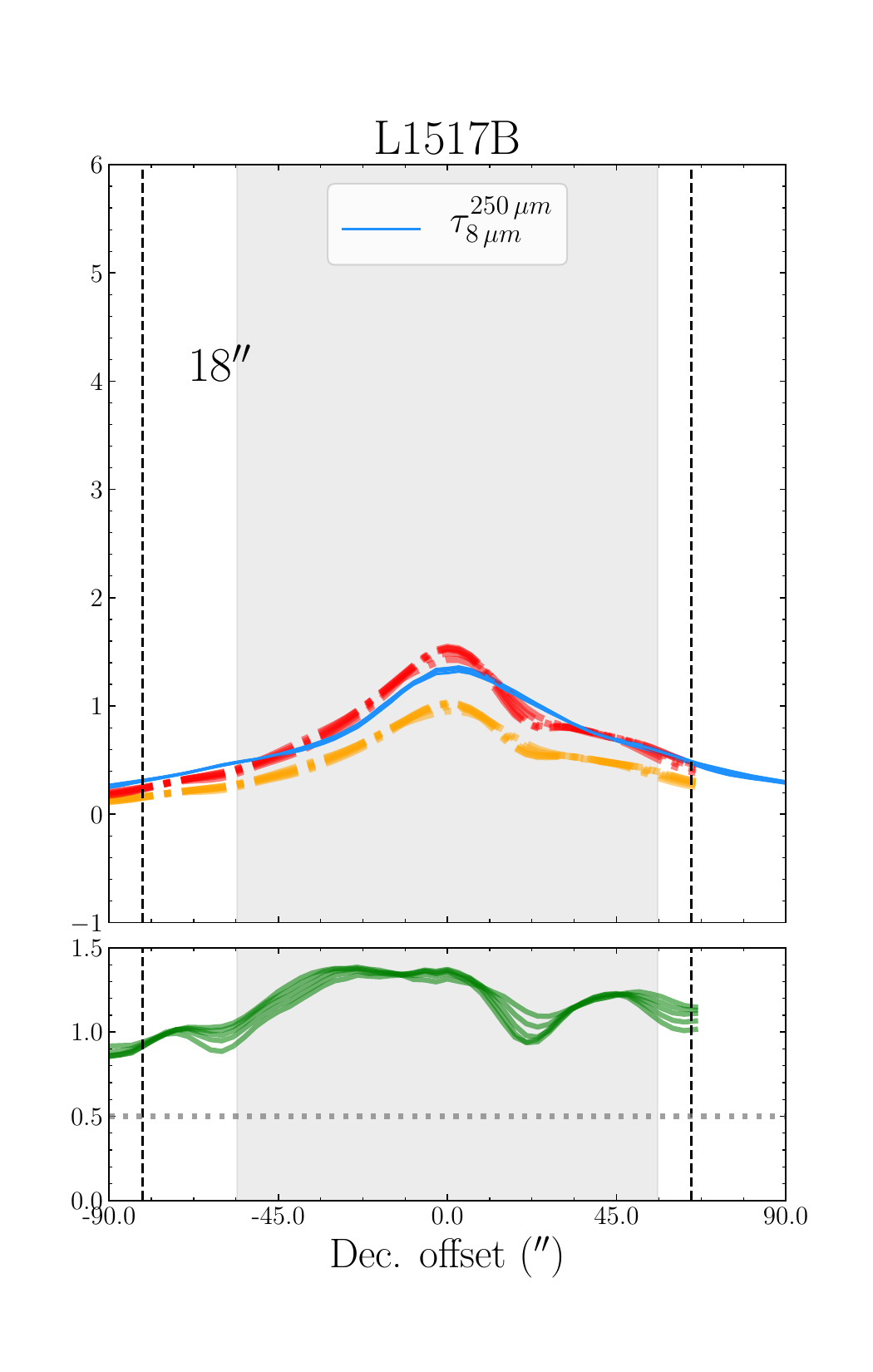}
\caption{Comparison between profiles (along Declination) of opacity maps, extracted from original $\tau_{\mathrm{8\,\mu m}}$ map, and that predicted from $\tau_{\mathrm{850\,\mu m}}$ (denoted as $\tau_{\mathrm{8\,\mu m}}^{\mathrm{850\,\mu m}}$) or $\tau_{\mathrm{250\,\mu m}}$ in the case of L1517B, and the scaled-up $\tau_{\mathrm{8\,\mu m}}$ map. The two versions of $\tau_{\mathrm{8\,\mu m}}$ map are both smoothed to 14$''$ (18$''$) to match the angular resolution of the $\tau_{\mathrm{850\,\mu m}}$ ($\tau_{\mathrm{250\,\mu m}}$) map. The lower panel shows the ratios between scaled-up $\tau_{\mathrm{8\,\mu m}}$ and $\tau_{\mathrm{8\,\mu m}}^{\mathrm{850\,\mu m}}$ ($\tau_{\mathrm{8\,\mu m}}^{\mathrm{250\,\mu m}}$). Gray shaded region indicates the effective radius of the final trimmed parameter maps of NH$_{3}$. Vertical dotted lines show the offset where the ratio drops below 0.5 (indicated by the horizontal dotted line). }
\label{fig:tau_comp}
\end{figure*}

Practically, the assumption of constant foreground emission level can cause underestimation of the column densities particularly in the outer parts of the core. By inspecting the 8$\mu m$ map and the contours of $N(\mathrm{H_{2}})$ map from {\it{Herschel}} it is obvious that some bright 8$\mu m$ emission is associated with the foreground of the source. The small-scale median filter method effectively removes all contribution from compact bright sources for interpolation at the core area, ignoring the fact that there may be a bunch of stars that compose the background of the cores. The uncertainties associated with back- and foreground emission level estimates are hard, if not impossible, to assess.  We resort to matching the $\tau$ profiles from the extinction method and from that the SED by a constant scaling factor. This is based on essentially the assumption that the morphological features of the $N(\mathrm{H_{2}})$ map derived by the extinction method, preserve the real profiles of the core structure without spatially varying degrees of deficit. This is largely based on what we find empirically, the constant scaling up can match the column density cuts along the core for the inner part and outer part simultaneously. The comparison between the profiles of original $\tau_{\mathrm{8\,\mu m}}$ and that predicted from  $\tau_{\mathrm{850\,\mu m}}$ are shown in Fig. \ref{fig:tau_comp}, together with the scaled-up $\tau_{\mathrm{8\,\mu m}}$.  Comparing the ratios between the scaled-up $\tau_{\mathrm{8\,\mu m}}$ and that predicted from $\tau_{\mathrm{850\,\mu m}}$, the consistency of $\sim$1 can be achieved for most of our interested core area. 
Practically, the assumption of constant foreground emission level can cause underestimation of the column densities particularly in the outer parts of the core. By inspecting the 8$\mu m$ map and the contours of $N(\mathrm{H_{2}})$ map from {\it{Herschel}} it is obvious that some bright 8$\mu m$ emission is associated with the foreground of the source. The column densities in the whole core may also be underestimated if the background level estimate is too low. The small-scale median filter method effectively removes all contribution from compact bright sources for interpolation at the core area, ignoring the fact that there may be a bunch of stars that compose the background of the cores. The uncertainties associated with back- and foreground emission level estimates are hard, if not impossible, to assess.  We resort to matching the $\tau$ profiles from the extinction method and from that the SED by a constant scaling factor. This is based on essentially the assumption that the morphological features of the $N(\mathrm{H_{2}})$ map derived by the extinction method, preserve the real profiles of the core structure without spatially varying degrees of deficit. This is largely based on what we find empirically, the constant scaling up can match the column density cuts along the core for the inner part and outer part simultaneously. The comparison between the profiles of original $\tau_{\mathrm{8\,\mu m}}$ and that predicted from  $\tau_{\mathrm{850\,\mu m}}$ are shown in Fig. \ref{fig:tau_comp}, together with the scaled-up $\tau_{\mathrm{8\,\mu m}}$.  Comparing the ratios between the scaled-up $\tau_{\mathrm{8\,\mu m}}$ and that predicted from $\tau_{\mathrm{850\,\mu m}}$, the consistency of $\sim$1 can be achieved for most of our interested core area. 

As a simple sanity check, we derive the foreground image of 8$\,\mu$m, using the scale-up 2$''$ column density map, and assume the background image estimated from SMF is relatively robust. The level of the derived foreground image, compared to the constant $I_{\mathrm{fg}}$ we assumed, is a factor of 1.0-1.6 times higher for the pixels in the core area. This is a reasonable difference in terms of possible underestimates of the foreground emission level.  

In fact, the deficit of the derived $N(\mathrm{H_{2}})$ (after smoothing) compared to that derived from SED of {\it{Herschel}} data can have different origins. For the outer parts of the core, it is mainly the intrinsic problem of extinction method being not sensitive to lower column density regime due to the way fore- and background emission level are estimated; for the central part of the core, there is possible variation of the dust opacities due to grain growth at high density. Specifically, we use the thin ice mantle without coagulation according to the models of \citet{OH94}, while the ratio of $\frac{\kappa_{\mathrm{\scriptsize{8\,\mu m}}}}{\kappa_{\mathrm{\scriptsize{850\,\mu m}}}}$ can be lower to 50$\%$ with dust coagulation and thicker ice layers {\citealt{OH94}}. This change of dust opacities can reconcile the less than 1.5 times underestimated $N(\mathrm{H_{2}})$ from 8$\mu m$ extinction, but is expected to be most significant in the central $\sim$1000 au region (e.g. \citealt{CT19a}). 

The scaling factor of L1517B and L694-2 are 1.5 and 1.8, for L429 it is $\sim$2.5. This suggests that L429 is missing the most gas column densities from the extinction method. Indeed, the ambient cloud structure of L429 shows highest level of gas column densities, reaching 5$\times$10$^{22}$\,cm$^{-3}$ as seen from the {\it{Herschel}} map.

The uncertainty of $N(\mathrm{H_{2}})$ which comes from the dust grain scattering (main aggregates) and possible emission at 8\,$\mu$m (\citealt{Lef16}, \citealt{Steinacker05}) is negligible ($\sim$0.1 MJy\,sr$^{-1}$) compared to the uncertainty of the background and foreground emission level estimates (\citealt{Pineda22a}), for a source with extended sky brightness of several times of 1\,MJy\,sr$^{-1}$ at 8\,$\mu$m. We note that this is the case for all the three cores in our sample.

\section{$T_{\mathrm{kin}}$ derived from VLA-only datacubes}\label{app:tkin_interf}
\begin{figure}
\includegraphics[scale=0.47]
{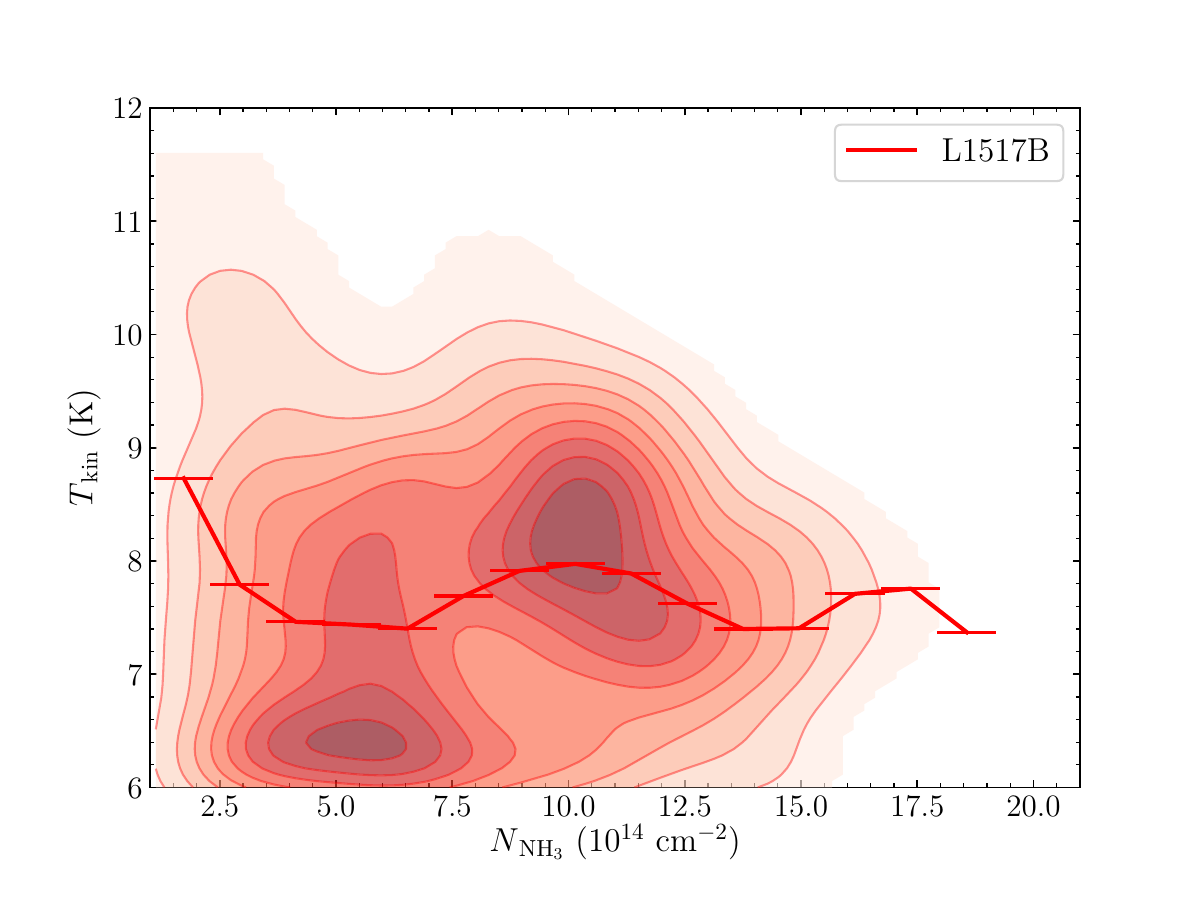}\\
\includegraphics[scale=0.47]
{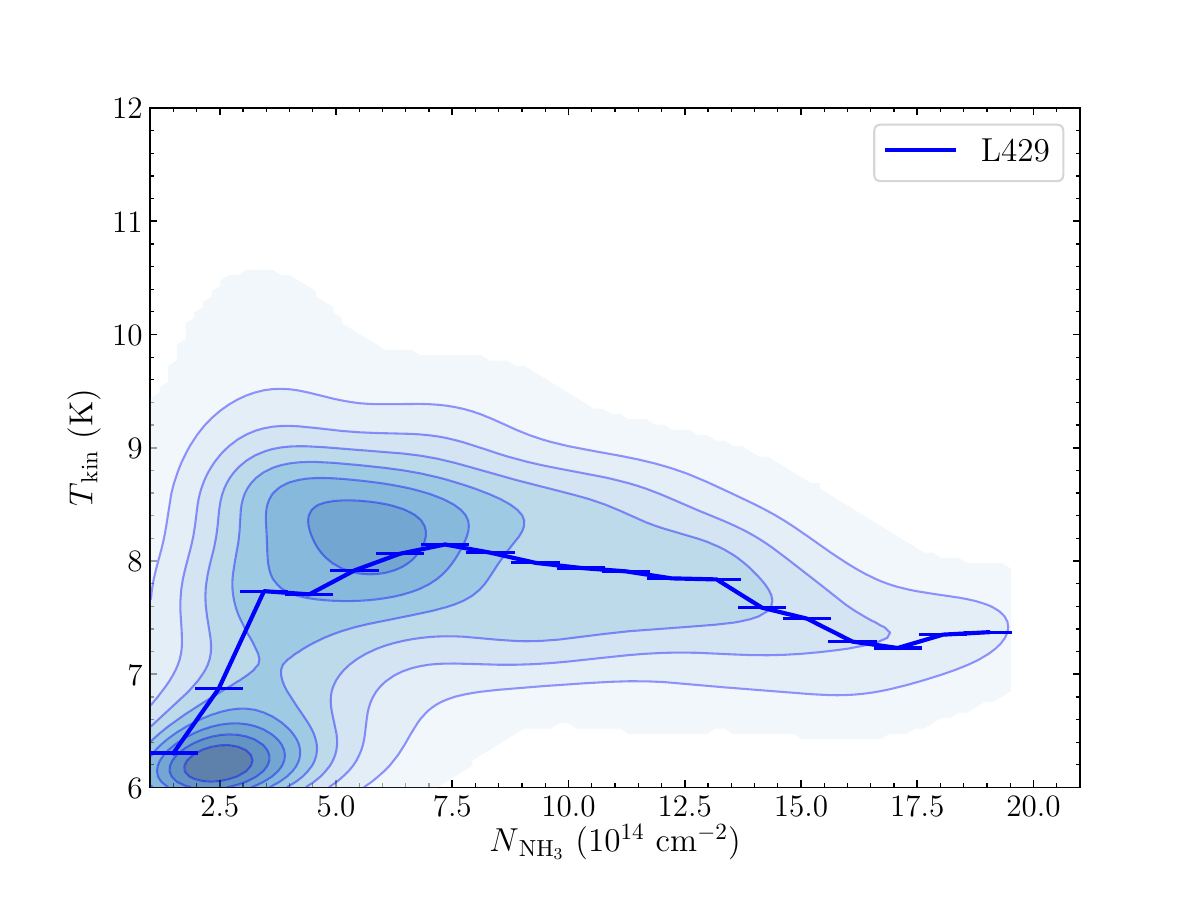}\\
\includegraphics[scale=0.47]
{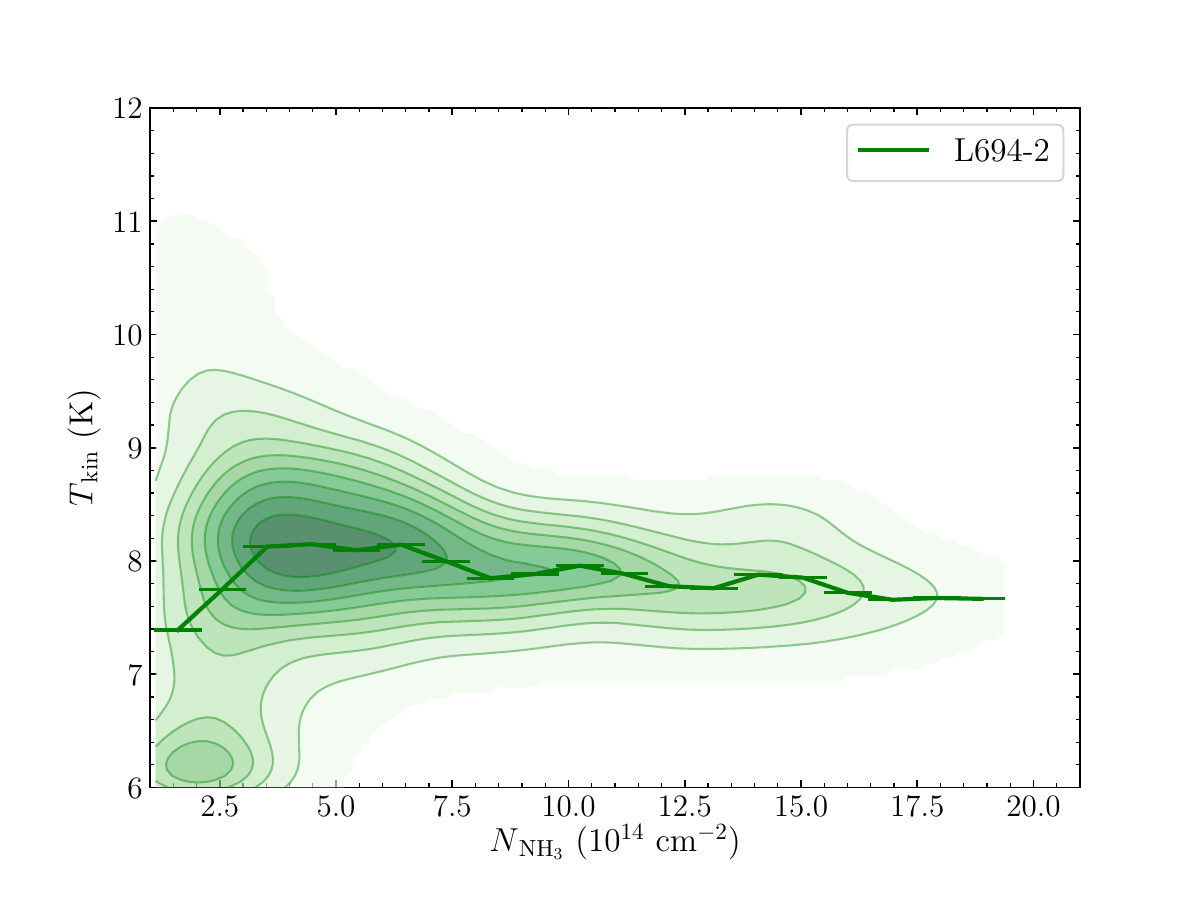}
\caption{Kinetic temperature as a function of NH$_{\mathrm{3}}$ column density for the three cores. Both quantities are derived from VLA datacube without combination with GBT data (as compared to right panel of Fig. \ref{fig:kde11}). The line segments correspond to mean values in  H$_{\mathrm{2}}$ column density bins, which are linked together. }
\label{fig:tkin_nnh3_vlaonly}
\end{figure}

In \citet{Crapsi07}, the temperature drop towards L1544 is revealed by the temperature map derived from VLA-only datacubes of NH$_{3}$ (1,1) and (2,2) lines. This means that the extended structure is missing, which presumably affects the emission of the low lying energy levels most, since these can be excited at lower density layers. The extent and degree of the missing extended structure is hard, if not impossible, to predict. We tested fitting the NH$_{3}$ models with the VLA-only NH$_{3}$ (1,1) and (2,2) lines and compare the distribution of $T_{\mathrm{kin}}$ vs. $N(\mathrm{NH_{3}})$ (as in Fig. \ref{fig:kde11}, left panel) in Fig. \ref{fig:tkin_nnh3_vlaonly}. Compared to results derived from combined data cubes, a large difference is seen towards L1517B, which shows an overall lower temperature of $\lesssim$8 K, close to that towards L429 and L694-2. With VLA-only data, still there is not a significant temperature drop of the three cores as that revealed towards L1544. The temperature variations of L429 in the VLA-only data are even more flattened than that in the combined data. This shows that the slight temperature drops seen in the combined data towards L429 and L1517B are mainly associated with (the LOS weighting of) outer gas layers of the core, where the region under influence of radiation field is better preserved.

\section{Sub-regions with secondary velocity component}\label{app:2comp}

\begin{figure*}
\begin{tabular}{p{0.25\linewidth}p{0.25\linewidth}p{0.25\linewidth}p{0.25\linewidth}}
\hspace{-1.cm}\includegraphics[scale=0.48]{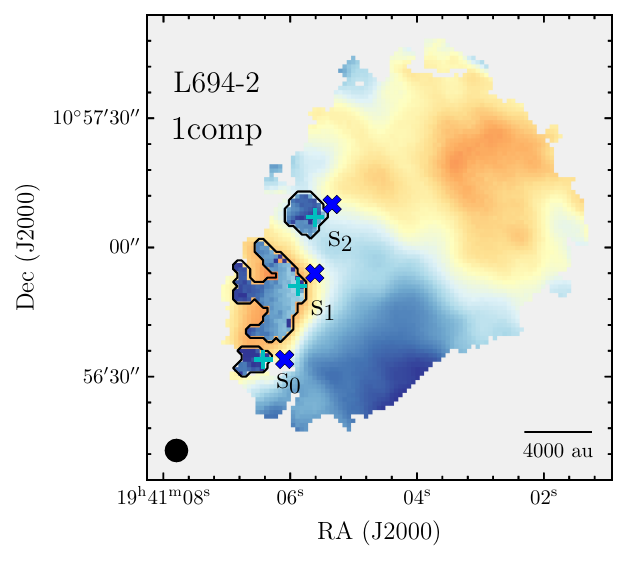}&\hspace{-0.9cm}\includegraphics[scale=0.515]{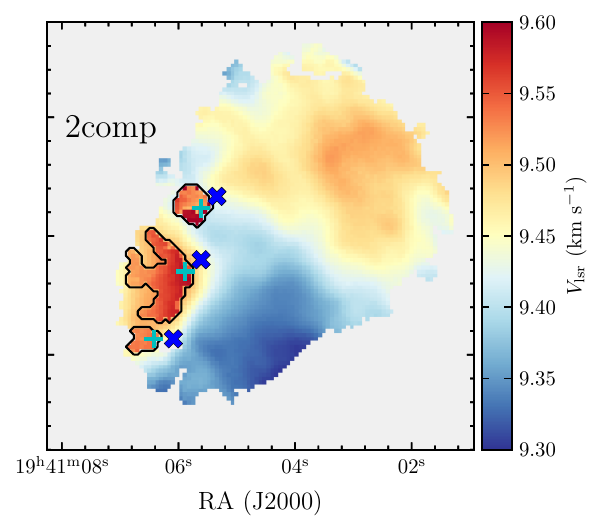}&\hspace{-0.85cm}\includegraphics[scale=0.48]{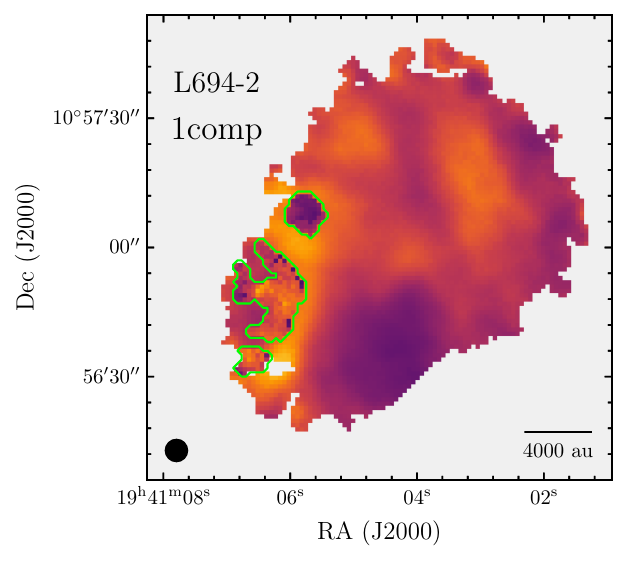}&\hspace{-0.85cm}\includegraphics[scale=0.515]{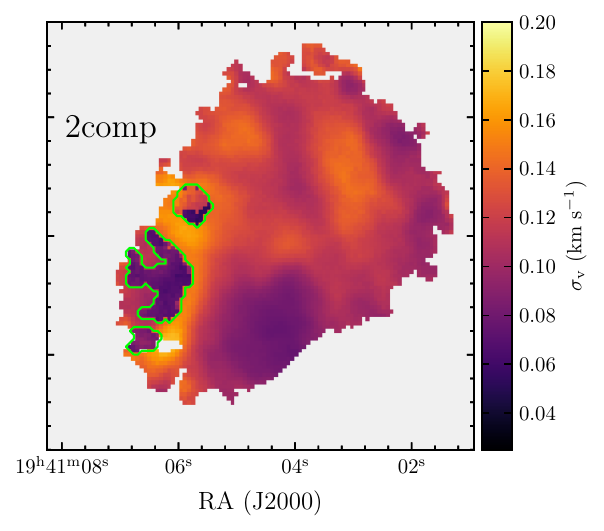}
\end{tabular}
\caption{The velocity and line-width maps of L694-2 combining sub-regions where a secondary velocity component exist and the rest regions showing only one velocity component. The two-component sub-regions are enclosed by contours, defined by $\ln K^{2}_{1}$ and goodness of fit.}
\label{fig:l694-2_2comp}
\end{figure*}

\begin{figure*}
\begin{tabular}{p{0.55\linewidth}p{0.45\linewidth}}
\hspace{-0.6cm}\includegraphics[scale=0.68]{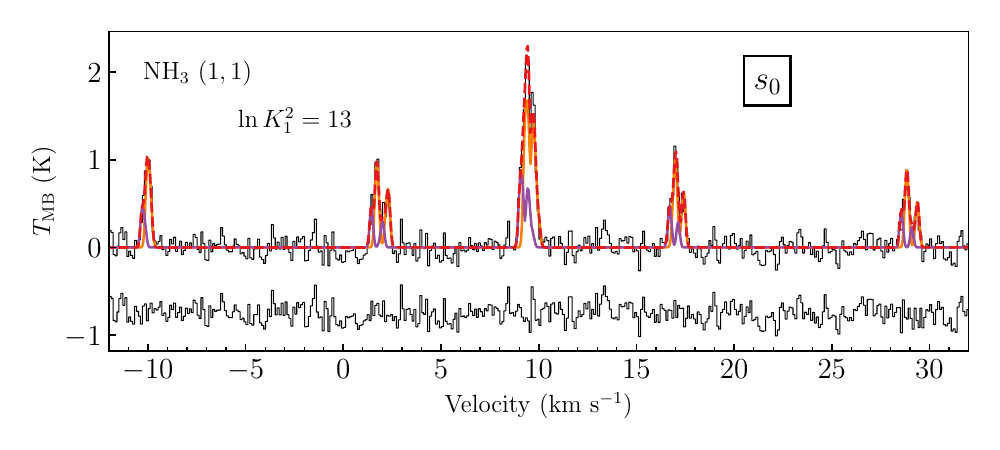}&\hspace{.1cm}\includegraphics[scale=0.68]{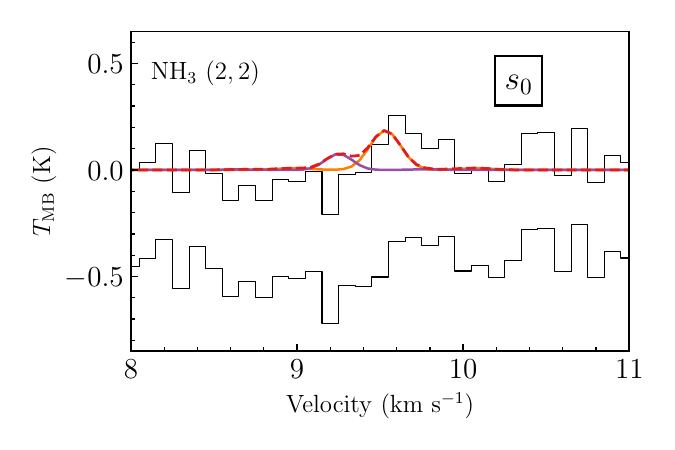}\\
\hspace{-0.6cm}\includegraphics[scale=0.68]{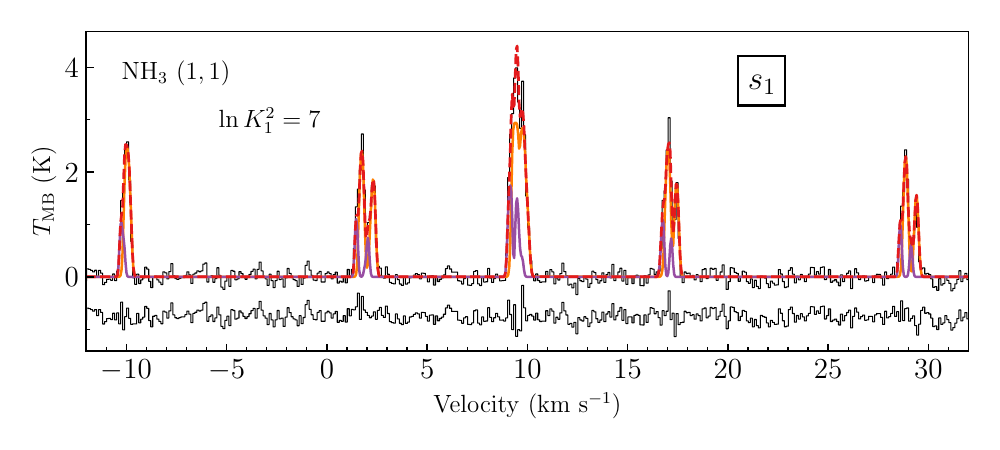}&\hspace{.1cm}\includegraphics[scale=0.68]{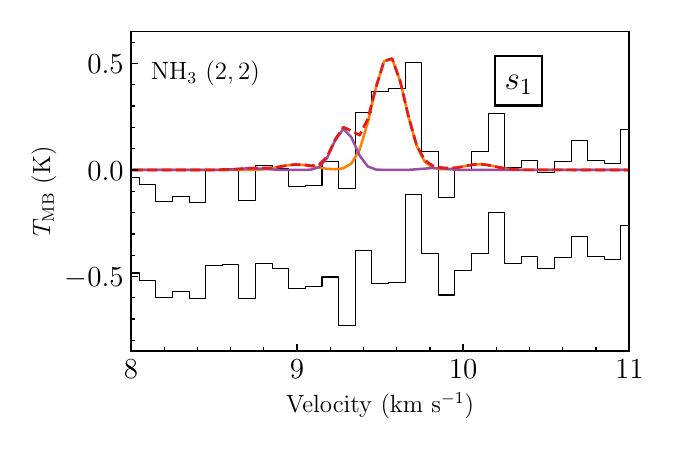}\\
\hspace{-0.6cm}\includegraphics[scale=0.68]{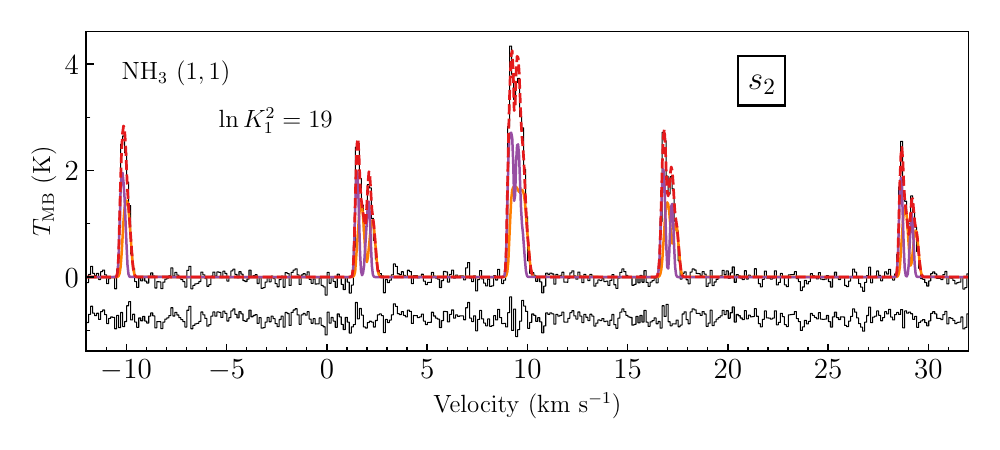}&\hspace{.1cm}\includegraphics[scale=0.68]{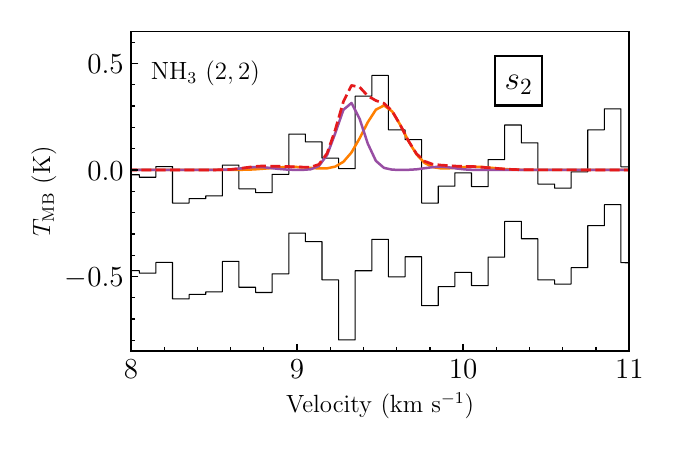}\\
\end{tabular}
\caption{Example of two-component spectra towards $s_{1}$, $s_{2}$ and $s_{3}$ regions of L694-2 as in Fig. \ref{fig:l694-2_2comp} (cyan pluses), of the NH$_{3}$ (1,1) and (2,2) lines. Red dashed lines show the overall two-component fit; orange and magenta lines shows the two component separately. The offset spectra show the residual of observed spectra minus the two-component model spectra.}
\label{fig:l694-2_2compsp}
\end{figure*}

\begin{figure*}
\begin{tabular}{p{0.55\linewidth}p{0.45\linewidth}}
\hspace{-0.6cm}\includegraphics[scale=0.68]{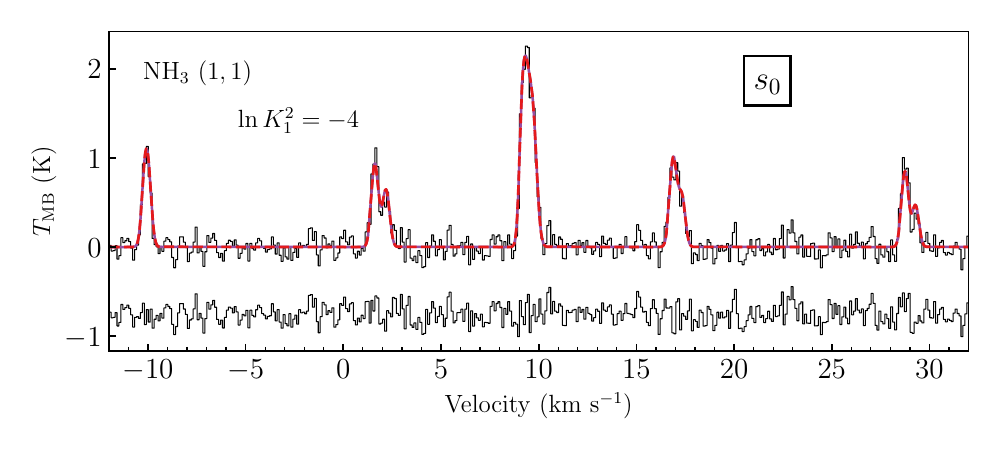}&\hspace{.1cm}\includegraphics[scale=0.68]{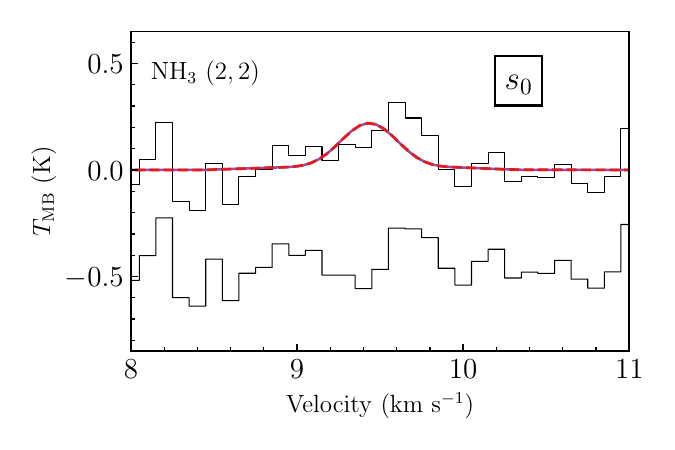}\\
\hspace{-0.6cm}\includegraphics[scale=0.68]{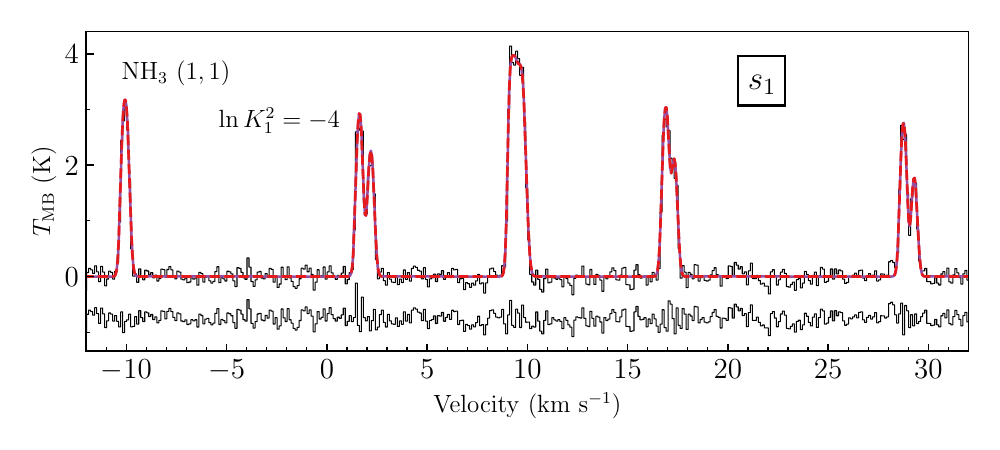}&\hspace{.1cm}\includegraphics[scale=0.68]{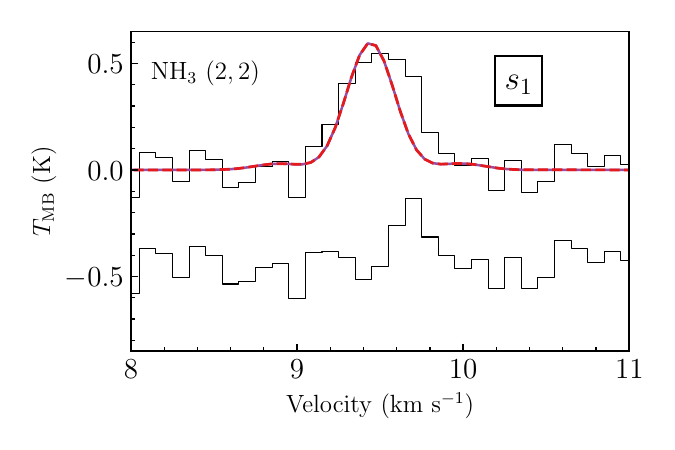}\\
\hspace{-0.6cm}\includegraphics[scale=0.68]{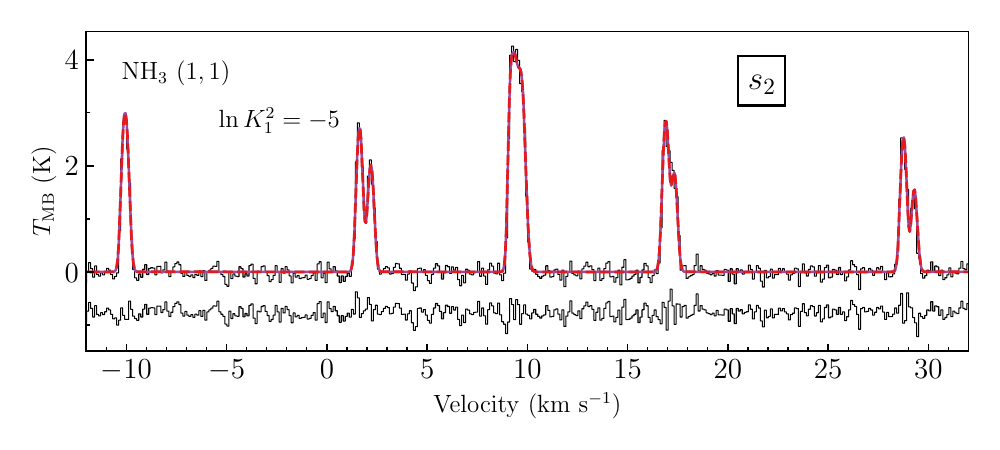}&\hspace{.1cm}\includegraphics[scale=0.68]{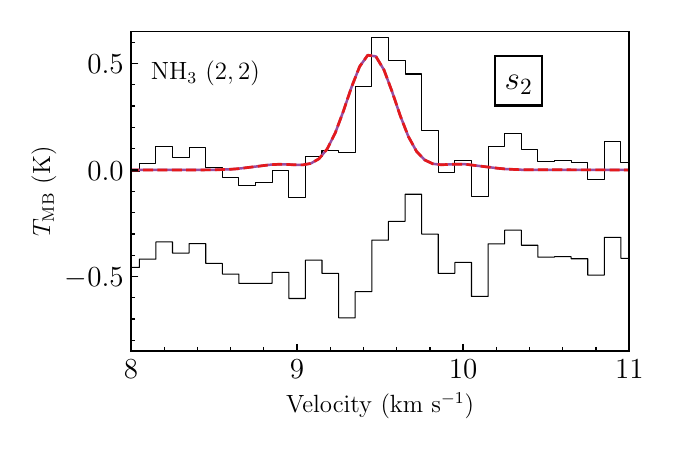}\\
\end{tabular}
\caption{Same as Fig. \ref{fig:l694-2_2compsp}, but of the spectra in the inner region of $s_{1}$, $s_{2}$ and $s_{3}$ (blue crosses in Fig. \ref{fig:l694-2_2comp}) that only show one velocity component.}
\label{fig:l694-2_1compsp}
\end{figure*}

\begin{figure*}
\begin{tabular}{p{0.25\linewidth}p{0.25\linewidth}p{0.25\linewidth}p{0.25\linewidth}}
\hspace{-1.cm}\includegraphics[scale=0.48]{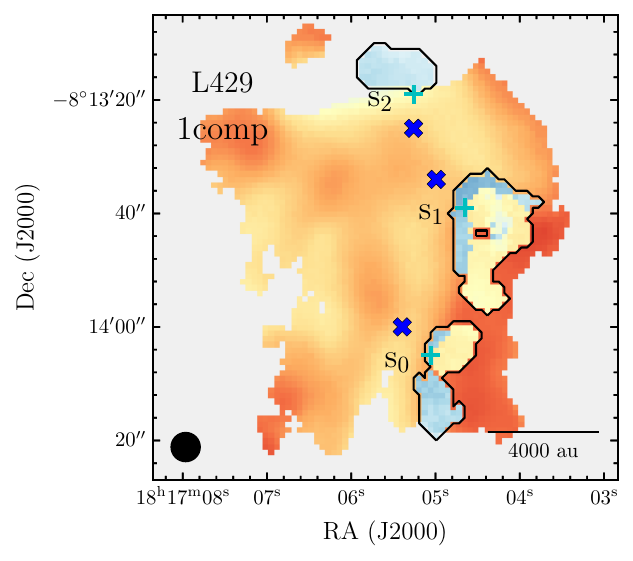}&\hspace{-0.9cm}\includegraphics[scale=0.515]{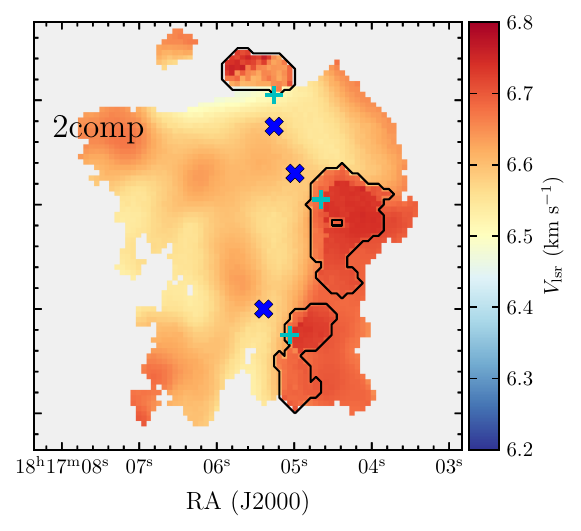}&\hspace{-0.85cm}\includegraphics[scale=0.48]{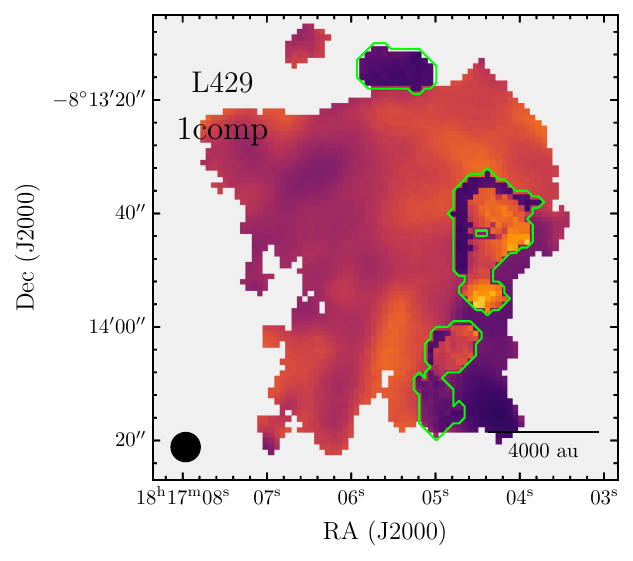}&\hspace{-0.85cm}\includegraphics[scale=0.515]{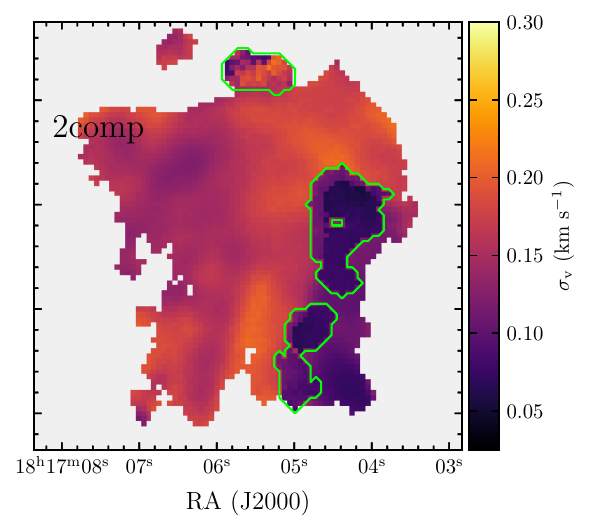}
\end{tabular}
\caption{Same as Fig. \ref{fig:l694-2_2comp}, but for core L429.}
\label{fig:l429_2comp}
\end{figure*}

\begin{figure*}
\begin{tabular}{p{0.55\linewidth}p{0.45\linewidth}}
\hspace{-0.6cm}\includegraphics[scale=0.68]{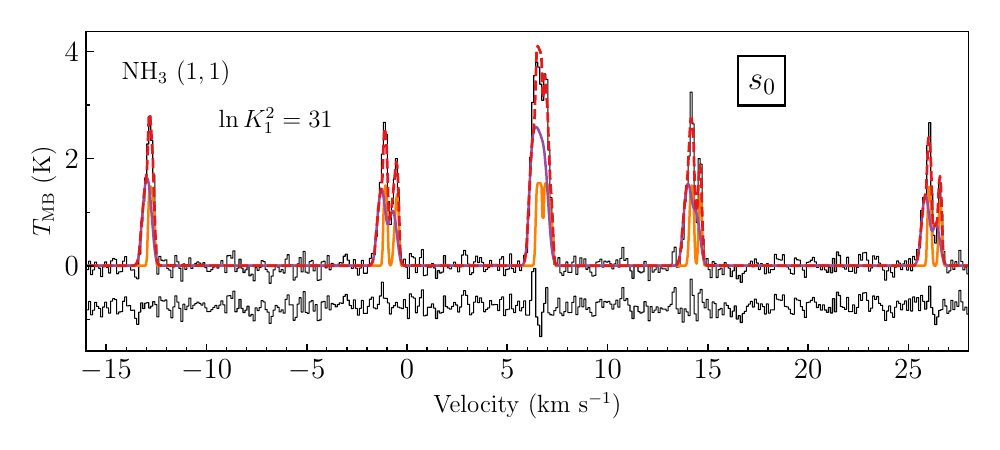}&\hspace{.1cm}\includegraphics[scale=0.68]{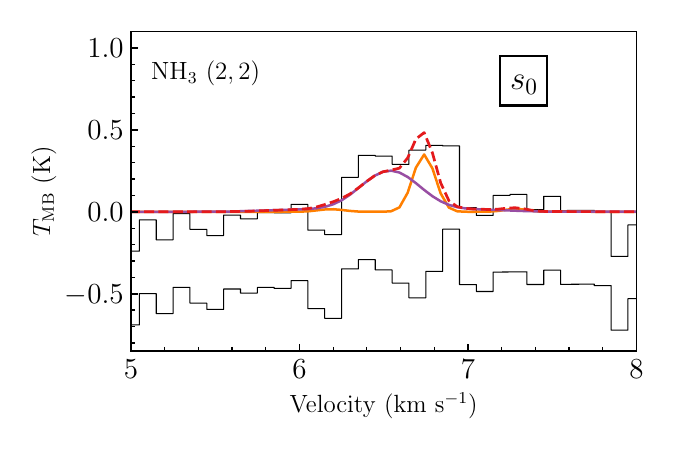}\\
\hspace{-0.6cm}\includegraphics[scale=0.68]{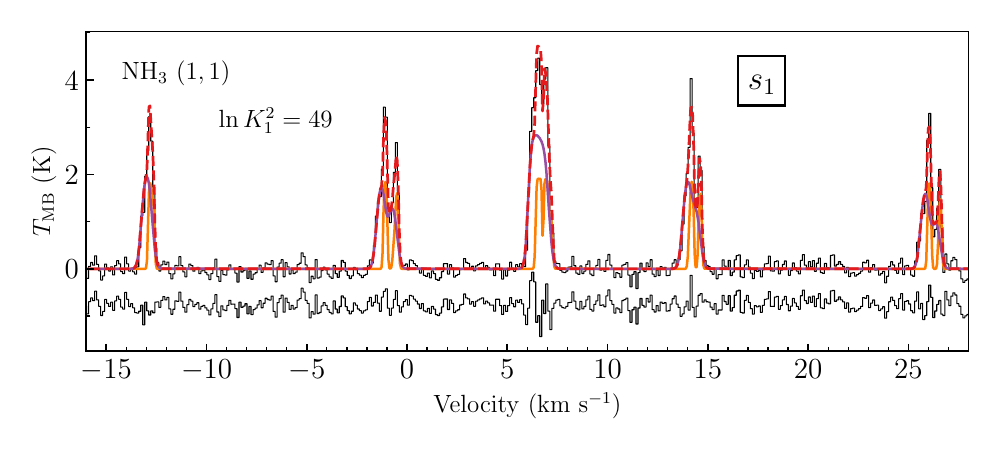}&\hspace{.1cm}\includegraphics[scale=0.68]{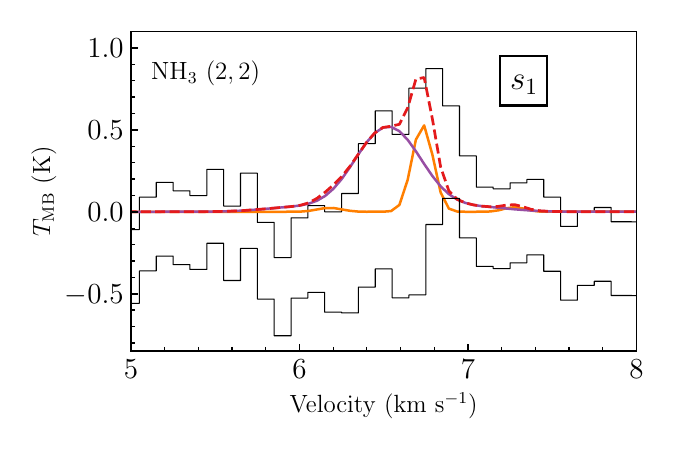}\\
\hspace{-0.6cm}\includegraphics[scale=0.68]{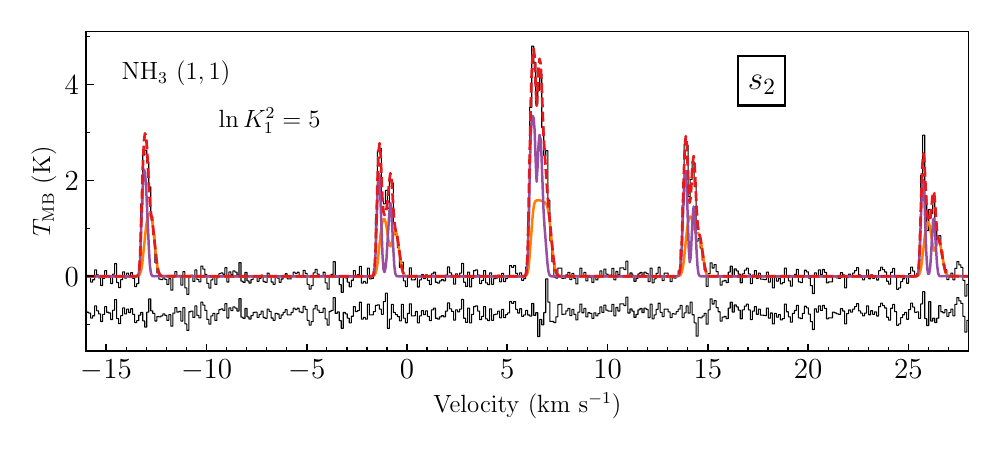}&\hspace{.1cm}\includegraphics[scale=0.68]{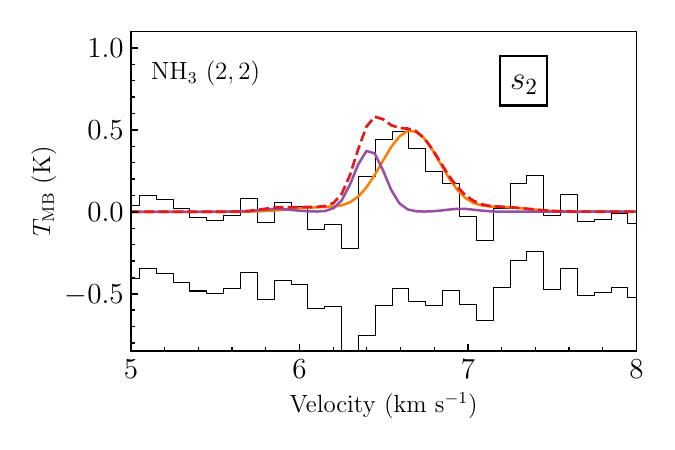}\\
\end{tabular}
\caption{Same as Fig. \ref{fig:l694-2_2compsp}, but of core L694-2.}
\label{fig:l429_2compsp}
\end{figure*}

As stated in Sect. \ref{sec:nh3_model}, for sub-regions of L694-2 and L429, there is evidence of a secondary velocity component.
We present the two-component velocity and line-width maps in Figs. \ref{fig:l694-2_2comp} and \ref{fig:l429_2comp} and the example spectra across the sub-regions that have two or one velocity component in Figs. \ref{fig:l694-2_2compsp}, \ref{fig:l694-2_1compsp}, \ref{fig:l429_2compsp}, \ref{fig:l429_2compsp}. When generating the parameter maps, for the vast region where one-component fit is statistically preferred, the fitted values from this fit are assigned to corresponding pixels. In Figs. \ref{fig:l694-2_2comp}-\ref{fig:l429_2comp} sub-regions showing two velocity components are marked by closed contours. These sub-regions are selected based on ln$K^{2}_{1}$ threshold and trimmed according to fitted errors. Compared to trimming the one-component fit in Sect. \ref{sec:nh3_model}, we used a less tight criteria: uncertainties of $T_{\mathrm{kin}}$ or $T_{\mathrm{ex}}$ larger than 2 K, $\rm{v}_{\mathrm{LSR}}$ or $\sigma$ larger than four channel widths (0.4 km\,s$^{-1}$). 

Examining Fig. \ref{fig:l429_2comp} and Fig. \ref{fig:l694-2_2comp}, it is clear that the sub-regions showing a secondary velocity component are where there are abrupt changes of the one-component velocity map. Comparing the two-component velocities with the one-component velocity of the main regions across the core, mostly, the velocity component that has a larger line-width seems to show a better consistency with main regions.  The red-shifted, smaller line-width sub-regions connect to the outermost region of the core, which is clearer in the case of L429.

\begin{figure*}
\begin{tabular}{p{0.55\linewidth}p{0.45\linewidth}}
\hspace{-0.6cm}\includegraphics[scale=0.68]{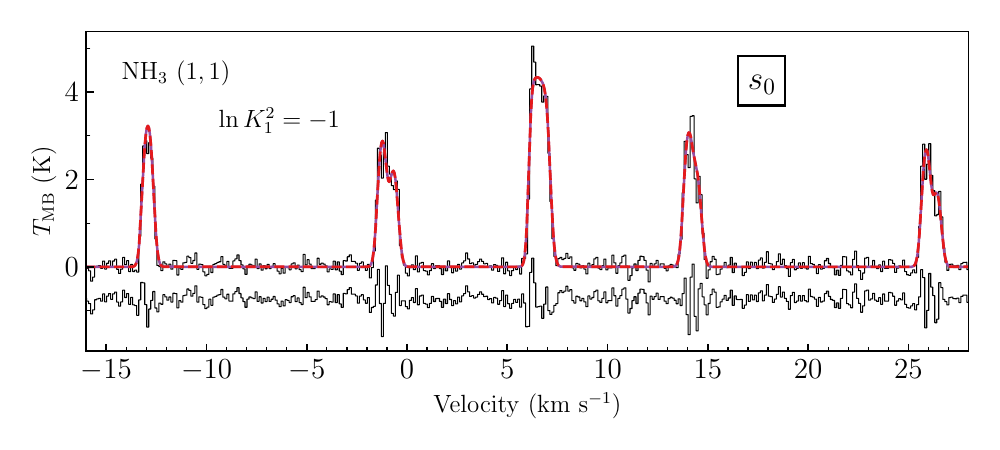}&\hspace{.1cm}\includegraphics[scale=0.68]{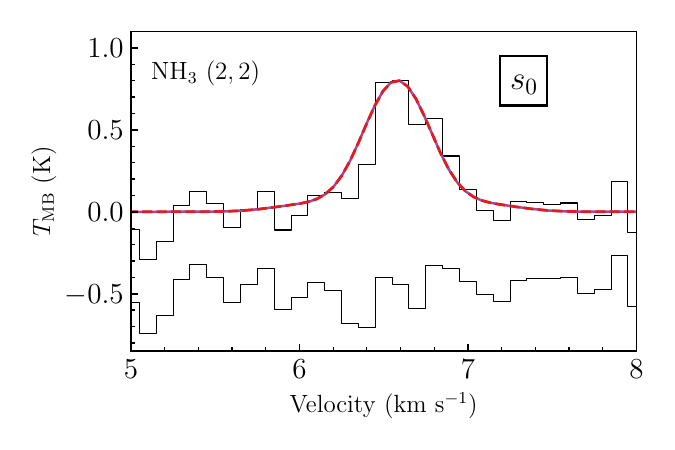}\\
\hspace{-0.6cm}\includegraphics[scale=0.68]{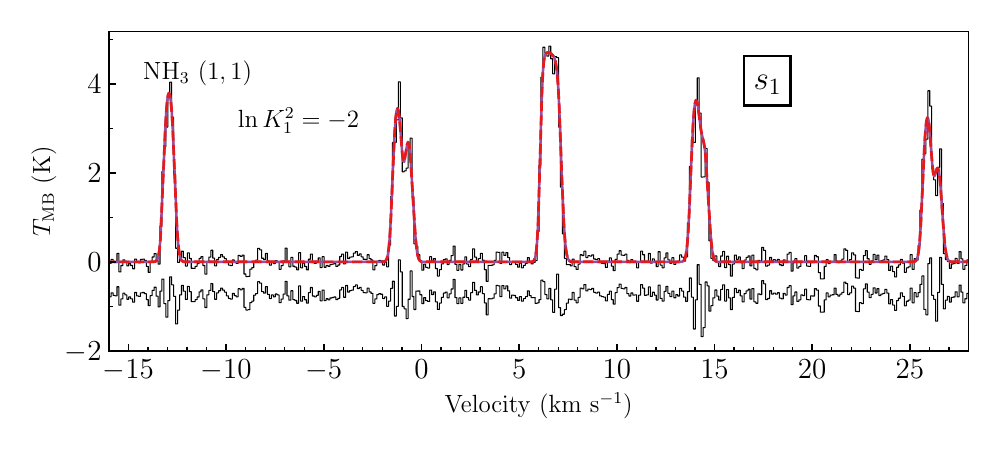}&\hspace{.1cm}\includegraphics[scale=0.68]{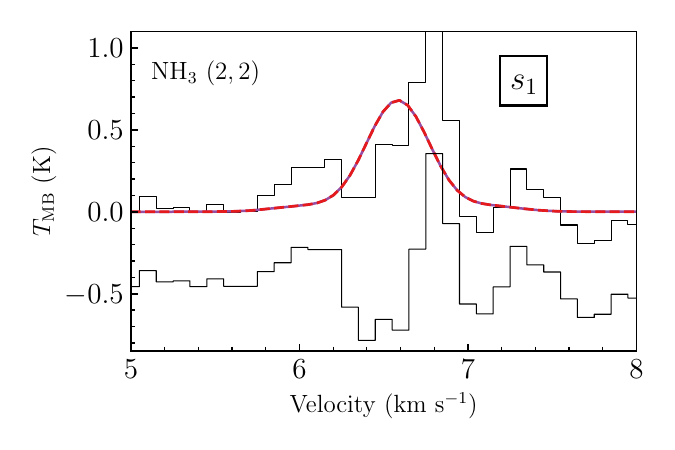}\\
\hspace{-0.6cm}\includegraphics[scale=0.68]{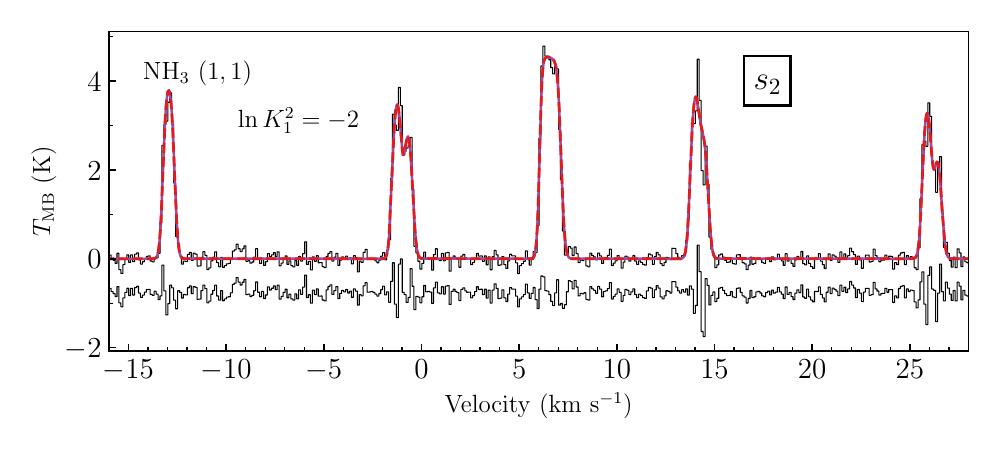}&\hspace{.1cm}\includegraphics[scale=0.68]{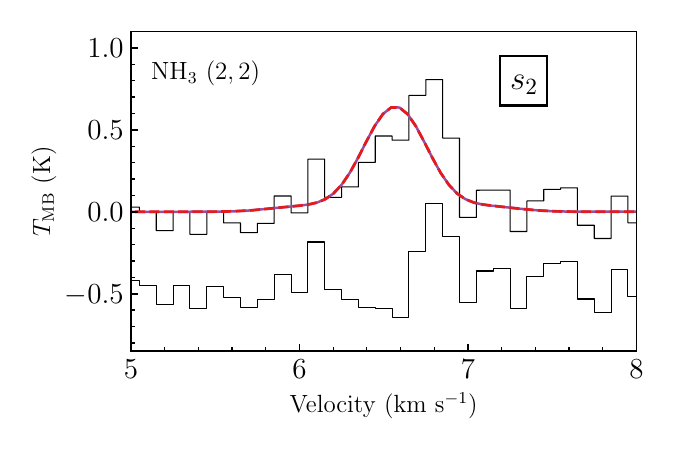}\\
\end{tabular}
\caption{Same as Fig. \ref{fig:l694-2_1compsp}, but of the core L429.}
\label{fig:l429_1compsp}
\end{figure*}

\end{appendix}
\end{document}